\documentclass[twocolumn]{aastex7}
\usepackage{amsmath}
\usepackage{graphicx,subcaption}
\usepackage{hyperref,comment}
\newcommand{\Rs}{$\rm{R}_{\odot}$}

\newcommand{\alf}{Alfv\'{e}n }

\submitjournal{ApJ}
\shortauthors{Sachdeva et al.}

\begin{document}

\title{Evolution of Coronal Mass Ejections in Different Data-Driven Solar Wind Conditions}

\author[orcid=0000-0001-9114-6133,sname='Sachdeva']{Nishtha Sachdeva}
\affiliation{Department of Climate and Space Sciences and Engineering, University of Michigan, Ann Arbor, MI 48109, USA}
\email[show]{nishthas@umich.edu}

\author[orcid=0000-0003-1674-0647]{Zhenguang Huang}
\email[hide]{zghuang@umich.edu}
\affiliation{Department of Climate and Space Sciences and Engineering, University of Michigan, Ann Arbor, MI 48109, USA}

\author[orcid=0000-0001-8459-2100]{Gabor Toth}
\email[hide]{gtoth@umich.edu}
\affiliation{Department of Climate and Space Sciences and Engineering, University of Michigan, Ann Arbor, MI 48109, USA}

\author[orcid=0000-0001-7075-6530]{Hongfan Chen}
\email[hide]{chenhf@umich.edu}
\affiliation{Department of Mechanical Engineering, University of Michigan, Ann Arbor, MI 48109, USA}

\author[orcid=0000-0003-0472-9408]{Ward B. Manchester}
\email[hide]{chipm@umich.edu}
\affiliation{Department of Climate and Space Sciences and Engineering, University of Michigan, Ann Arbor, MI 48109, USA}

\author[orcid=0000-0001-5260-3944
]{Bart van der Holst}
\email[hide]{bartvand@bu.edu}
\affiliation{Astronomy Department, Boston University, Boston, MA 02215, USA}

\begin{abstract}
Numerical models of the solar wind and coronal mass ejections (CMEs) utilize photospheric magnetic field observations to prescribe the inner boundary conditions for the plasma solutions. These magnetic field data are available to the community through various observational instruments, prepared via different methodologies and/or flux-transport models. The solar wind solution driven by these maps provides the ambient plasma environment into which CMEs travel, coupling, and interacting with the surrounding plasma and governing the CME evolution and propagation in the solar corona and inner heliosphere. In this work, we use different input magnetic field maps for the same time period to drive the global \alf Wave Solar atmosphere Model (AWSoM). We obtain the ambient solar wind conditions and compare the plasma properties and magnetic morphology in the coronal domain to study the influence of the input maps. To understand how the resulting coronal solutions impact CMEs, we launch eruptions described by analytical flux ropes into these data-driven solutions and compare their evolution in the coronal domain (up to 24 \Rs\ radially). The CMEs achieve varying speeds, deceleration rates, propagation directions, mass and energies while coupling with the background solar wind. We quantify these differences to show that the different input driving maps can significantly impact the simulated CME propagation in the solar wind plasma. This also highlights the importance of understanding the uncertainties associated with data-driven modeling that become increasingly important in operational models and space weather prediction.
\end{abstract}


\section{Introduction}\label{sec:Intro}
Space weather prediction is largely dependent on how quickly and accurately models can simulate and predict the solar wind conditions, and the arrival and impact of Coronal Mass Ejections (CMEs) at Earth (and beyond). CMEs are major drivers of geomagnetic activity \citep{Gosling:1991,Howard:2006,Gopalswamy:2016,Lamy:2019,Gopalswamy:2018}. 
As they propagate out from the solar corona, CMEs evolve by interacting with the interplanetary medium \citep{Gopalswamy:2000,WangCME:2004,Sachdeva:2015,Sachdeva:2017}, sometimes driving shocks \citep{Vourlidas:2003,Gopalswamy:2001} that accelerate solar energetic particles (SEPs) that can be observed at Earth  and throughout the heliosphere \citep{Zhang:2017, Richardson:2015,Gopalswamy:2017}. Therefore, a comprehensive understanding of the evolution of CMEs and their interaction with the solar wind plasma they propagate through is crucial for improving the space weather prediction capabilities.

Significant efforts have been made to model the solar wind and CMEs using first-principles-based numerical models \citep{Mikic:1999,Cohen:2007,Lionello:2013,Downs:2010,Jin:2017a,Torok:2018, Linan:2025, Provornikova:2024, Mayank:2022} and (semi-) empirical approaches \citep{Cargill:2004,Temmer:2011,Lugaz:2013,Vrsnak:2013,Kay:2022}. Some modeling techniques solve the magnetohydrodynamics (MHD) equations to simulate the solar wind and CMEs in the solar corona and couple with the heliosphere \citep{Lionello:2009,Riley:2001,Toth:2012swmf, Manchester:2006, Usmanov:2018,vanderHolst:2014awsom,Linker:2024} while others may introduce a CME into a solar wind background prescribed by empirical models \citep{Wang:1995} or use a global MHD coronal solution \citep{Odstrcil:2003enlil}. Many flux rope models have been developed to estimate the erupting structure of CMEs \citep{Gibson:1998, Titov:1999, Titov:2014, Titov:2018} along with models of sheared arcade formed by helicity condensation \citep{Antiochos:2013,Dahlin:2019} and flux emergence \citep{Manchester:2003, Manchester:2004b}. These configurations have been widely used to model CMEs in MHD simulations by initiating a flux rope near the Sun \citep{Jin:2017b,Torok:2018,Manchester:2008b,Dahlin:2022,Liu:2025} or in the inner heliosphere \citep{Pomoell:2018,Verbeke:2019, Baratashvili:2025, Provornikova:2024, Mayank:2024}.

Most physics-based as well as (semi-) empirical models of the solar wind plasma and CMEs utilize observations of the radial component of the photospheric magnetic fields of the Sun to specify the inner boundary conditions, and drive the solar wind solutions. These observations are however limited, often incomplete and require assumptions to be made for the regions where data are unavailable or unreliable \citep{Temmer:2011, Bertello:2014}. Several factors contribute to this uncertainty, including instrumental noise, variable observational conditions, and limited coverage of the solar surface — especially near the poles. Additionally, as the radial field becomes aligned with the plane of the sky, field measurements near the solar limb are significantly reduced. These limitations can introduce considerable uncertainty into model results leading to unreliable predictions of the global solar wind and CME structure, including the orientation and variation of the interplanetary magnetic field that plays a dominant role in determining the geoeffectiveness of Earth-impacting space weather events. Therefore, it is important to assess the impact of the magnetic field observations on the solar wind structure and the properties of the CME propagating through it.

As a CME propagates through the solar wind, the magnetic morphology of the solar wind and the plasma speed and density impact the evolution of the CME through complex interactions. These impacts have been studied in both observations \citep{Temmer:2011, Temmer:2021} and simulations \citep{Riley:1997, Odstrcil:2004, Jin:2022}. The magnetic forces on an erupting flux rope in the corona due to active regions or large coronal holes can cause strong deflections from their radial trajectories \citep{Kay:2015, Lugaz:2012, Shen:2011, Mostl:2015, Cremades:2006, Gopalswamy:2016}. The solar wind can accelerate or decelerate a CME depending on their relative speeds\citep{Sachdeva:2015, Sachdeva:2017, Cargill:2004, Vrsnak:2013, Subramanian:2012}. The surrounding solar wind can also cause distortion/deformation and inhibit expansion of the propagating flux rope structure \citep{Manchester:2004,Riley:2004, Kilpua:2009, Shen:2011,WangCME:2004,Manchester:2017,Isavnin:2014,Kay:2015}. Numerical simulations have also been used to study key features during CME propagation, including magnetic erosion \citep{Manchester:2014b, Lugaz:2013,Hosteaux:2021}, deflection \citep{Zuccarello:2012, Zhou:2013, Kay:2015} and rotation \citep{Lynch:2009, Kliem:2012, Fan:2004}. \cite{Scolini:2021} presented the first evidence that CME structures propagating through different solar wind backgrounds develop different complexity evolutionary patterns, based on numerical simulations employing a spheromak flux rope model.

In this work, we study how the simulated properties of the solar wind plasma and CME evolution vary when a numerical model is driven by different magnetic field observations for the same time period. Specifically, we use different magnetic field maps for Carrington rotation (CR) CR2123, centered on 2012-05-12 to prescribe the inner boundary conditions for the three-dimensional (3D) global \alf Wave Solar atmosphere Model (AWSoM, \citet{vanderHolst:2014awsom, Gombosi:2021}) and simulate the ambient solar wind conditions. Then we launch a flux rope CME into these solar wind plasma solutions (driven by various magnetic field maps) to study the differences in their propagation and evolutionary properties. Sections \ref{sec:Model} and \ref{sec:Maps} describe the numerical models and magnetic field data used in this study. Section \ref{sec:Sim} outlines the simulation set up followed by the results of the comparative analysis of solar wind and CME evolution in the different data-driven cases in Section \ref{sec:Results}. Finally, Section \ref{sec:conc} outlines the conclusions and summarizes the work.
\begin{figure*}[th!]
\centering
\includegraphics[trim=10 150 100 65, clip, width=8cm,height=4.cm]{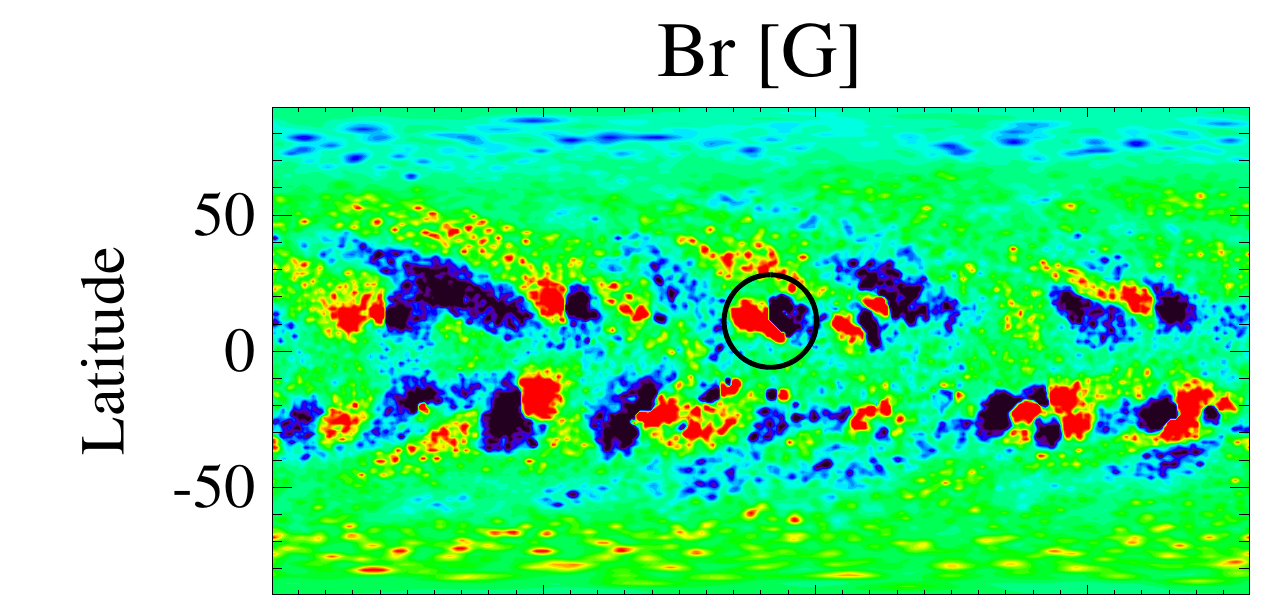}\hspace{0.7cm}
\includegraphics[trim=135 150 30 65, clip, width=7cm,height=4.cm]{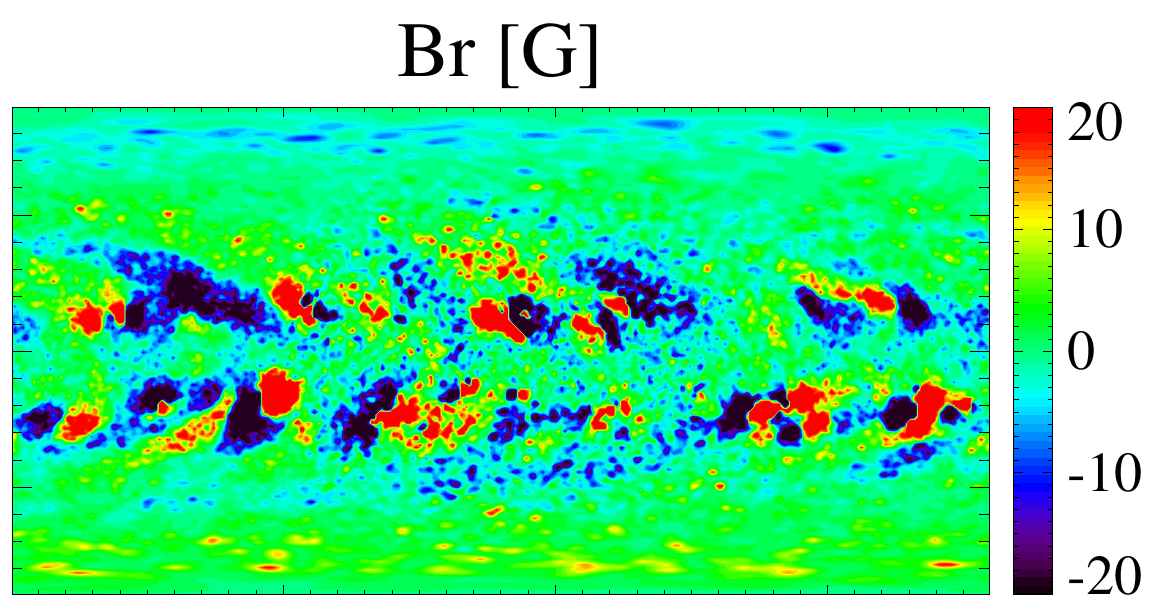}\\
\hspace{1.0cm}(a) Case 1: ADAPT GONG 
\hspace{3.6cm}(b) Case 2: ADAPT HMI\\
\hspace{0.05cm}\includegraphics[trim=10 30 100 108, clip, width=8cm,height=5.cm]
{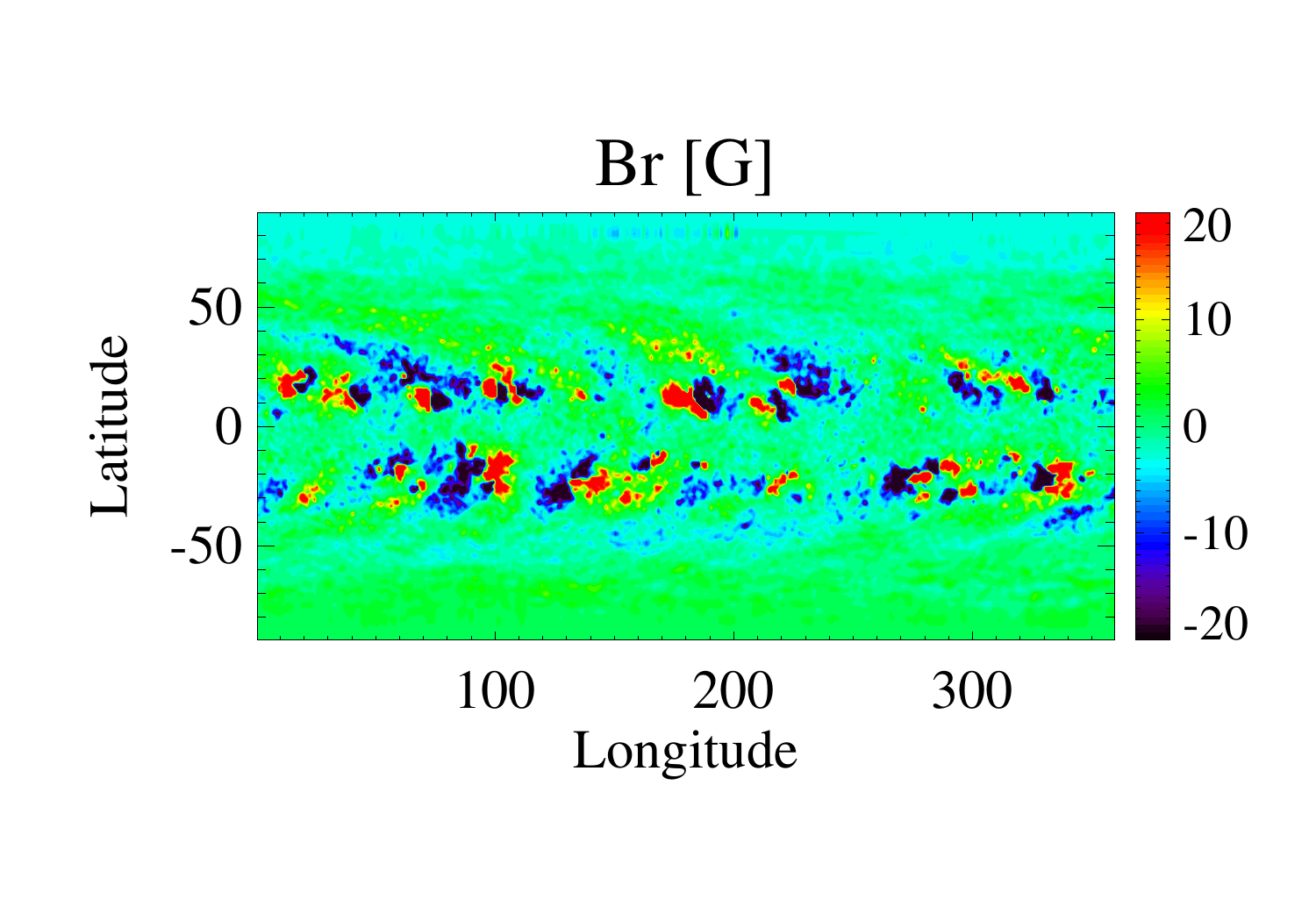}\hspace{0.7cm}
\includegraphics[trim=135 30 30 108, clip, width=7cm,height=5.cm]{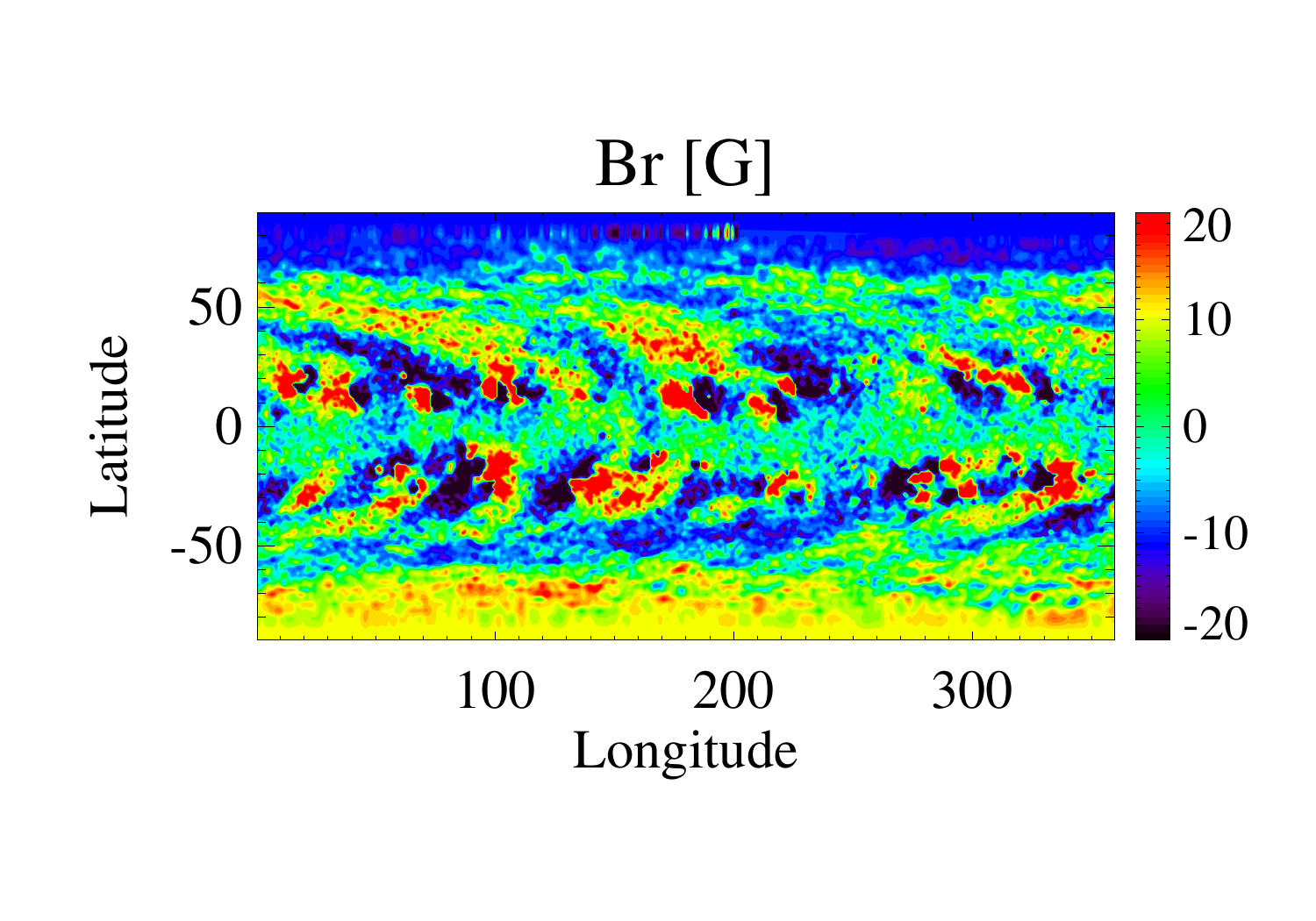}\\
\vspace{-0.5cm}
\hspace{1.5cm}(c) Case 3: GONG 
\hspace{4.cm}(d) Case 4: polar enhanced GONG 
\caption{ADAPT GONG, ADAPT HMI, GONG and polar enhanced GONG magnetic field maps showing the radial photospheric magnetic field observations for CR2123 in panels a, b, c and d, respectively. The $B_{r}$ field range of $\pm$ 20\,G is chosen to highlight the features on the maps. The active region circled in panel (a) indicates the location of flux rope CME insertion, which is same for all the maps.}\label{fig:Maps}
\end{figure*}
\begin{table*}[t!]
\centering
\caption{Comparison of input magnetic field maps.}\label{tbl0}
\begin{tabular}{|c|c|c|c|c|c|}
\hline
\multicolumn{4}{|c|}{} & \multicolumn{2}{c|}{Total Unsigned Flux}\\
\hline
Case & Input Map & $B_{r}$ Range & Dipole Moment & r=1 \Rs &  r=2.5 Rs\\
 &  & [G]& [$10^{26}$\,G\,m$^{3}$] & \multicolumn{2}{c|}{[$10^{23}$\rm{Mx}]}\\
 \hline
1 & ADAPT GONG & (-1153.9, 569.7) & 3.9 & 5.1 & 2.8\\
\hline
2 & ADAPT HMI & (-1589.2, 850.9)& 3.6 & 5.4 & 3.6\\
\hline
3 & GONG & (-781.5, 603.7) & 1.6 & 3.1 & 1.5\\
\hline
4 & polar enhanced GONG & (-834.3, 646.3) & 7.2 & 6.5 & 4.4\\
\hline
\end{tabular}
\end{table*}

\begin{figure*}
\centering
\begin{tabular}{cccc}
\includegraphics[trim=40 50 10 50, clip, width=0.2625\textwidth]{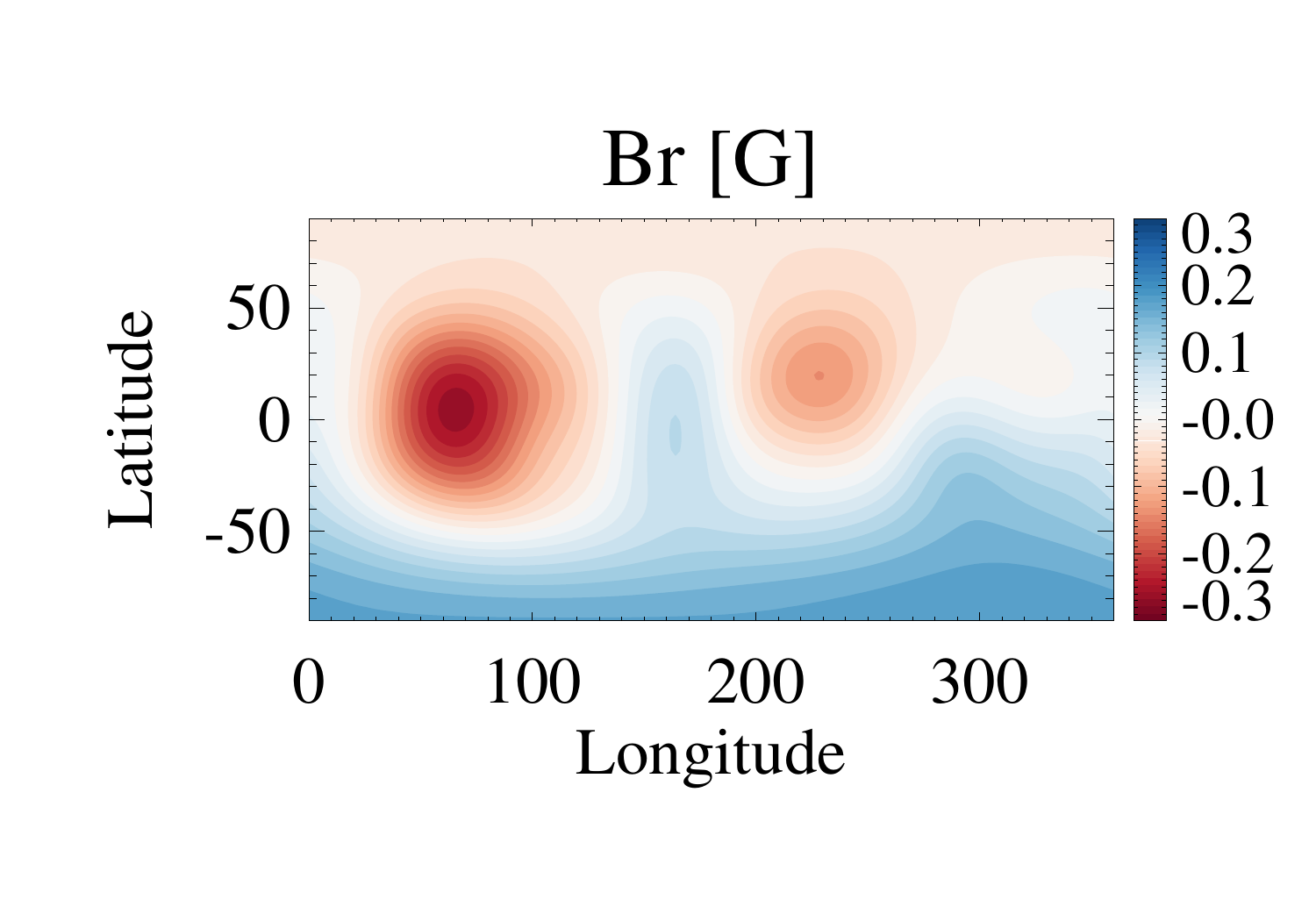} &
\hspace{-0.3cm}\includegraphics[trim=100 50 10 50, clip, width=0.24\textwidth]{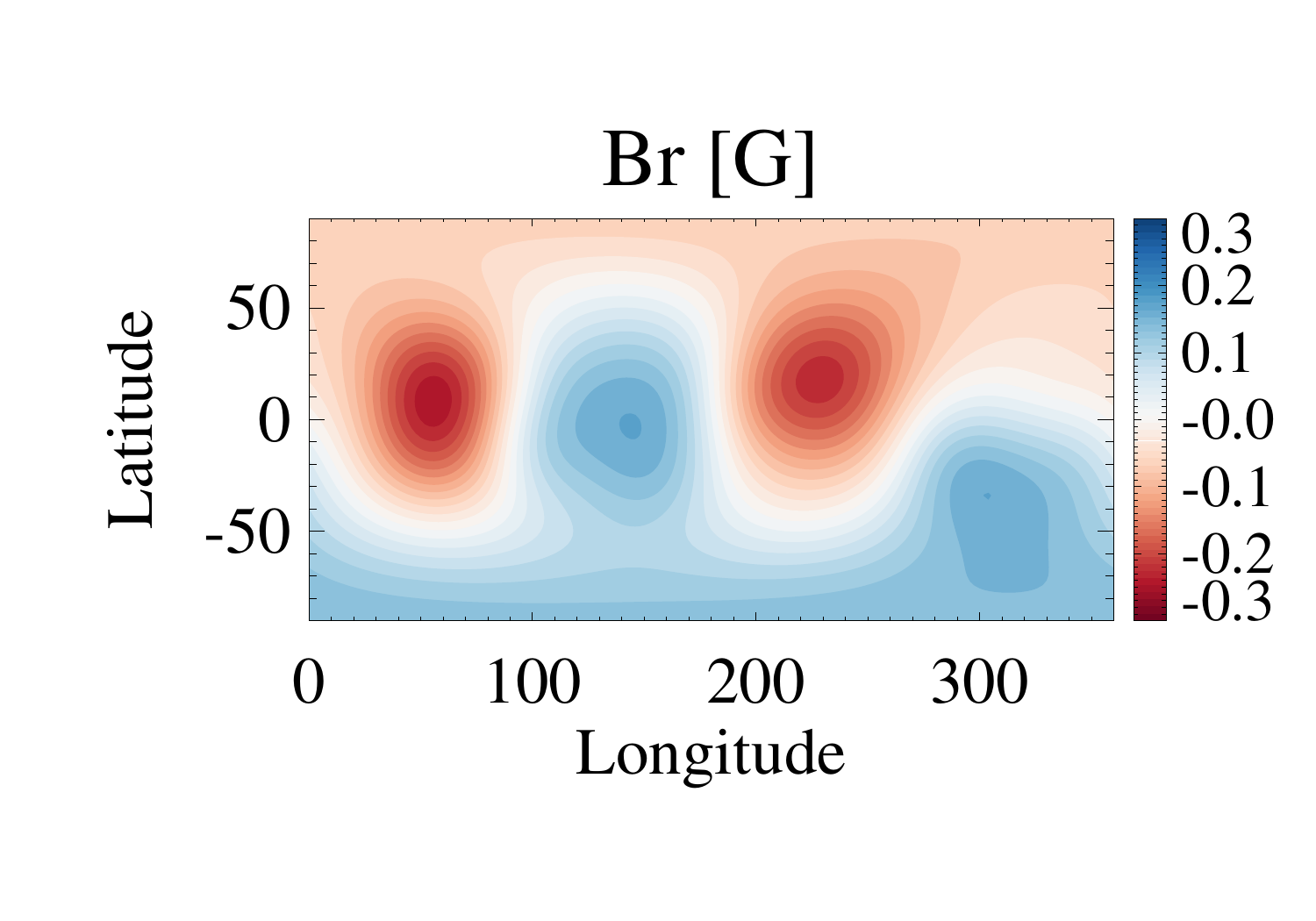} &
\hspace{-0.3cm}\includegraphics[trim=100 50 10 50, clip, width=0.24\textwidth]{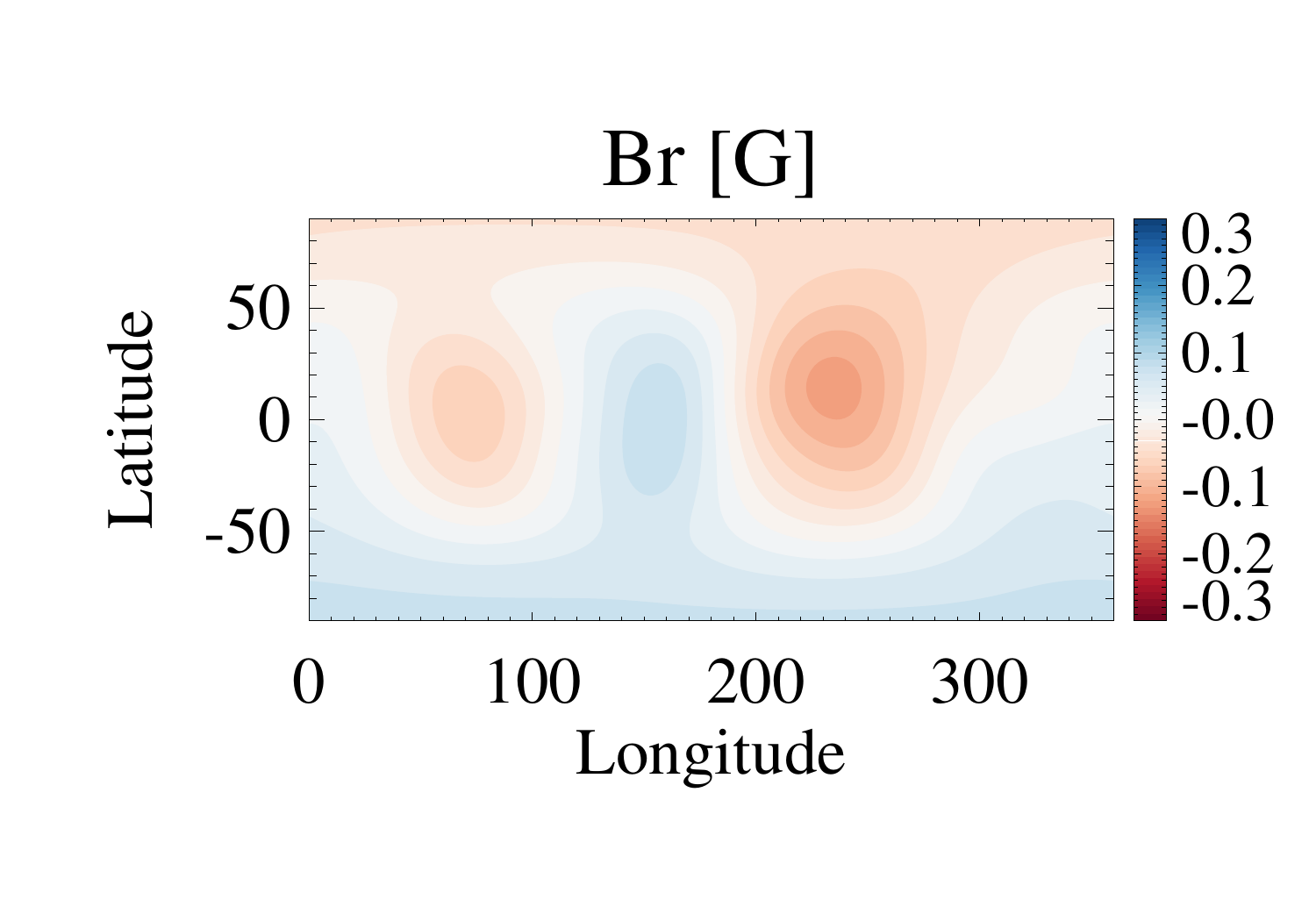} &
\hspace{-0.3cm}\includegraphics[trim=100 50 10 50, clip, width=0.24\textwidth]{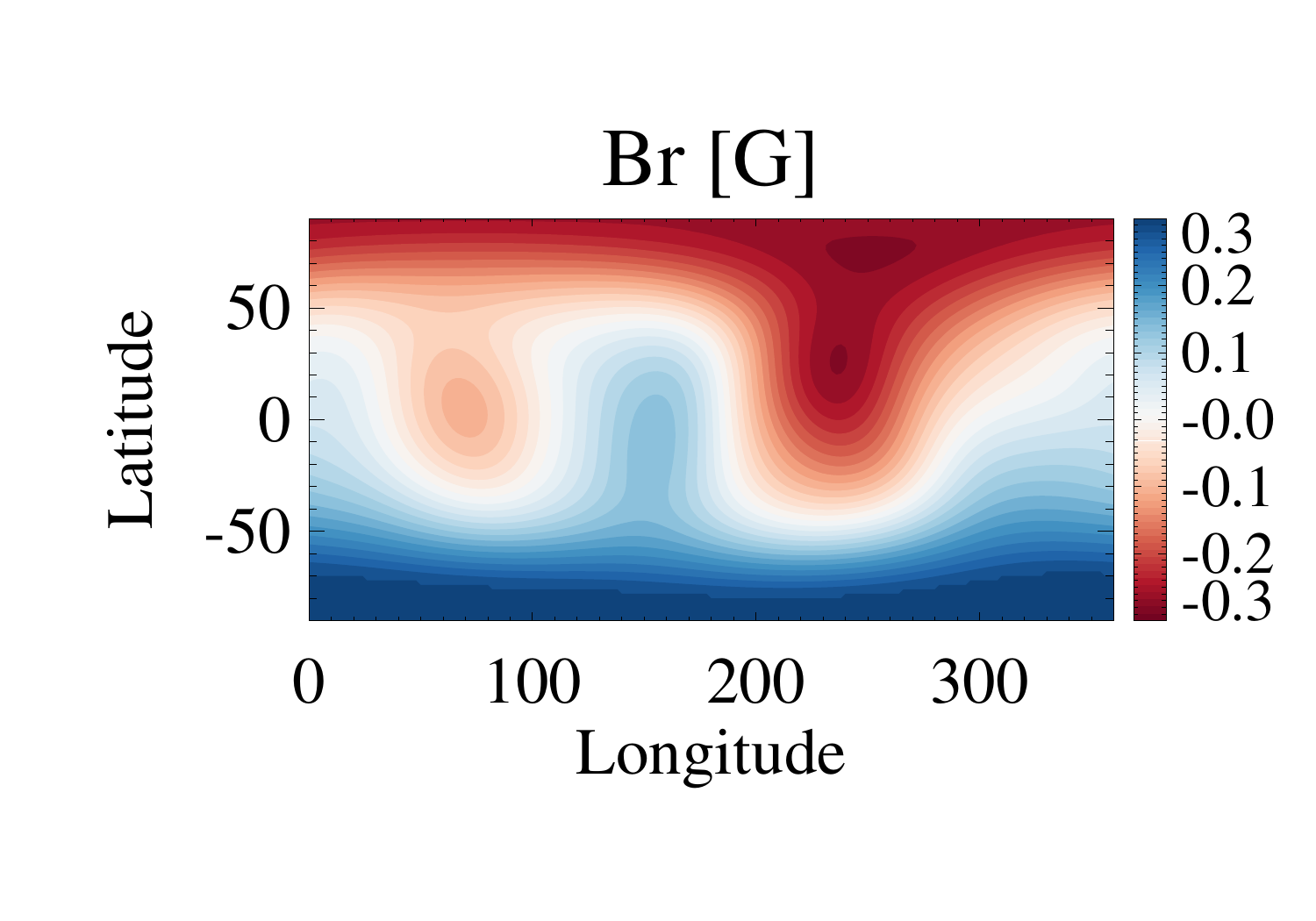} \\
\includegraphics[trim=40 50 10 50, clip, width=0.2325\textwidth]{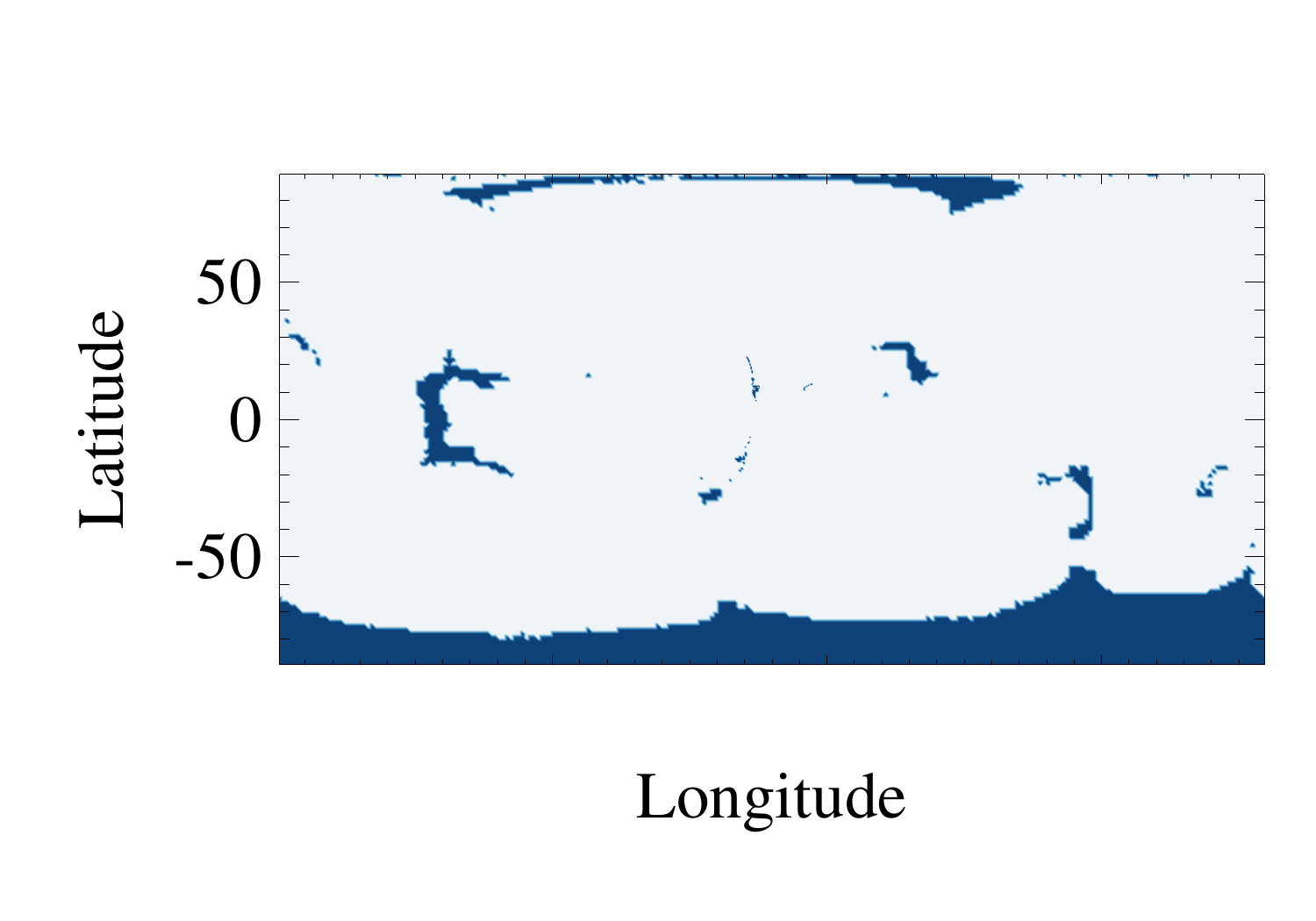} &
\hspace{-0.3cm}\includegraphics[trim=100 50 10 50, clip, width=0.21\textwidth]{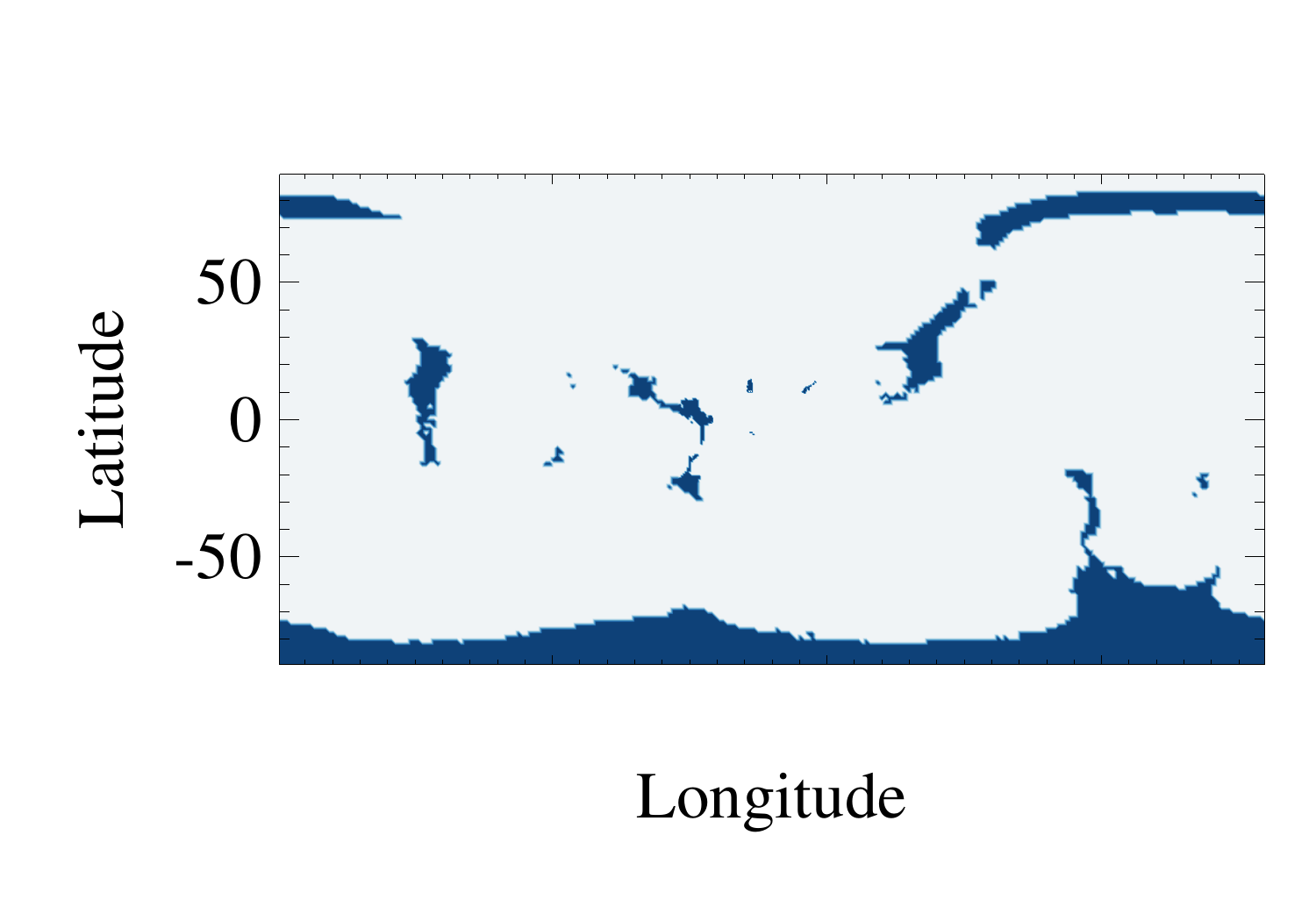} &
\hspace{-0.3cm}\includegraphics[trim=100 50 10 50, clip, width=0.21\textwidth]{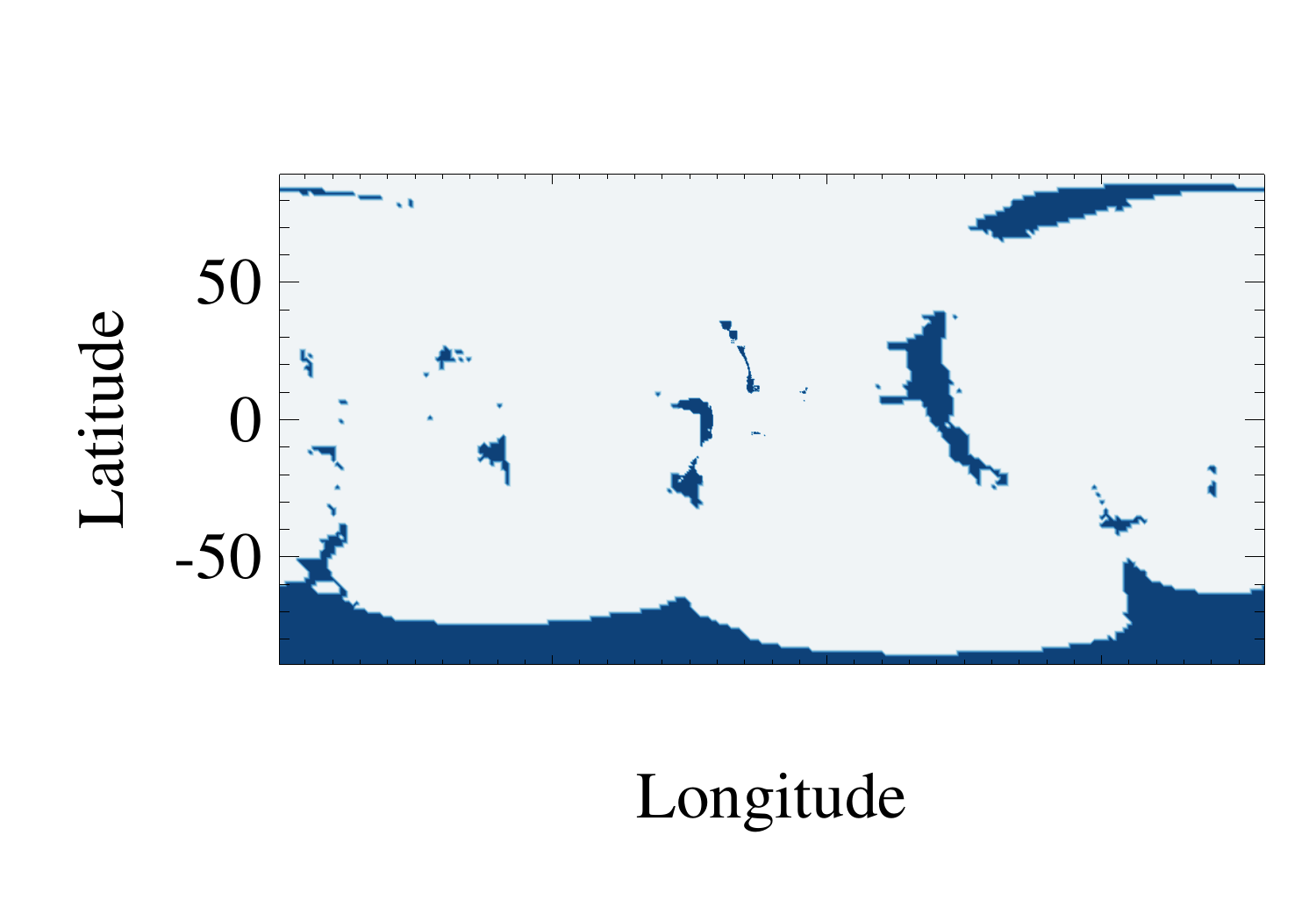} &
\hspace{-0.3cm}\includegraphics[trim=100 50 10 50, clip, width=0.21\textwidth]{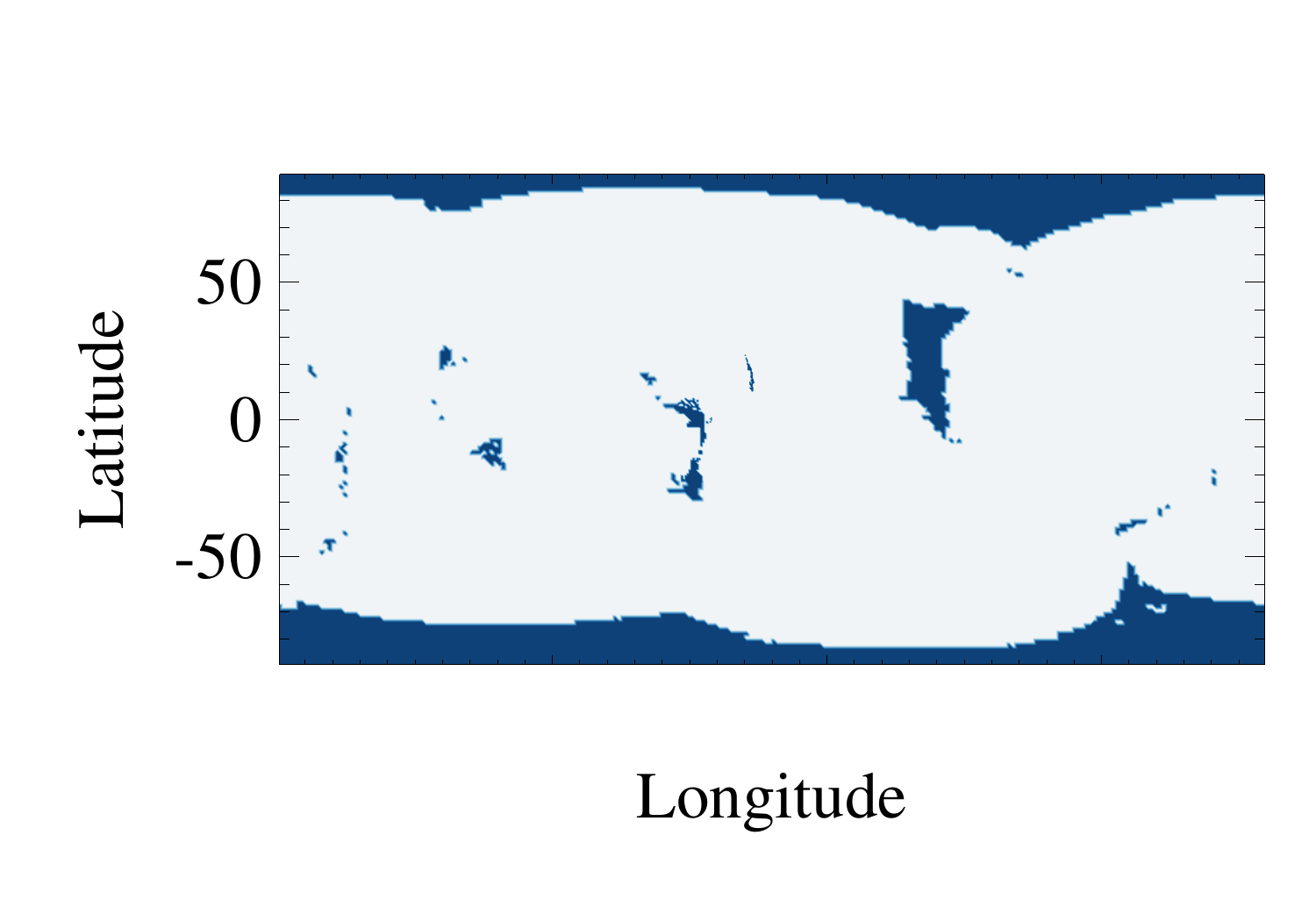} \\

\includegraphics[trim=10 10 10 10, clip, width=0.21\textwidth]{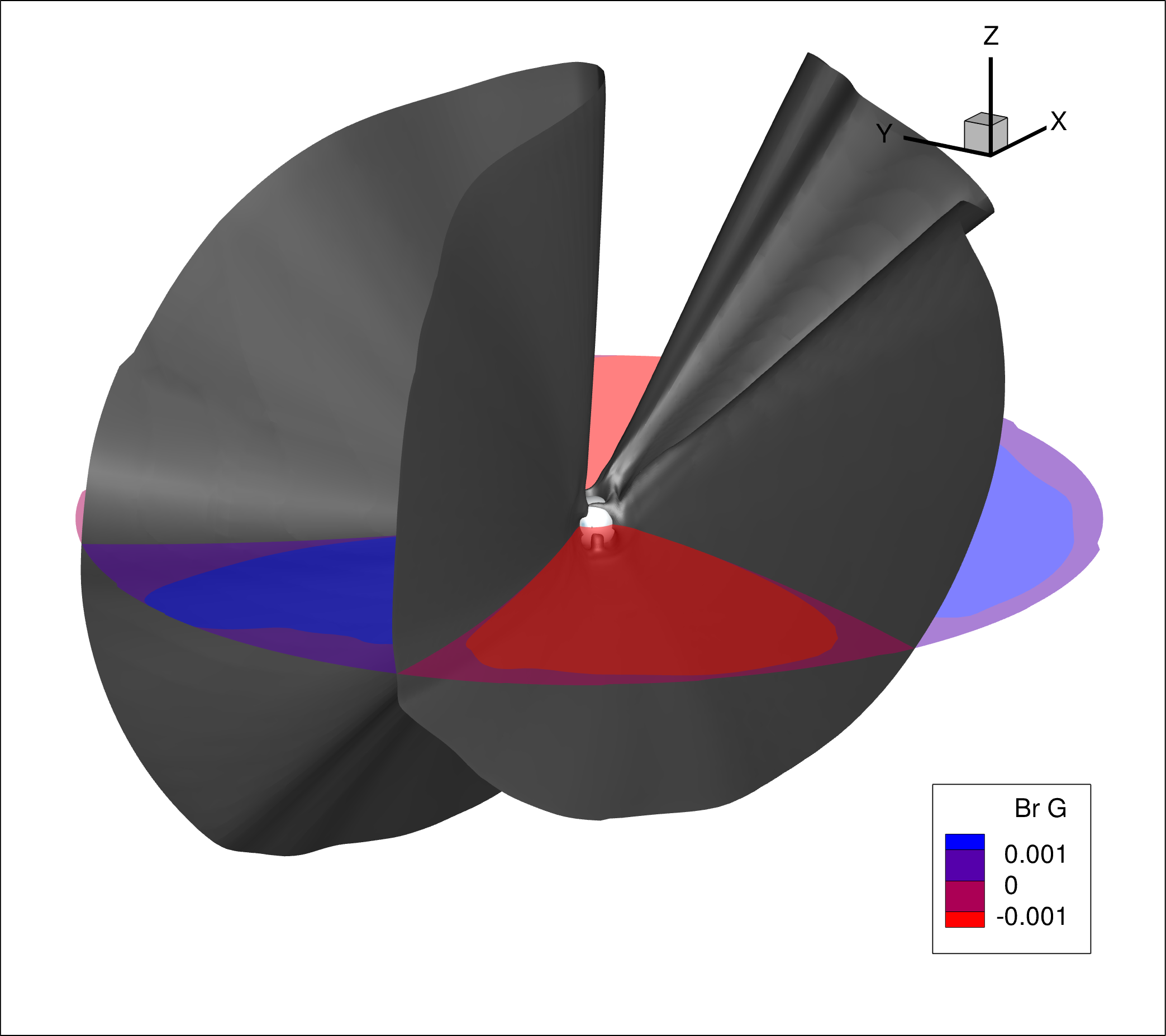} &
\includegraphics[trim=10 10 10 10, clip, width=0.21\textwidth]{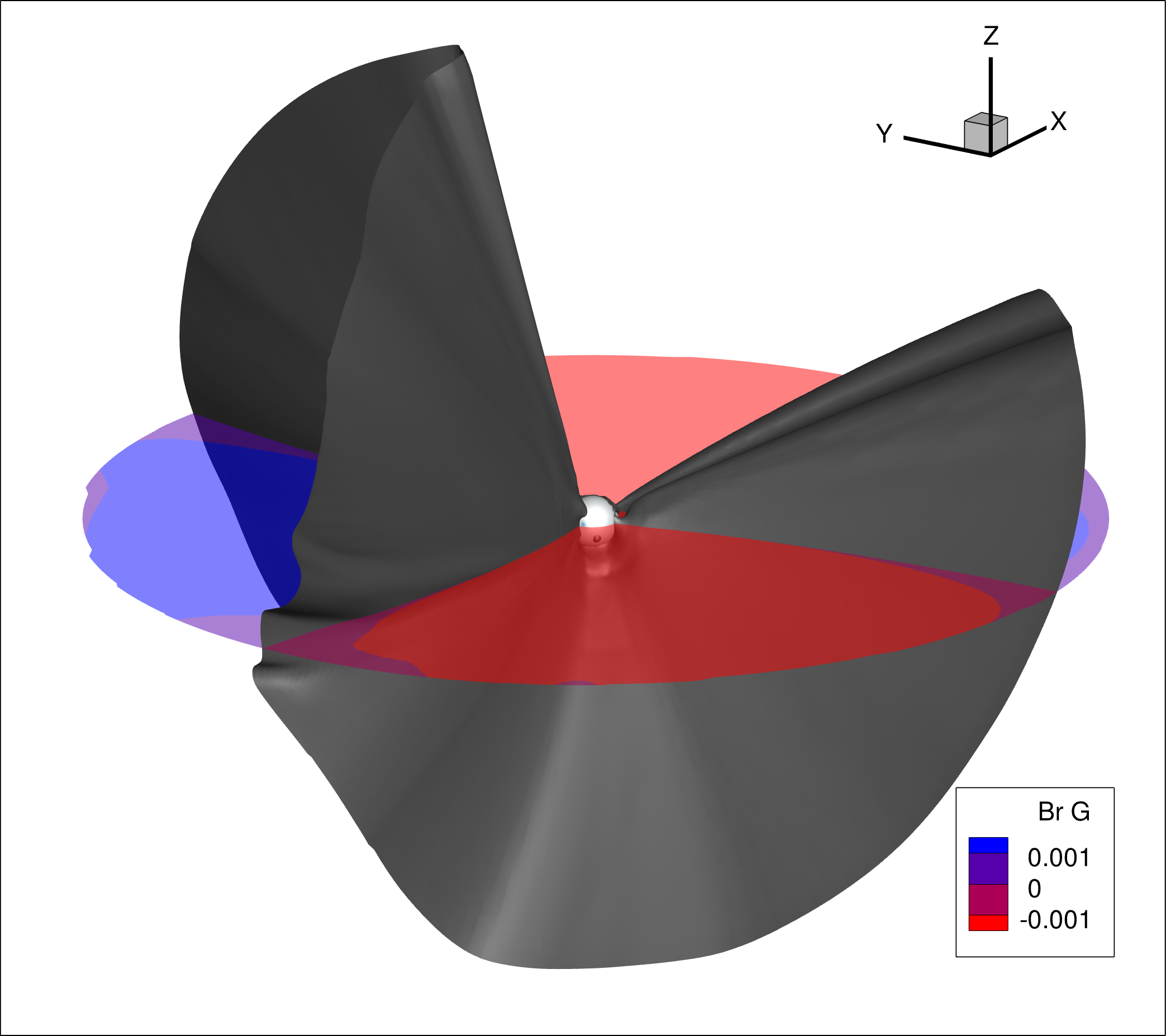} &
\includegraphics[trim=10 10 10 10, clip, width=0.21\textwidth]{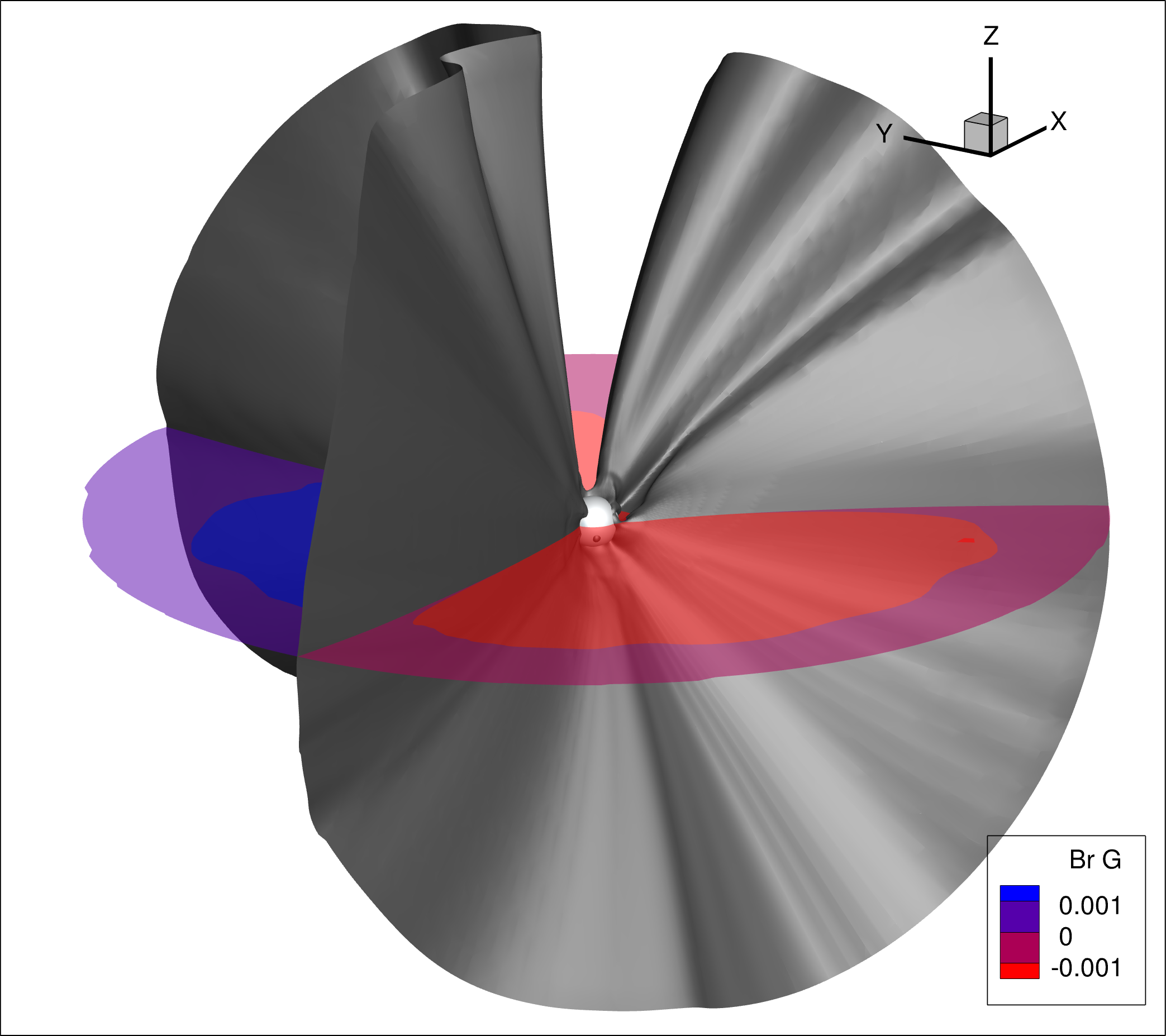} &
\includegraphics[trim=10 10 10 10, clip, width=0.21\textwidth]{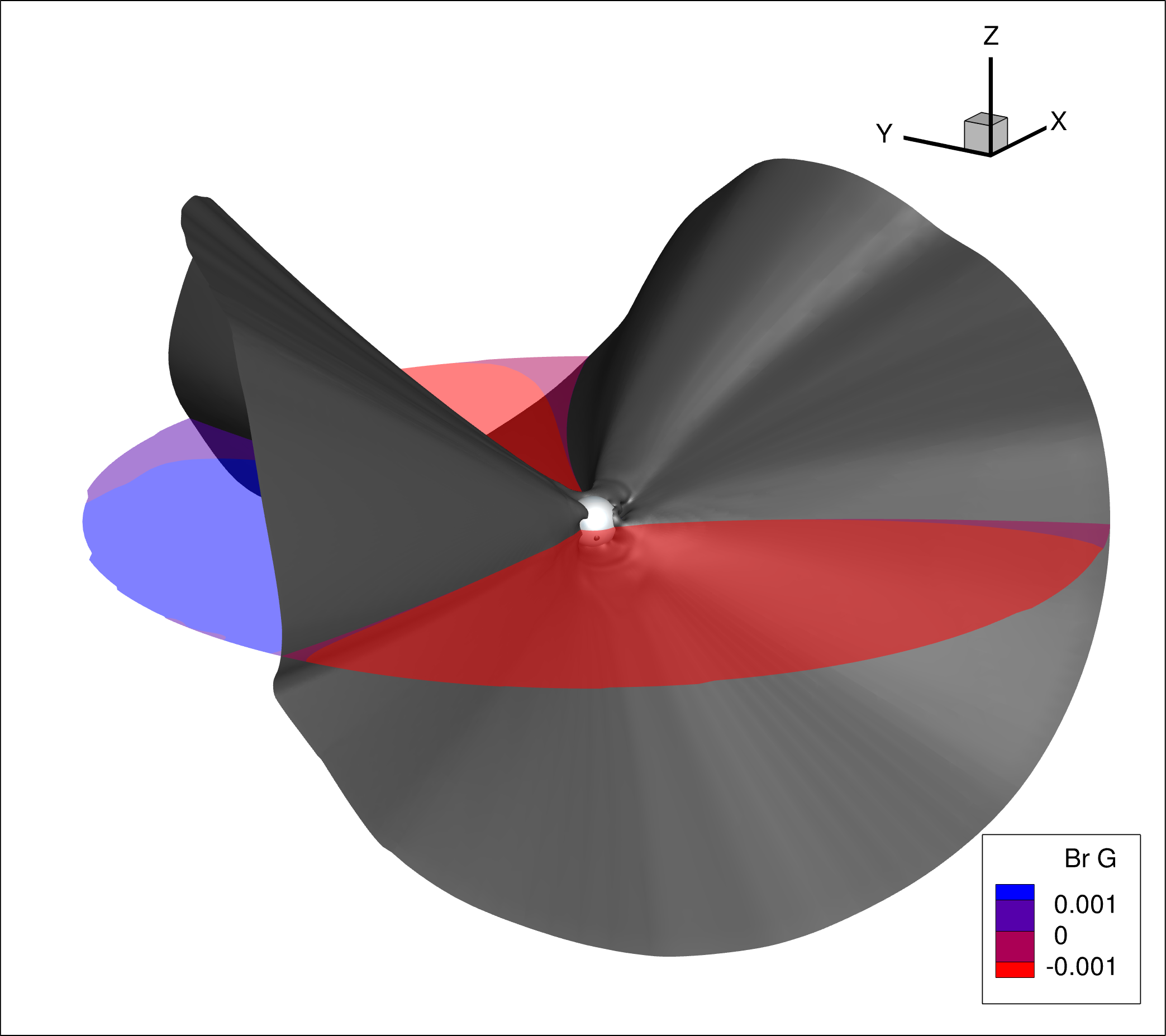} \\
\hspace{-0.6cm}(a) Case 1: ADAPT GONG & 
\hspace{-0.6cm}(b) Case 2: ADAPT HMI & 
\hspace{-0.8cm}(c) Case 3: GONG & 
\hspace{-0.8cm}(d) Case 4: polar enhanced GONG \\
\end{tabular}
\caption{Magnetic field morphology of the ambient solar wind. Results are shown for cases 1, 2, 3, 4 that correspond to solutions driven by ADAPT GONG, ADAPT HMI, GONG and polar enhanced GONG maps respectively, for CR2123. Top row: Radial magnetic field ($B_{r}$) at the source surface radius (2.5 \Rs) for the four cases.
Middle row: Open field regions at 1.01 \Rs for solar wind solutions driven by four maps in the same order as above. Deep blue regions represent the open field regions and the light blue regions depict the closed field regions. Bottom row: Heliospheric current sheet (gray) extracted from 3D solutions. The translucent equatorial slice (z=0) shows the radial magnetic field in Gauss.}\label{fig:pfss_openclose_hcs}
\end{figure*} 
\section{Solar Wind and CME Models} \label{sec:Model}
\subsection{\alf Wave Solar atmosphere Model} \label{sec:awsom}
The first step towards modeling the CME evolution is to simulate the global background solar wind plasma and magnetic field into which they can erupt and propagate. Within the Space Weather Modeling Framework (SWMF, \citet{Toth:2011, Gombosi:2021}), the solar corona (SC) and inner heliosphere (IH) domains are described together by the \alf Wave Solar atmosphere Model (AWSoM, \citep{vanderHolst:2014awsom}). AWSoM is a 3D, physics-based, extended MHD model that extends from the base of the solar transition region into the solar corona extending out into the inner heliosphere, going beyond 1 au. AWSoM uses the the Block-Adaptive-Tree-Solarwind-Roe-Upwind-Scheme (BATS-R-US; \citealp{Powell:1999, Toth:2012swmf}) to solve the extended MHD equations. Isotropic electron temperature and anisotropic proton temperatures (parallel and perpendicular), are included in AWSoM with \alf wave turbulence that heats and accelerates the solar wind. Also included are heat conduction for electrons and radiative cooling. Detailed descriptions of the model and its implementation are available in \cite{Sokolov:2013,vanderHolst:2014awsom,vanderHolst:2022psp}. Model inputs include the inner boundary conditions for the magnetic field (provided by the radial component of the observed magnetic field), density, temperature, and wave energy flux along with a few free parameters for the turbulence model.

AWSoM has been utilized to study the solar wind plasma properties during varying phases of the solar cycle \citep{Sachdeva:2021, Sachdeva:2019,Huang:2023,Huang:2024} and has been  extensively validated against remote-sensing \citep{Szente:2022, Szente:2023, Lloveras:2017, Koban:2025} and in-situ measurements \citep{Wraback:2024, Sachdeva:2023} including Parker Solar Probe \citep{vanderHolst:2019,vanderHolst:2022psp}. It has also been utilized to study the magnetic field evolution during the last solar cycle \citep{Huang:2024b}, simulation of solar energetic particle events \citep{Liu:2025} and operational testing of real time space weather predictions \citep{LiuW:2025b}.
\begin{figure*}[t!]
\centering
\gridline{\fig{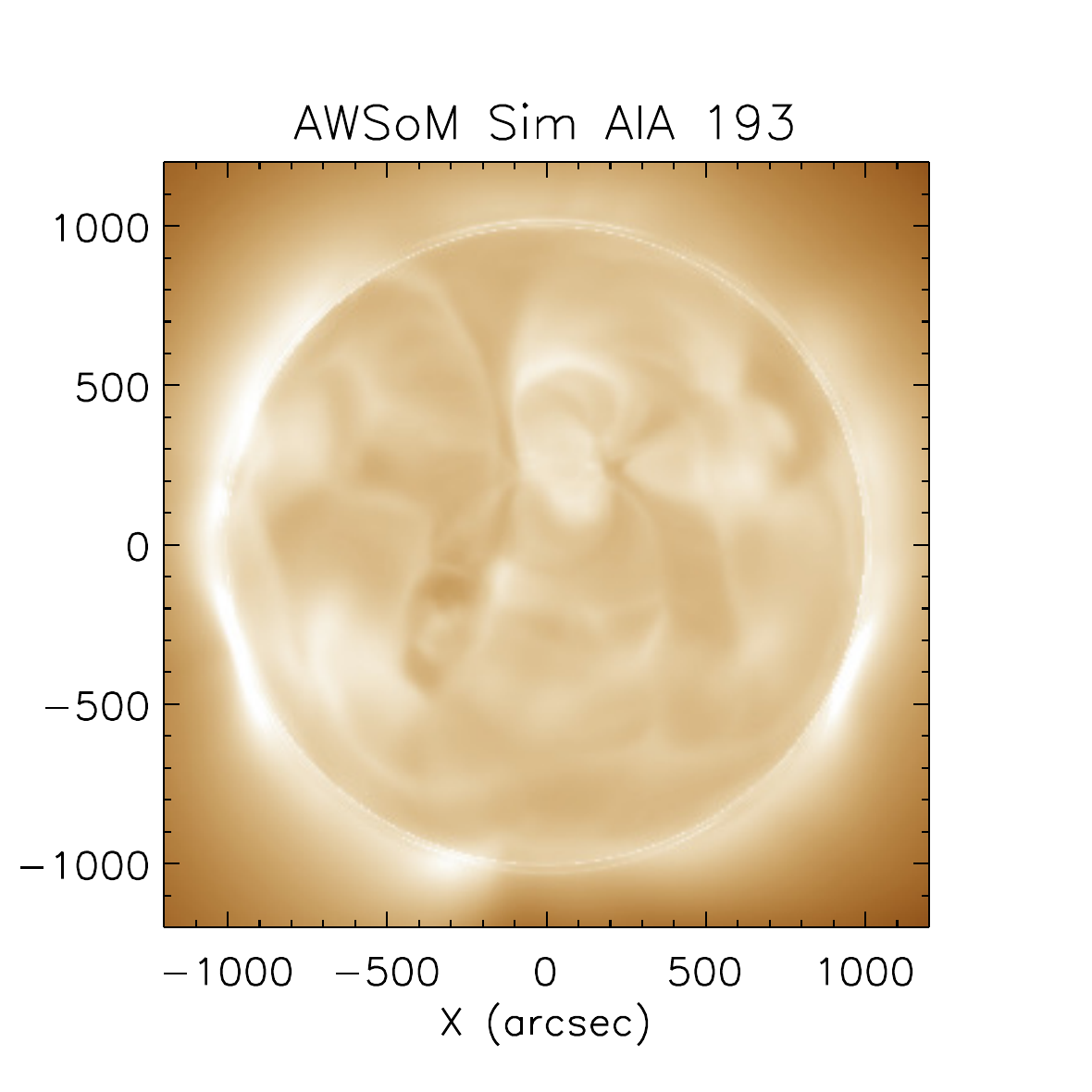} {0.25\textwidth} {}{\hspace{-1cm}}
        \fig{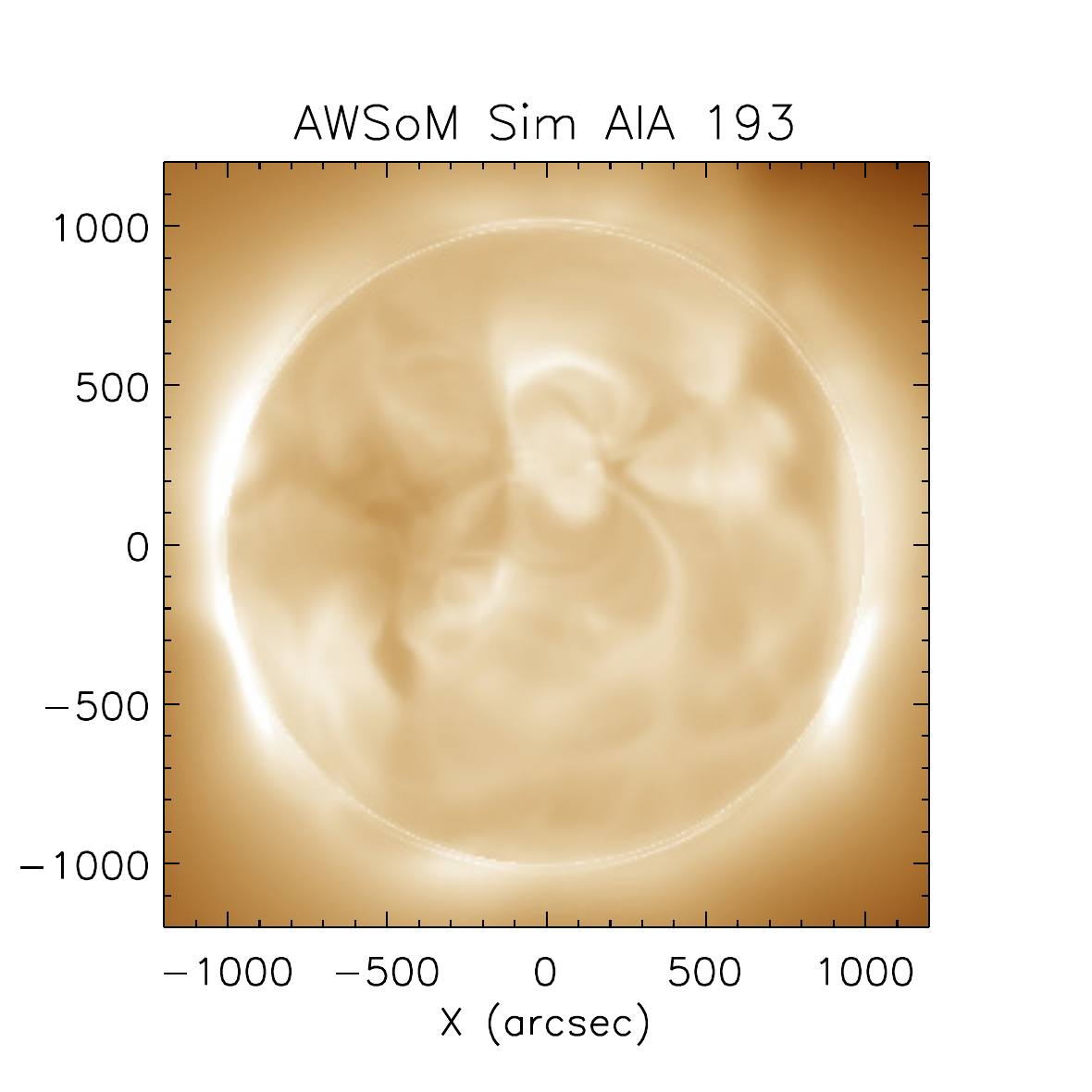}{0.25\textwidth}{}{\hspace{-1cm}}
        \fig{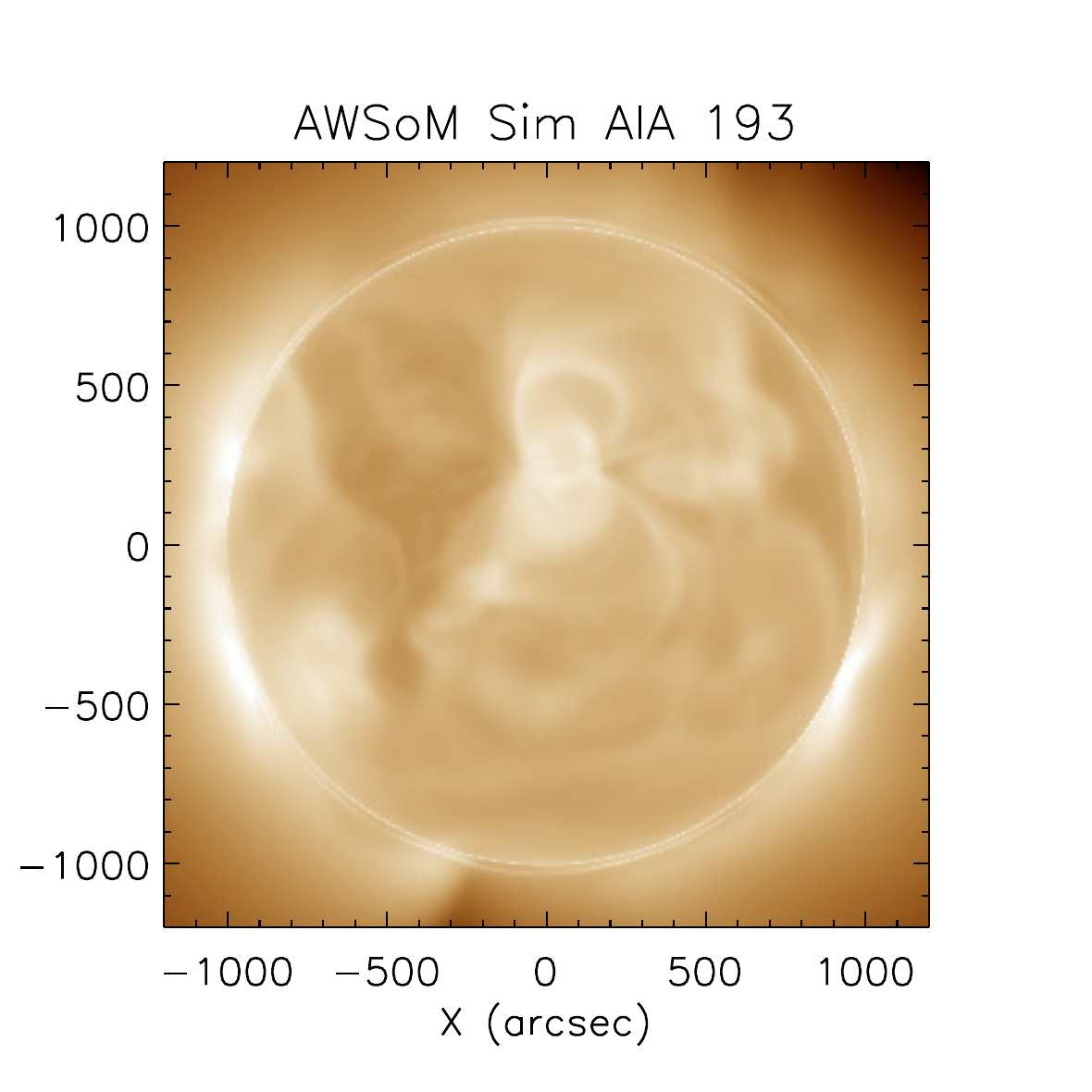} {0.25\textwidth}{}{\hspace{-1cm}}
        \fig{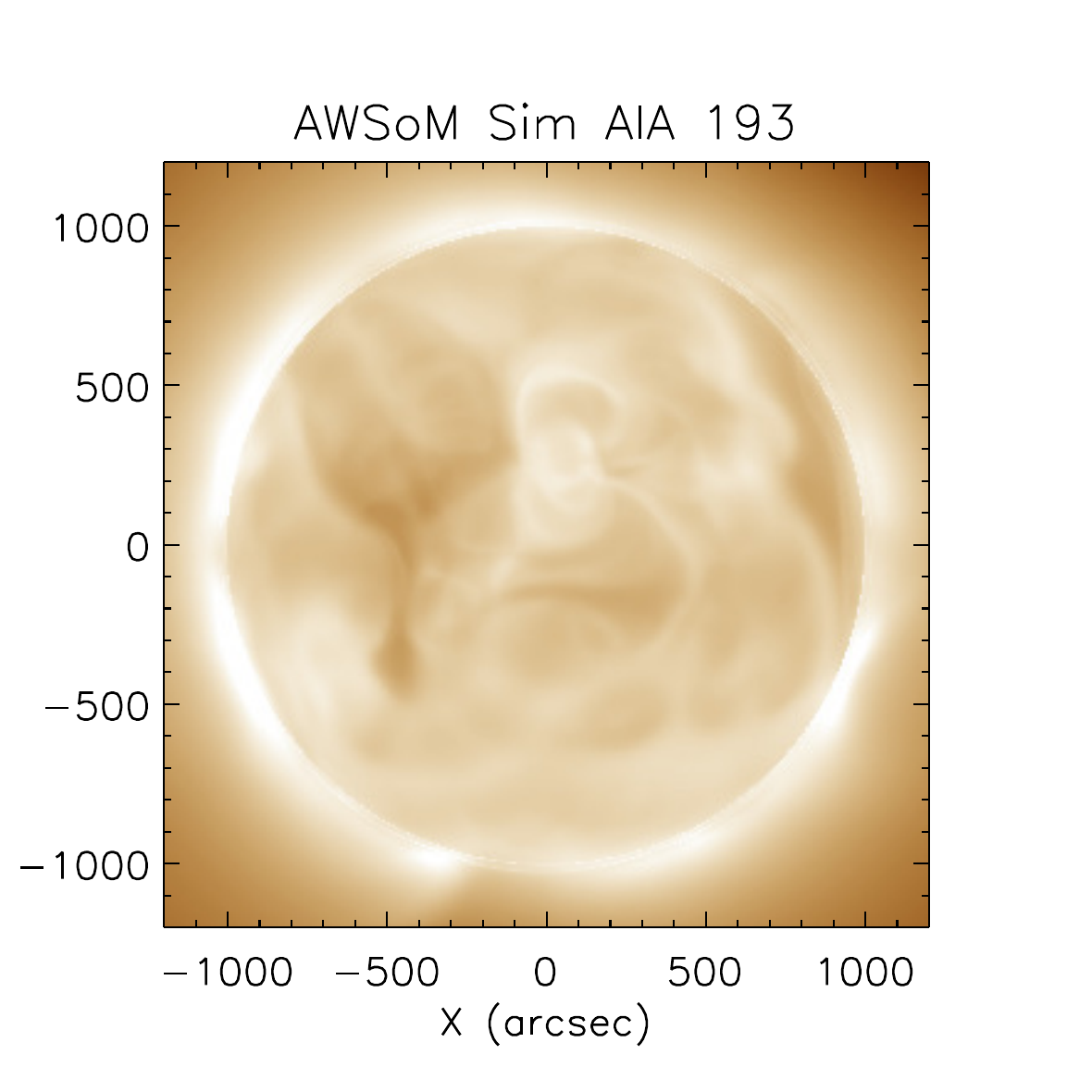}{0.25\textwidth}{}}
           \vspace{-1.3cm}
\gridline{\fig{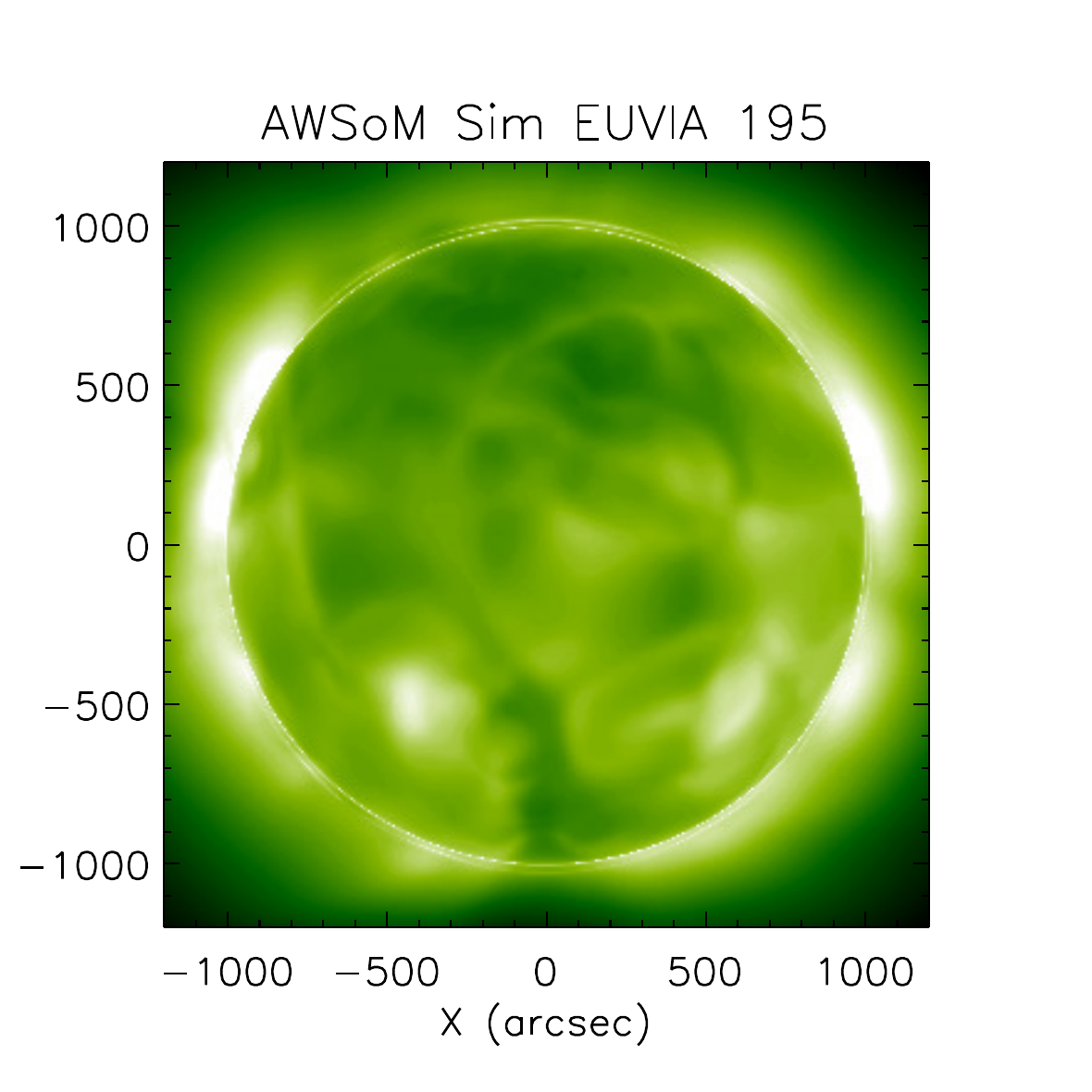} {0.25\textwidth} 
        {(a) Case 1}{\hspace{-1cm}}
        \fig{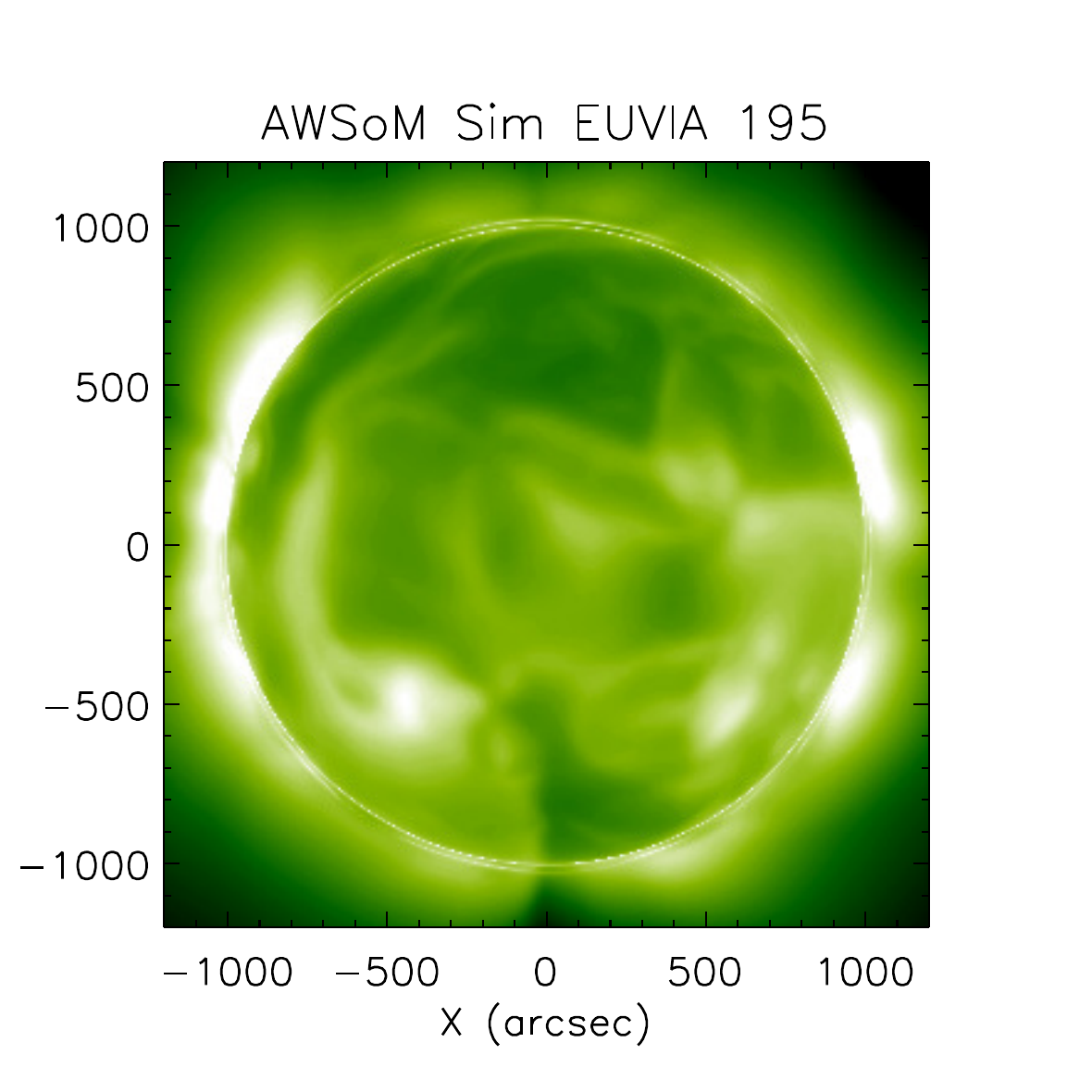}{0.25\textwidth}{(b) Case 2}{\hspace{-1cm}}
        \fig{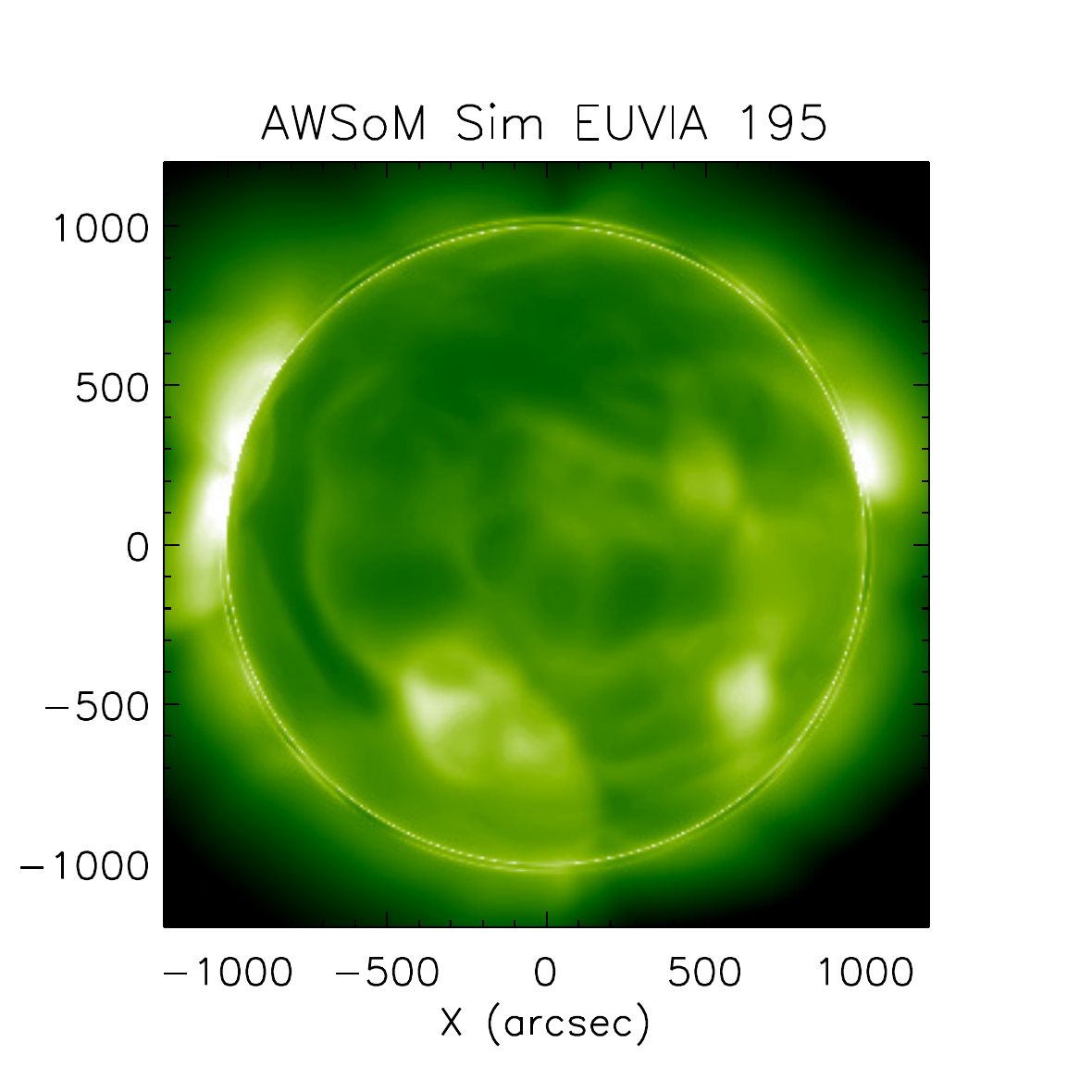} {0.25\textwidth}{(c) Case 3}{\hspace{-1cm}}
        \fig{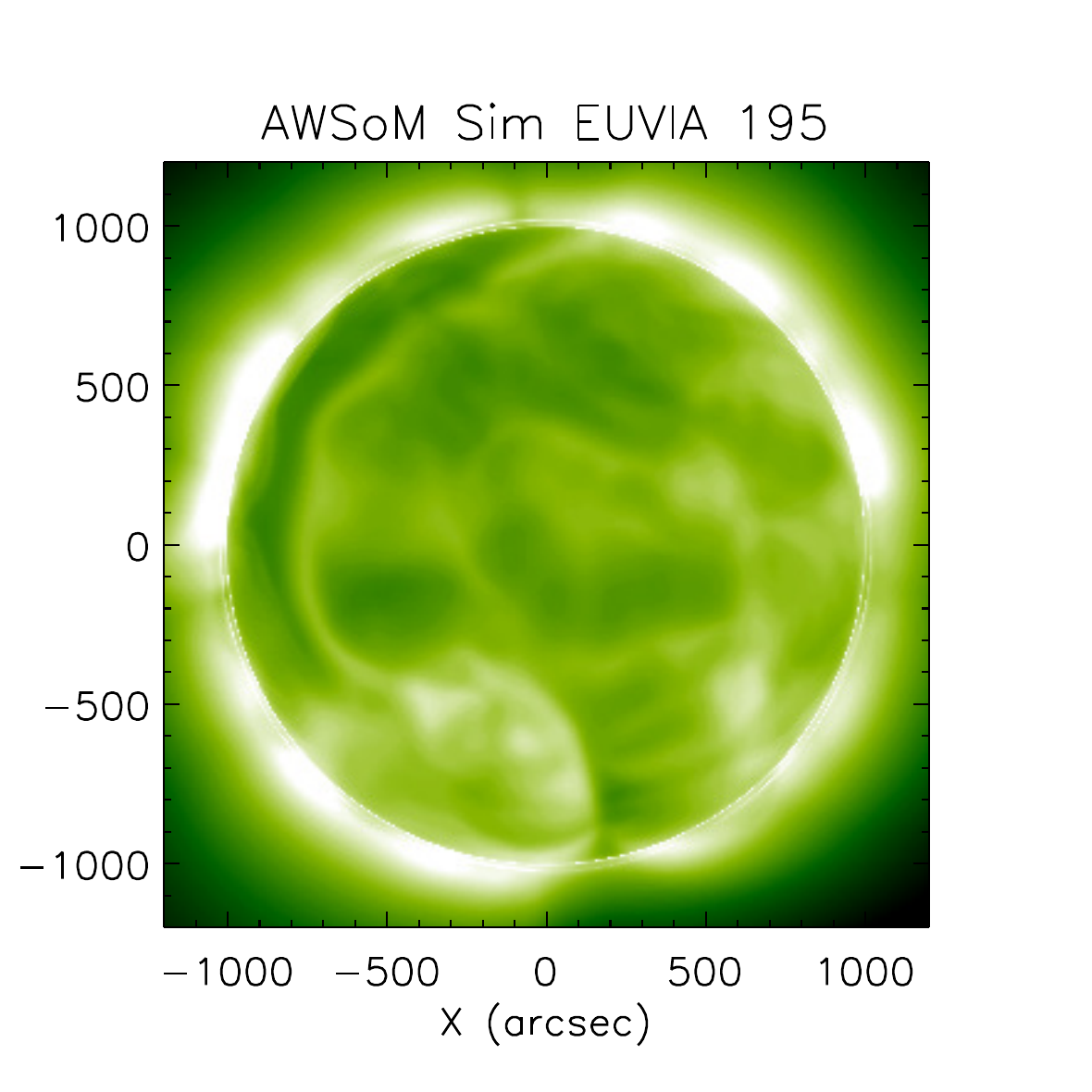}{0.25\textwidth}{(d) Case 4}}
        \vspace{-0.3cm}
\caption{AWSoM synthesized line of sight EUV images corresponding to SDO/AIA  193 \AA (top row) and STEREO-A EUV 195 \AA (bottom row) for all four cases.}\label{fig:los}
\end{figure*}
\subsection{CME Model}
The Eruptive Event Generator (EEG) module within SWMF facilitates both the initiation and eruption of magnetic flux ropes into the solar wind using AWSoM. EEG was originally developed by \cite{Jin:2017a, Borovikov:2017} to model CME eruptions by inserting the Gibson-Low (GL) \citep{Gibson:1998} flux rope structure into an active region (AR).  Examples of GL-based CME event studies include \citep{Manchester:2014a, Manchester:2014b, Jin:2017b, Jin:2018}. The GL flux rope with AWSoM has been utilized for spectral analysis of the cool filament material \citep{Wraback:2024spectra} and solar energetic particle events \citep{Zhao:2024,Liu:2025,XChen:2025}. EEG also includes the Titov-Demoulin (TD) model \citep{Titov:1999} based on the earlier work of \cite{Lugaz:2007, Manchester:2008, Manchester:2012}. Currently, EEG is being expanded to include CME initiation by adding a sheared arcade using the Statistical InjecTion of Condensed Helicity (STITCH) technique \citep{Antiochos:2013, Dahlin:2022} and may ultimately model CME initiation based on flux emergence \citep{Manchester:2004}. EEG with GL is now available through the Runs-on-Request service at NASA Community Coordinated Modeling Center (CCMC).

In this work, we use EEG to specify a GL magnetic flux rope to launch a CME from the solar surface by superimposing it onto the solar wind solution with AWSoM. The GL flux rope is an analytical flux rope model that also includes the 3-part density structure of CME coronagraph observations (core, cavity, outer shell) \citep{Illing:1986}. It is an analytical solution to the magnetohydrostatic equation, $(\nabla \times B) \times B -\nabla p - \rho g = 0$ 
assuming simple functional forms for density and pressure \citep{Borovikov:2017}.
The inserted flux rope is in a state of force imbalance and erupts immediately, propagating into and interacting with the background solar wind plasma.

\section{Magnetic Field Data} \label{sec:Maps}
The radial photospheric magnetic field observations are available as synoptic and synchronic maps to prescribe boundary conditions for numerical simulations of the solar corona and heliosphere \citep{Roussev:2003a, Usmanov:2003, Cohen:2007, vanderHolst:2010, Lionello:2013, Feng:2014, Feng:2015}. These maps are constructed using full-disk images of the solar photosphere over a period of solar rotation (usually 27 days or more), modified and assembled into synoptic magnetic field maps (or magnetograms). However, the drawback here is that portions of these maps can be as much as two weeks out of date, due to the solar rotation rate. An improvement over these maps is the methodology of producing global magnetic field maps using surface flux transport (SFT) models. These models include differential rotation, meridional and supergranulation flows and data-assimilation to update the maps by incorporating latest available observations producing synchronic maps \citep{Schrijver:2001, Upton:2014, Hickmann:2015, Schon:2022}. 
The Air Force Data Assimilative Photospheric flux Transport (ADAPT) model \citep{Worden:2000, Arge:2010, Arge:2013} provides an ensemble of maps of the photospheric magnetic field, each with associated uncertainty of the global flux distribution \citep{Silva:2023}. \cite{Barnes:2023} showed that the choice of the SFT model can greatly influence the global solar wind predictions. The variation in the solar wind results stems from the way in which different magnetogram products estimate the polar fields, or assimilate the latest magnetic field information from active regions. \cite{Jin:2022} showed the impact of Lockheed Martin synchronic maps and HMI synoptic maps on CME-driven shock connectivity and shock properties for the 2013-11 April event.

Information of emerging active regions on the far-side of the Sun may still be missing in these maps and can have major impact on solar wind prediction. Including far-side data using helioseismology has shown to improve the in-situ data-model comparisons \citep{Arge:2013}. Recent work by \cite{Perri:2024} utilized far-side magnetic field structures observed by Solar Orbiter/Photospheric Heliospheric Imager (SoLO/PHI) to study the impact on coronal and heliospheric wind simulations. They found that the inclusion of a single far-side active region had small local effect on the total magnetic flux, but global impact on the magnetic structure, altering the open and closed magnetic field line distributions.

\begin{figure*}[t!]
\centering
\gridline{\fig{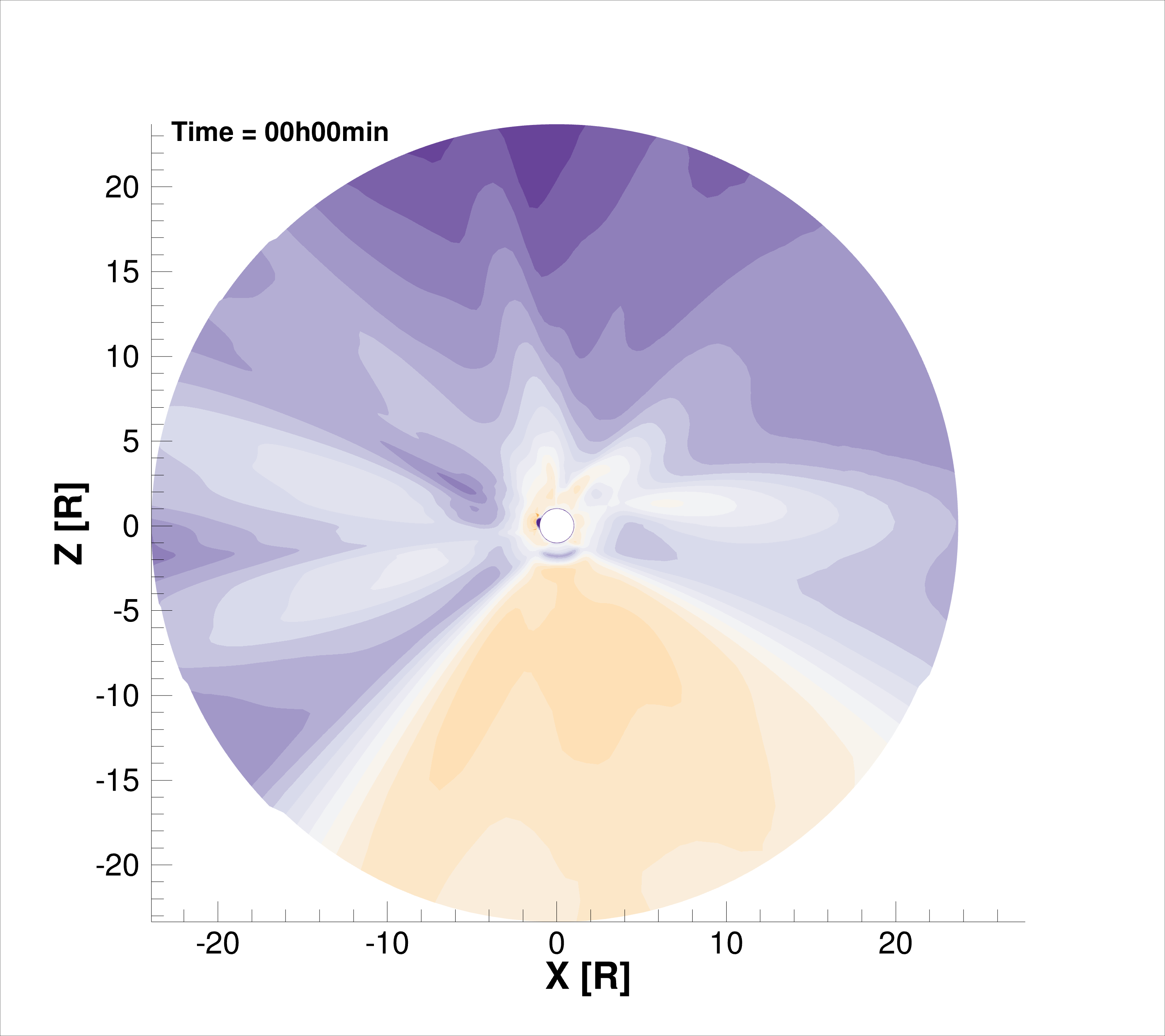} {0.265\textwidth} {}{\hspace{-1cm}}
           \fig{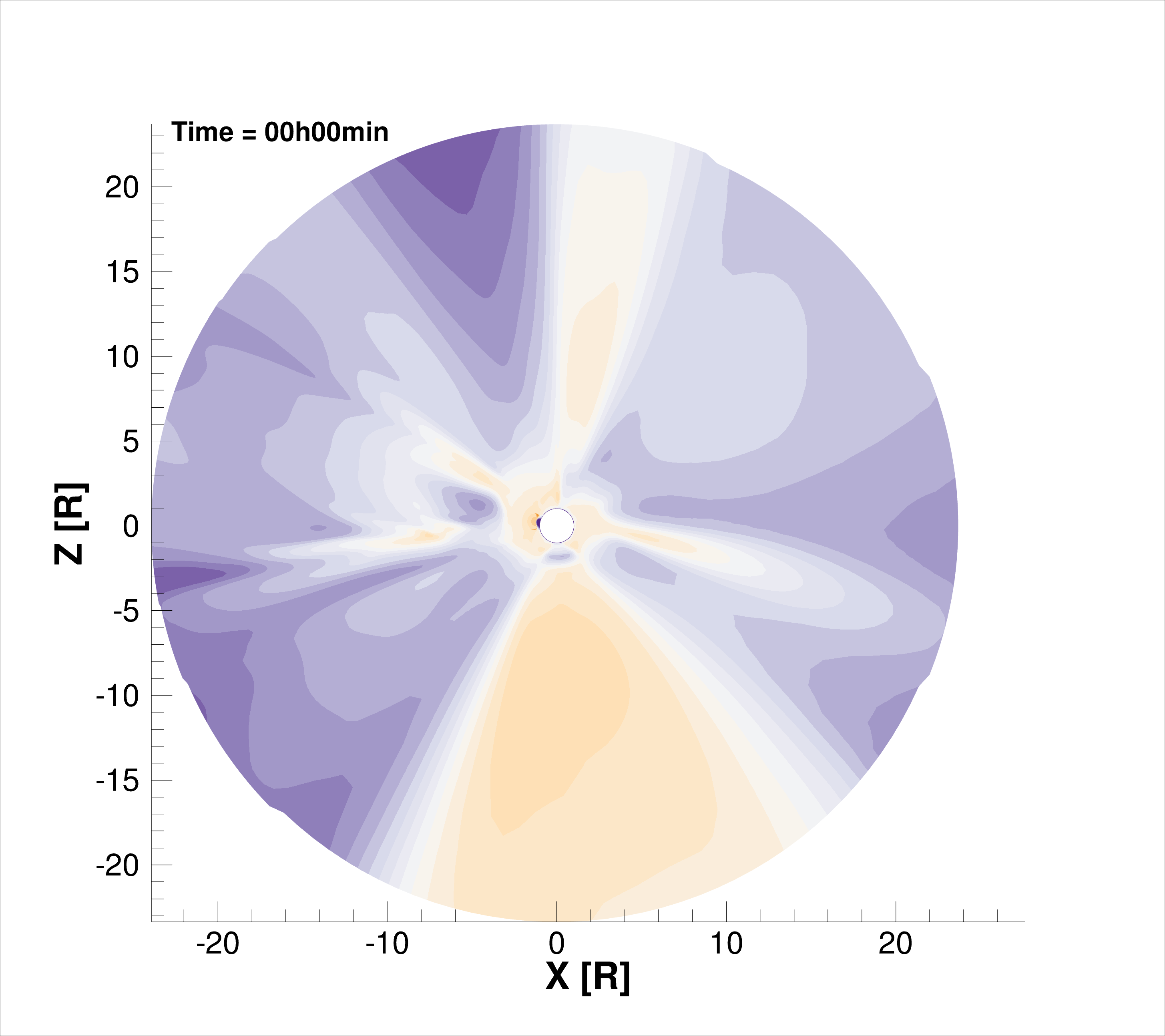}{0.265\textwidth}{}{\hspace{-1cm}}
           \fig{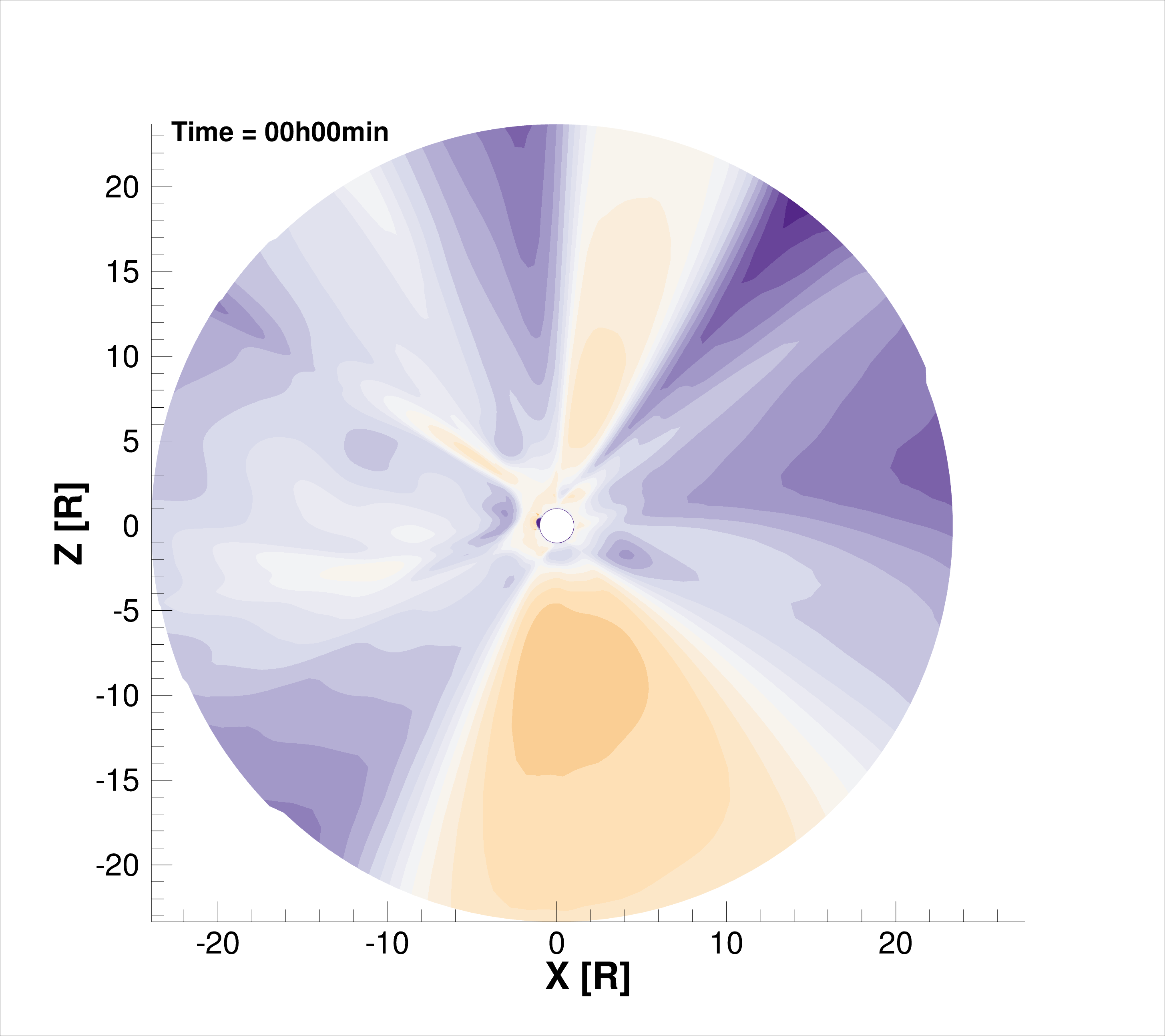} {0.265\textwidth}{}{\hspace{-1cm}}
          \fig{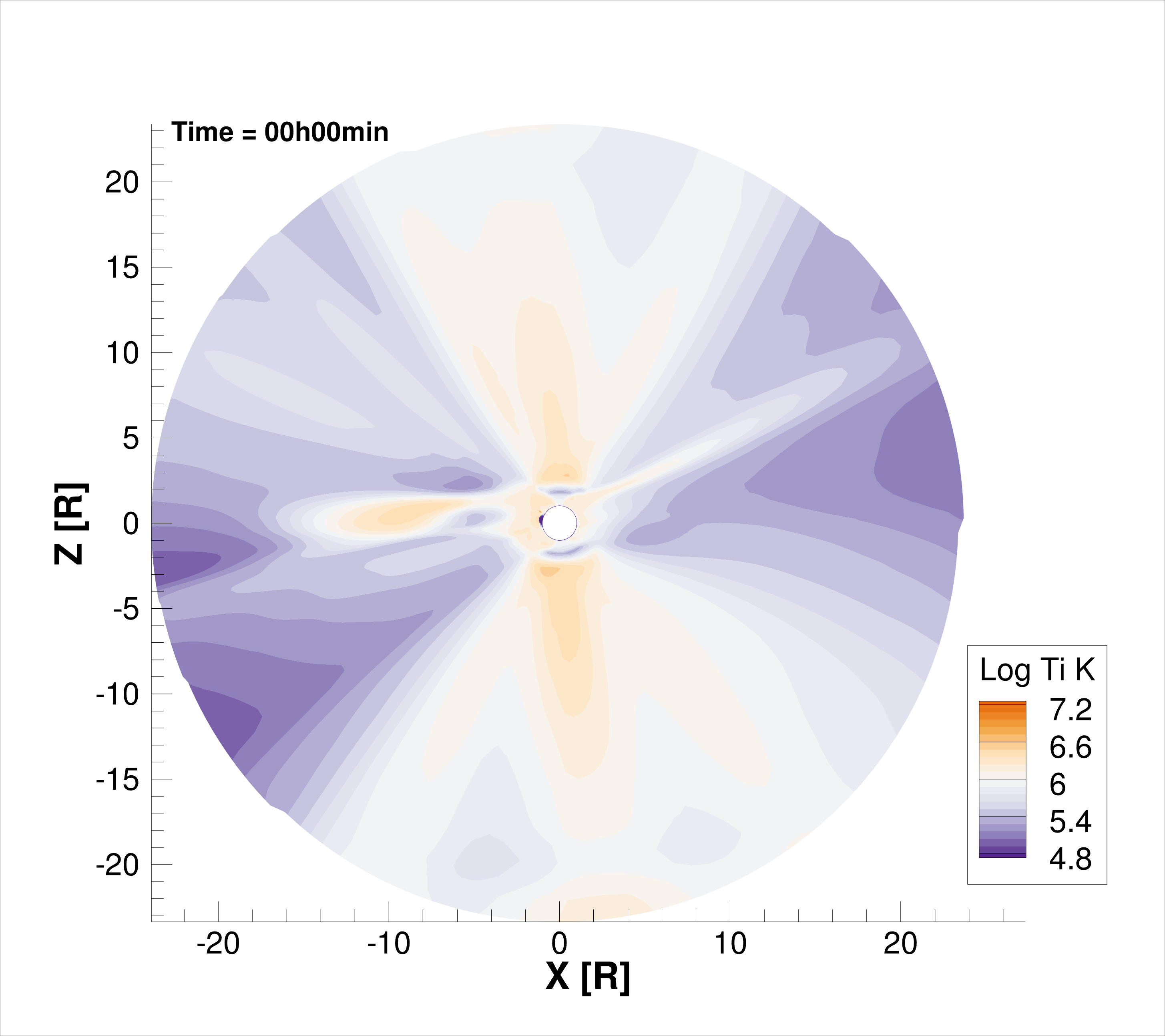}{0.265\textwidth}{}}
          \vspace{-1cm}
\gridline{\fig{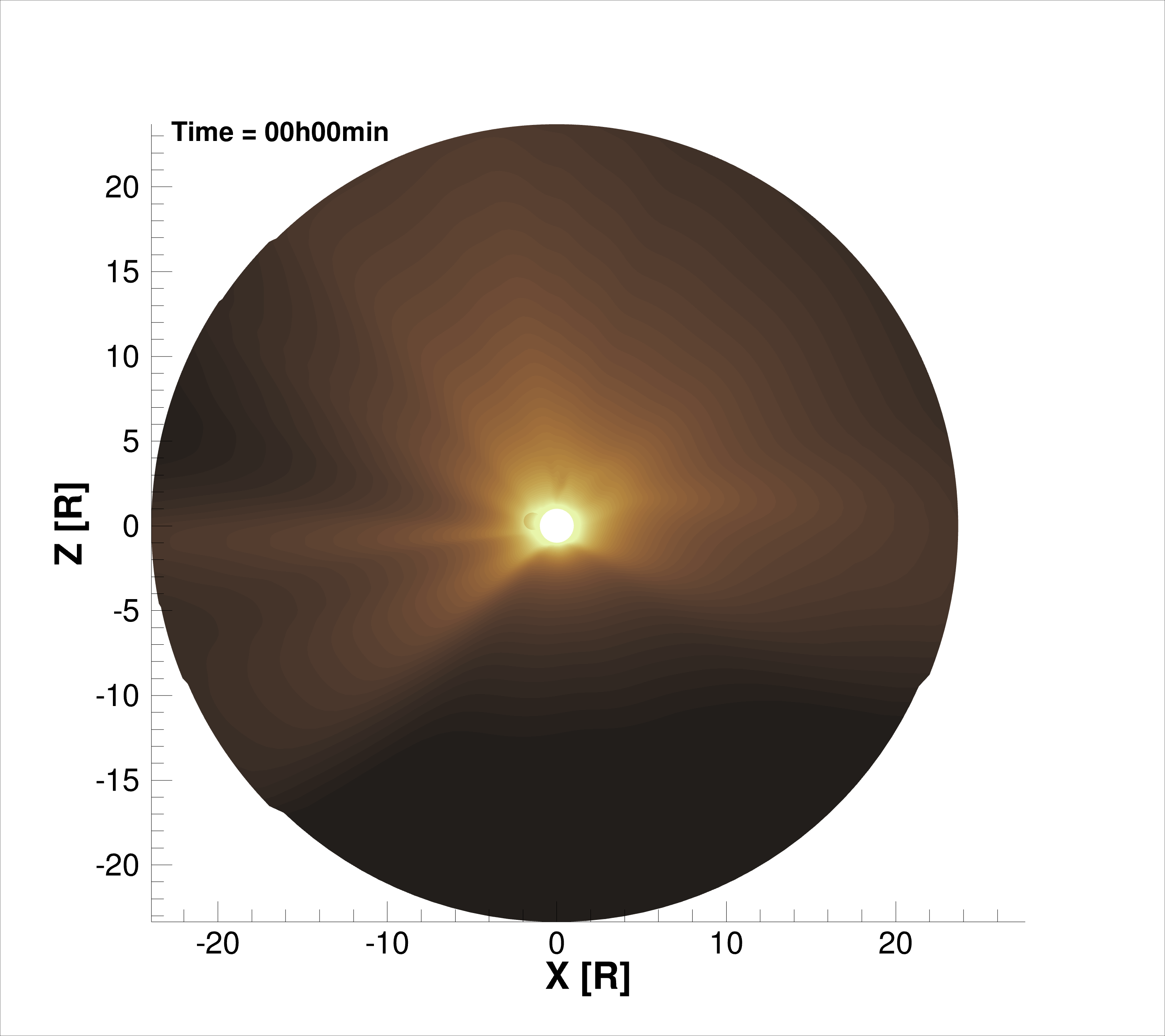} {0.265\textwidth} {}{\hspace{-1cm}}
           \fig{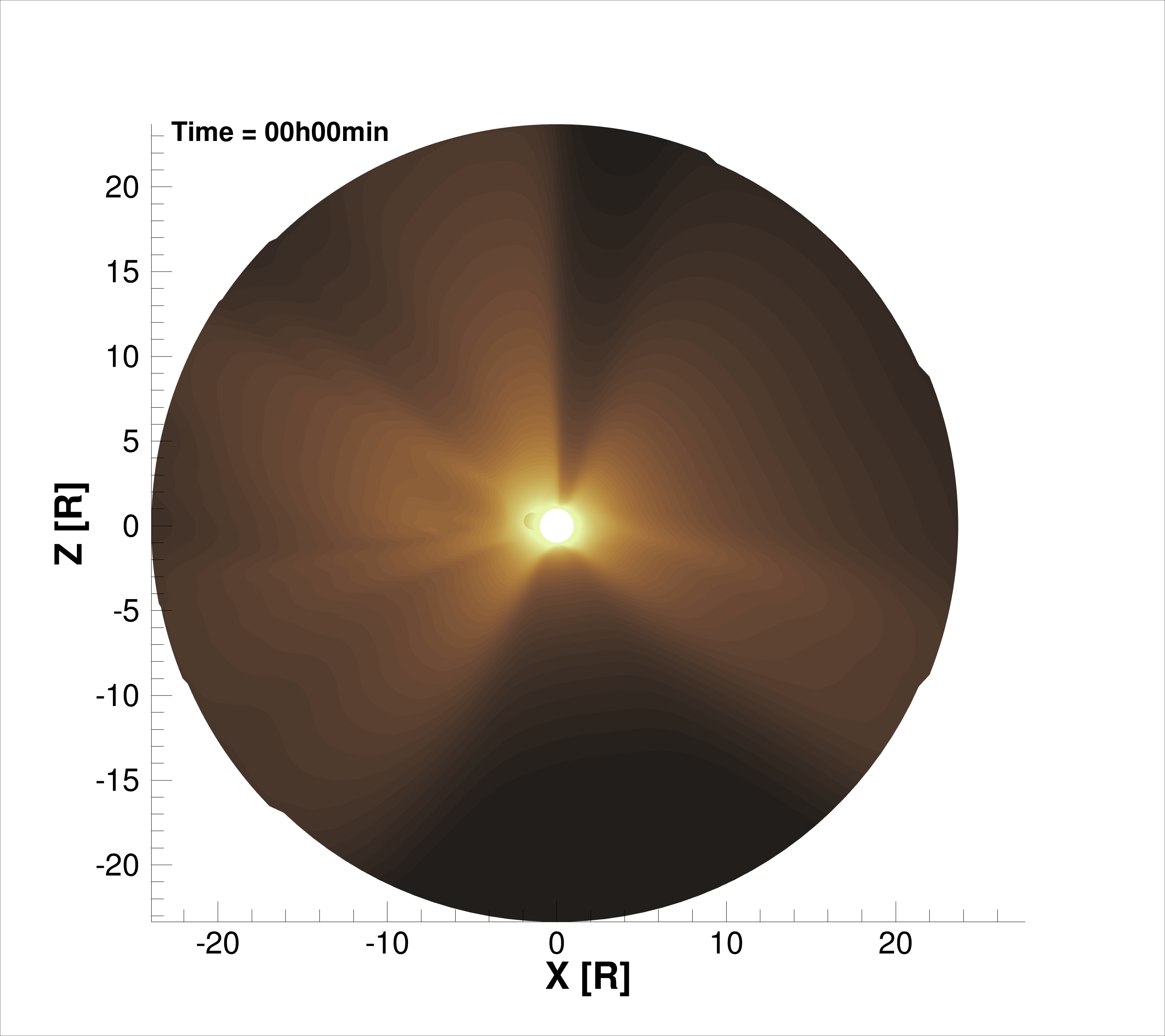}{0.265\textwidth}{}{\hspace{-1cm}}
           \fig{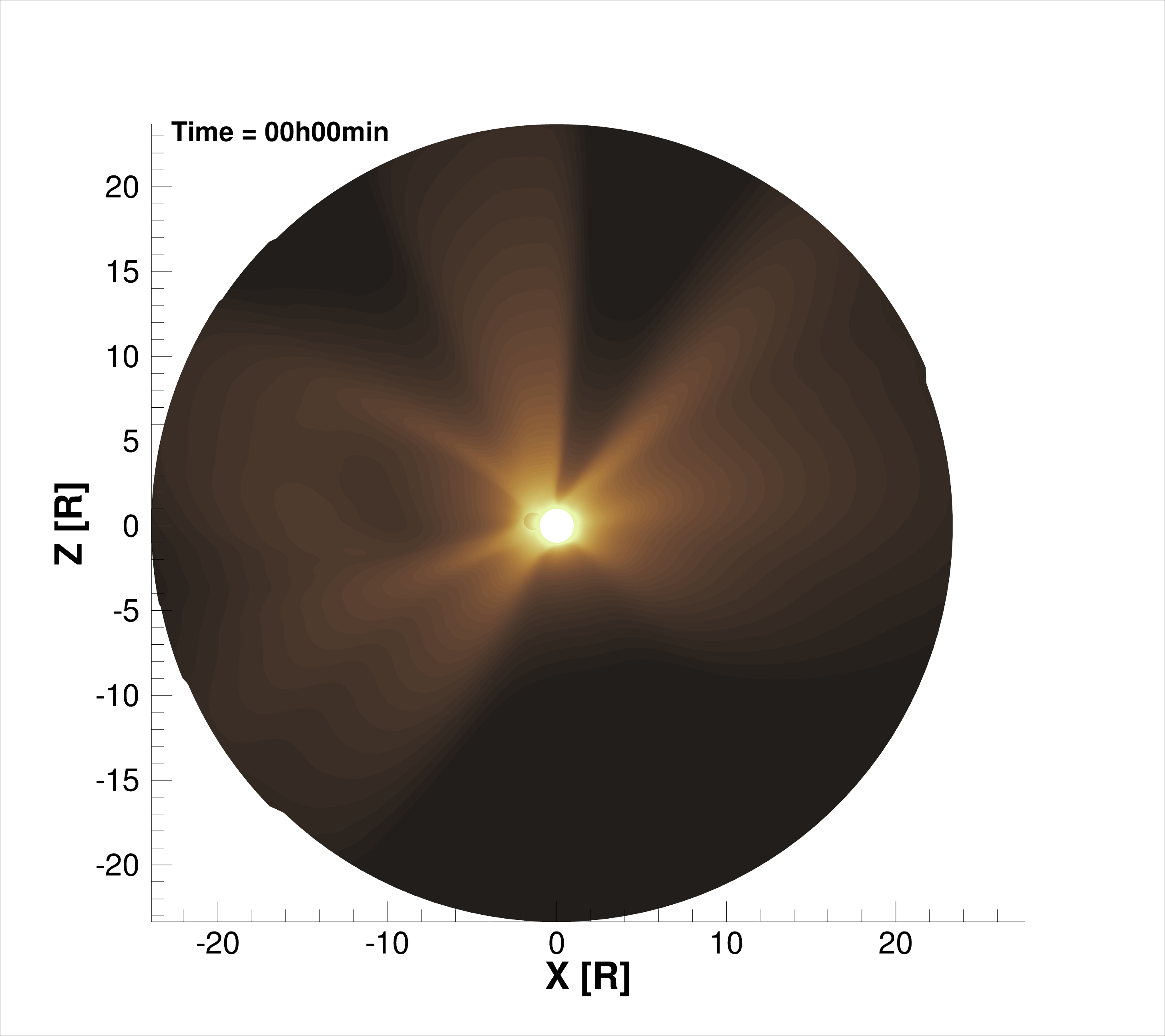} {0.265\textwidth}{}{\hspace{-1cm}}
          \fig{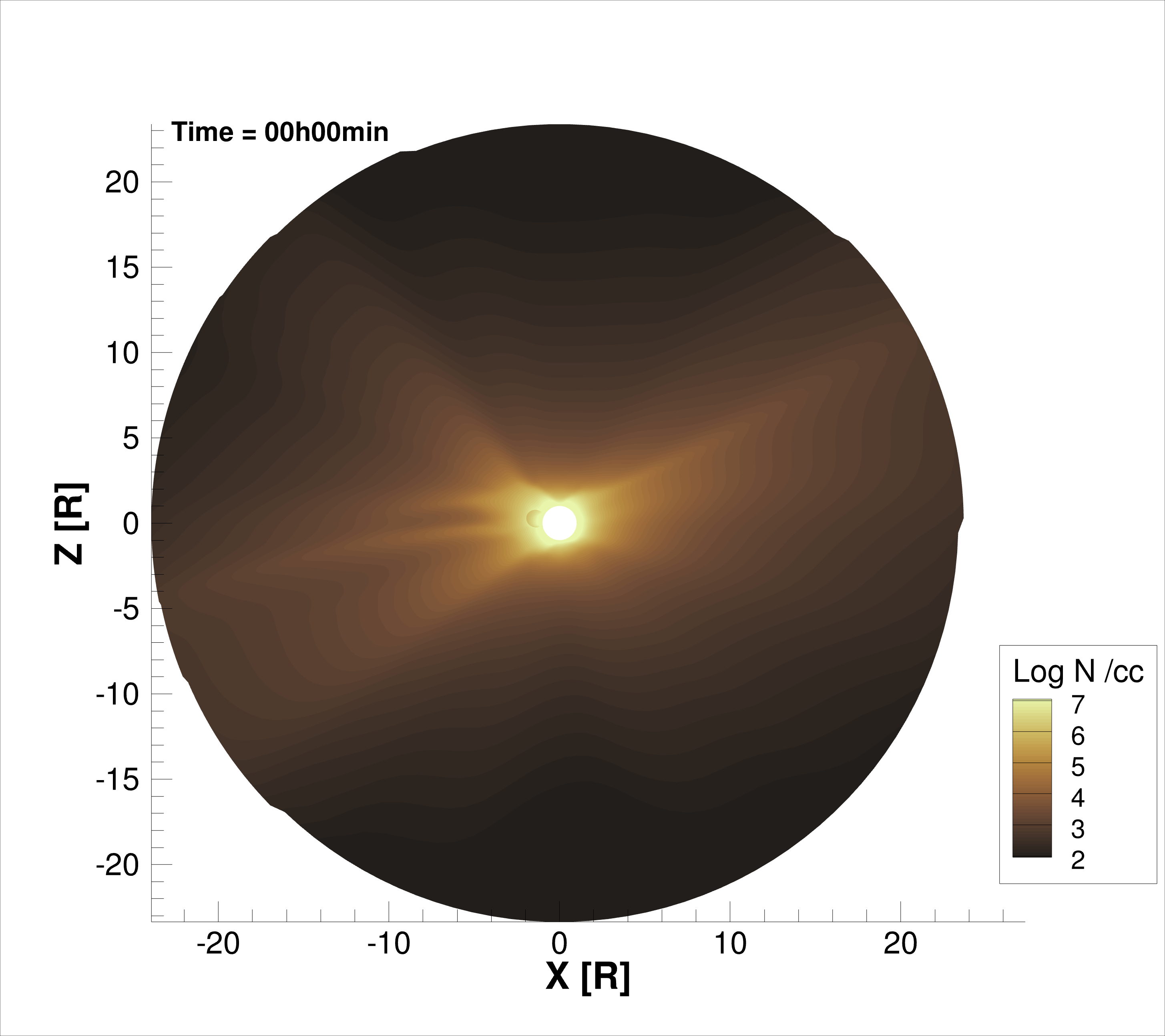}{0.265\textwidth}{}}
          \vspace{-1cm}
\gridline{\fig{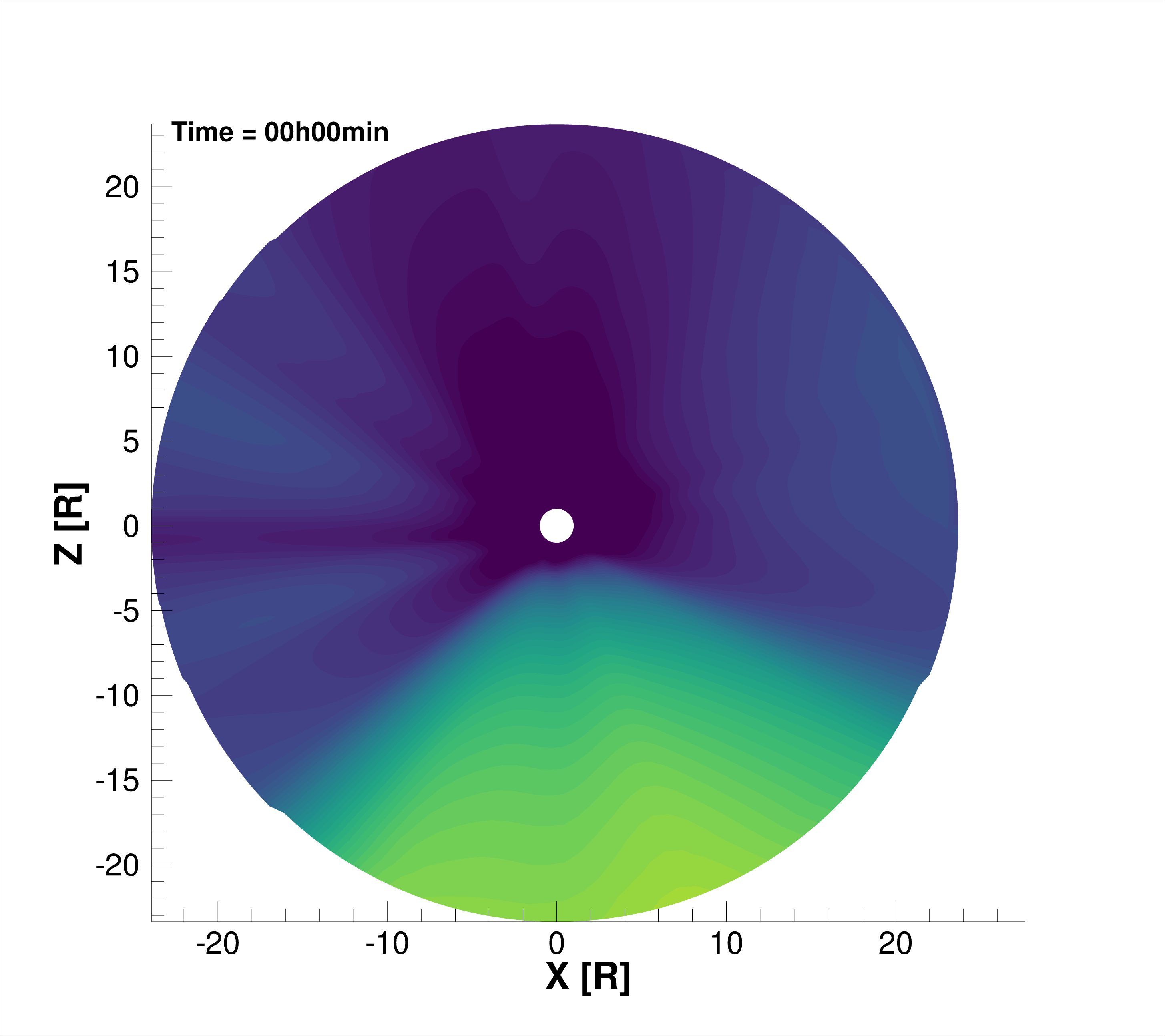} {0.265\textwidth} {(a) Case 1}{\hspace{-1cm}}
           \fig{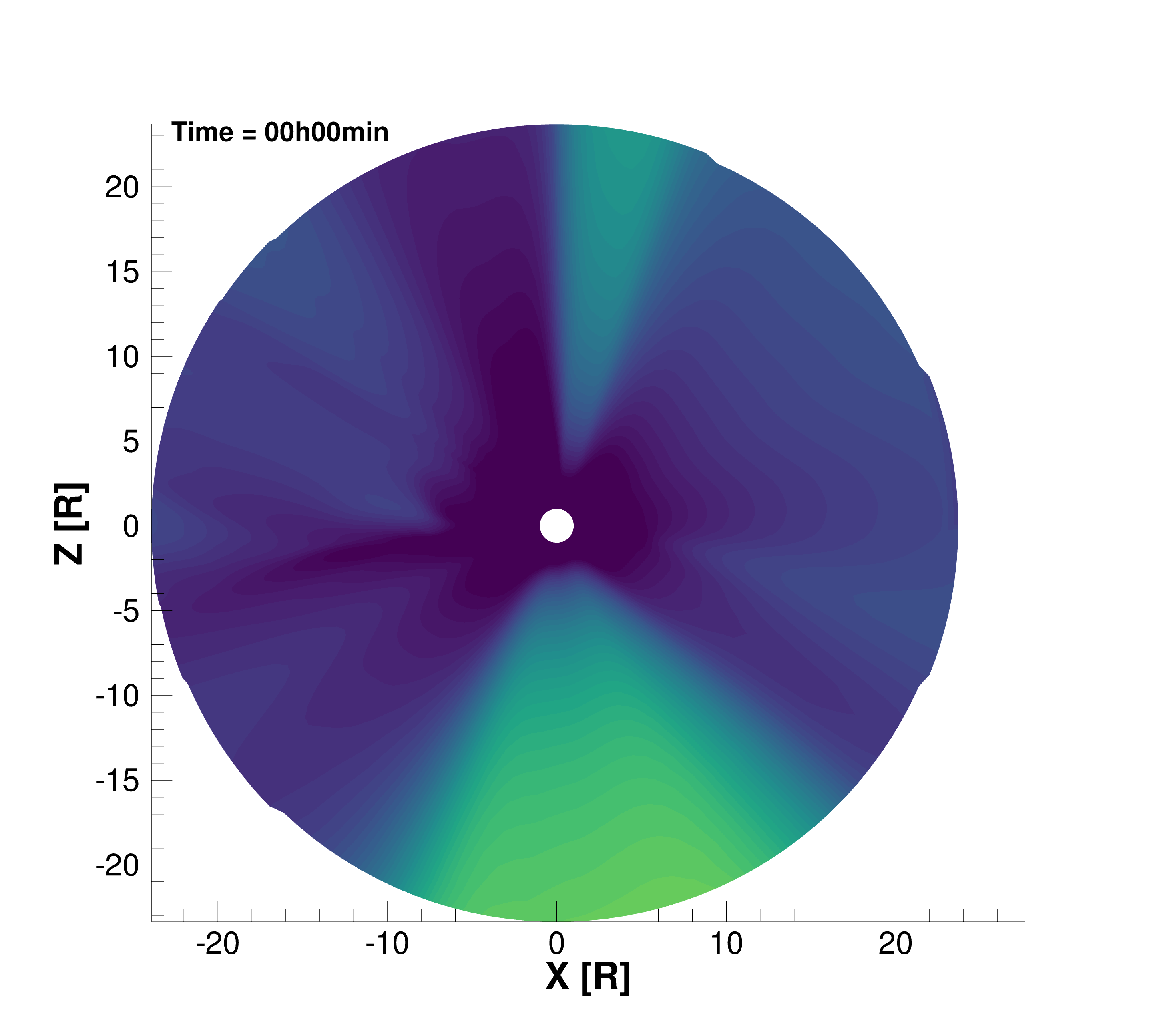}{0.265\textwidth}{(b) Case 2}{\hspace{-1cm}}
           \fig{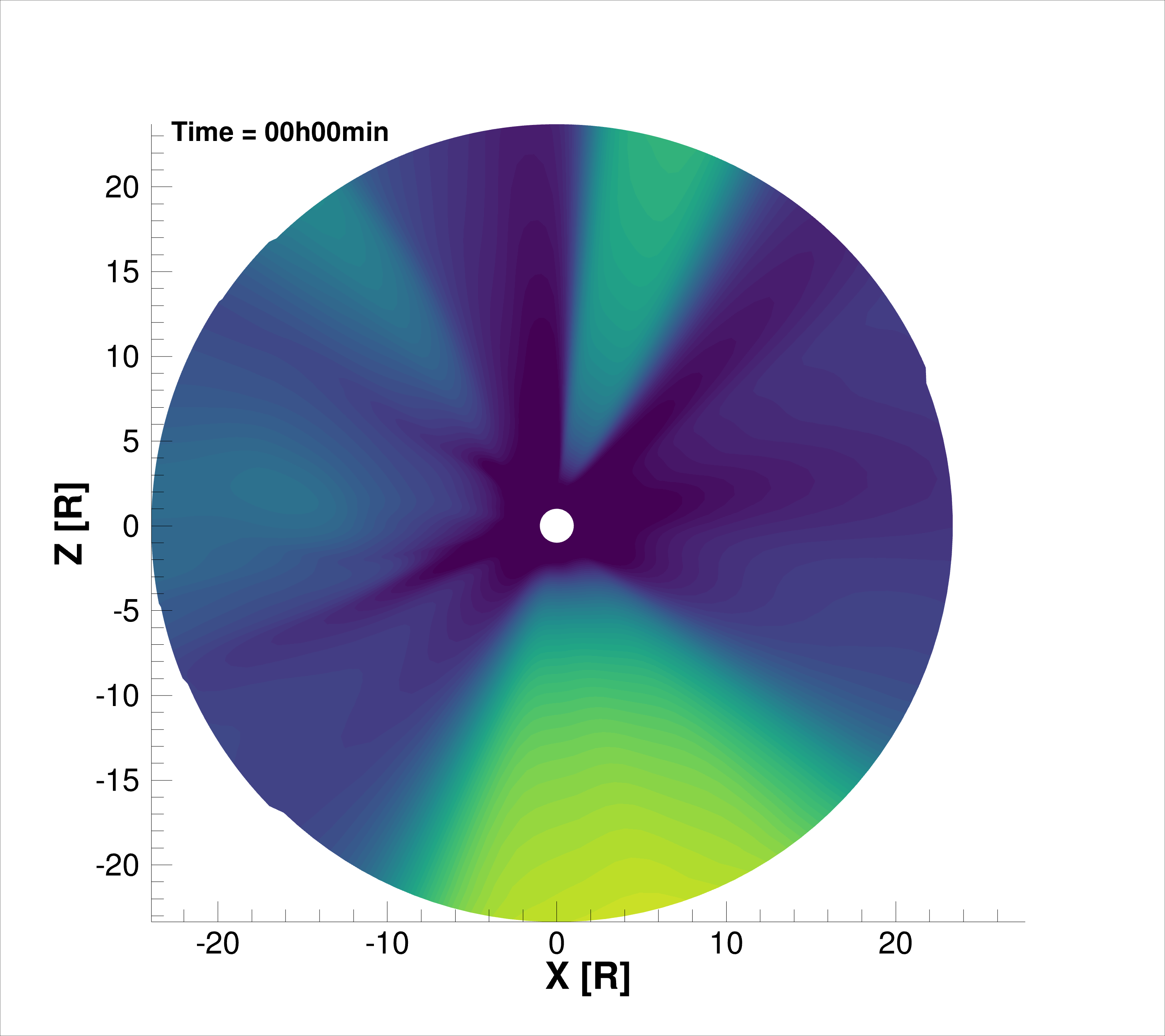} {0.265\textwidth}{(c) Case 3}{\hspace{-1cm}}
           \fig{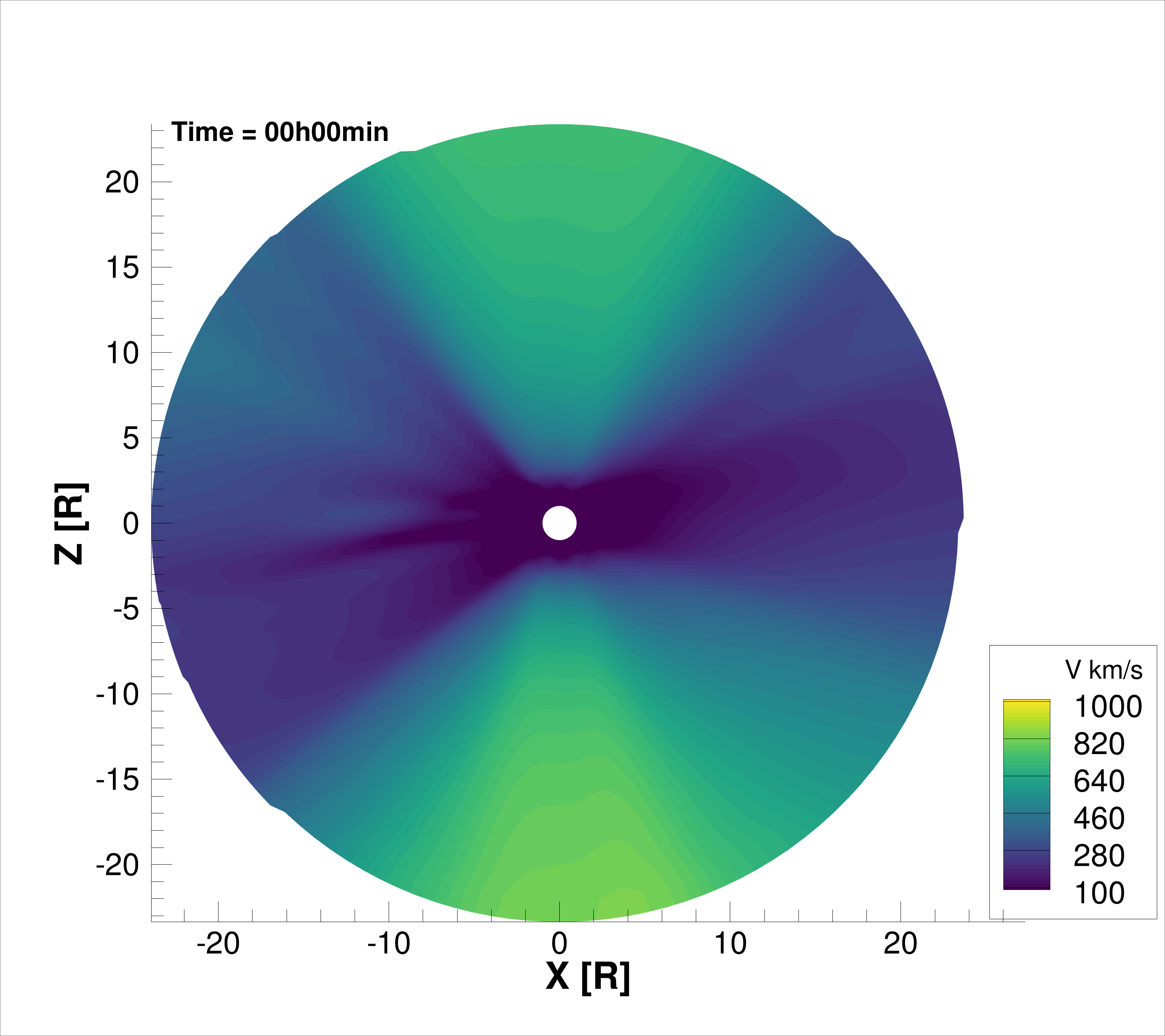}{0.265\textwidth}{(d) Case 4}}
 \vspace{-0.3cm}
\caption{
Ambient solar wind solution in the meridional plane at time t=0. From top to bottom: proton temperature (K, on log scale), density (cm$^{-3}$, on log scale) and speed (km\,s$^{-1}$) of the solar wind plasma at time t=0 when the flux rope is inserted in the background solution. Each column shows the AWSoM solution driven by the ADAPT GONG, ADAPT HMI, GONG and polar enhanced GONG map (cases 1, 2, 3, and 4), respectively.}\label{fig:SteadyState}
\end{figure*}
\begin{figure*}[t!]
\centering
\gridline{\fig{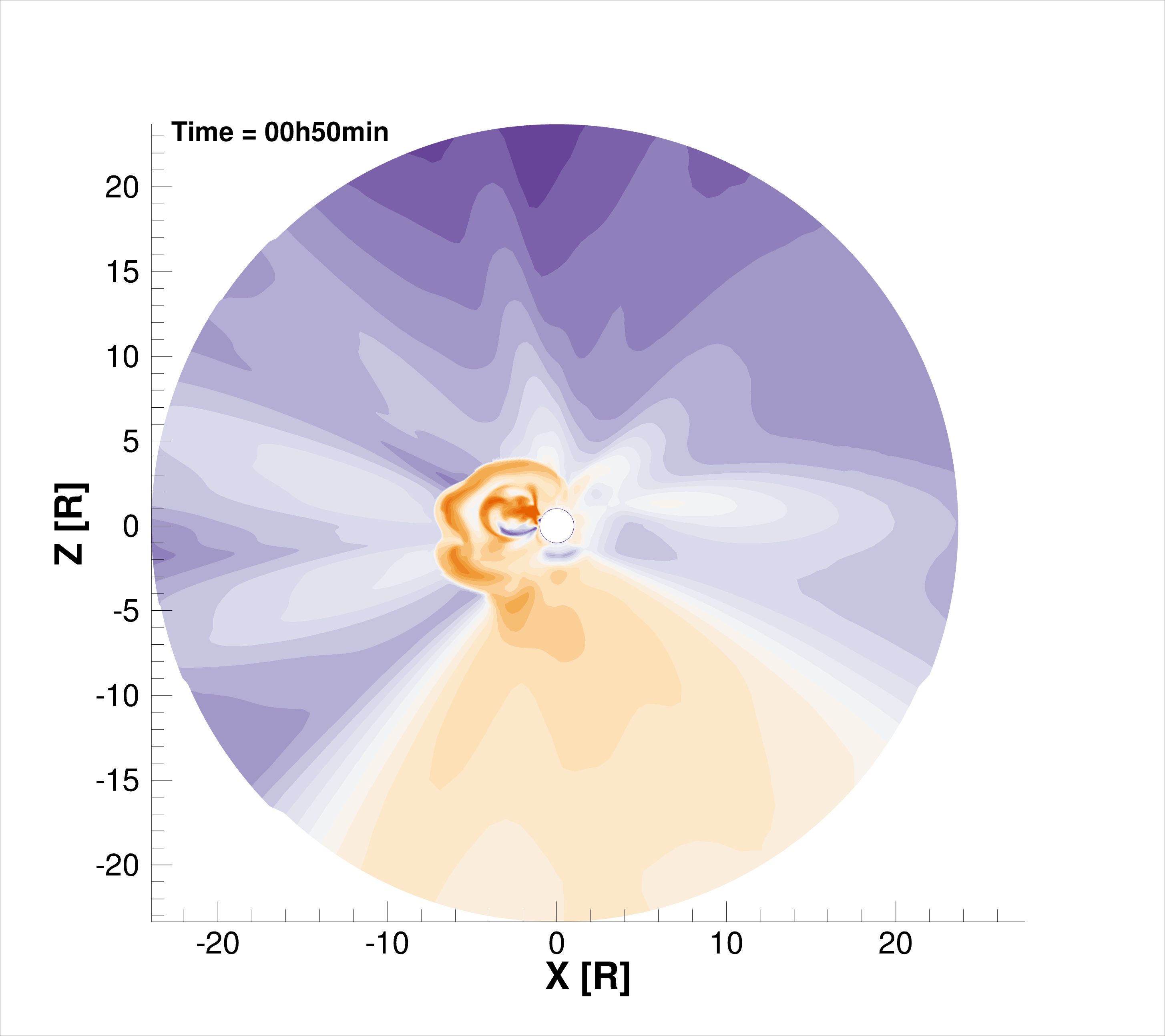}{0.245\textwidth} {}{\hspace{-1.5cm}}
           \fig{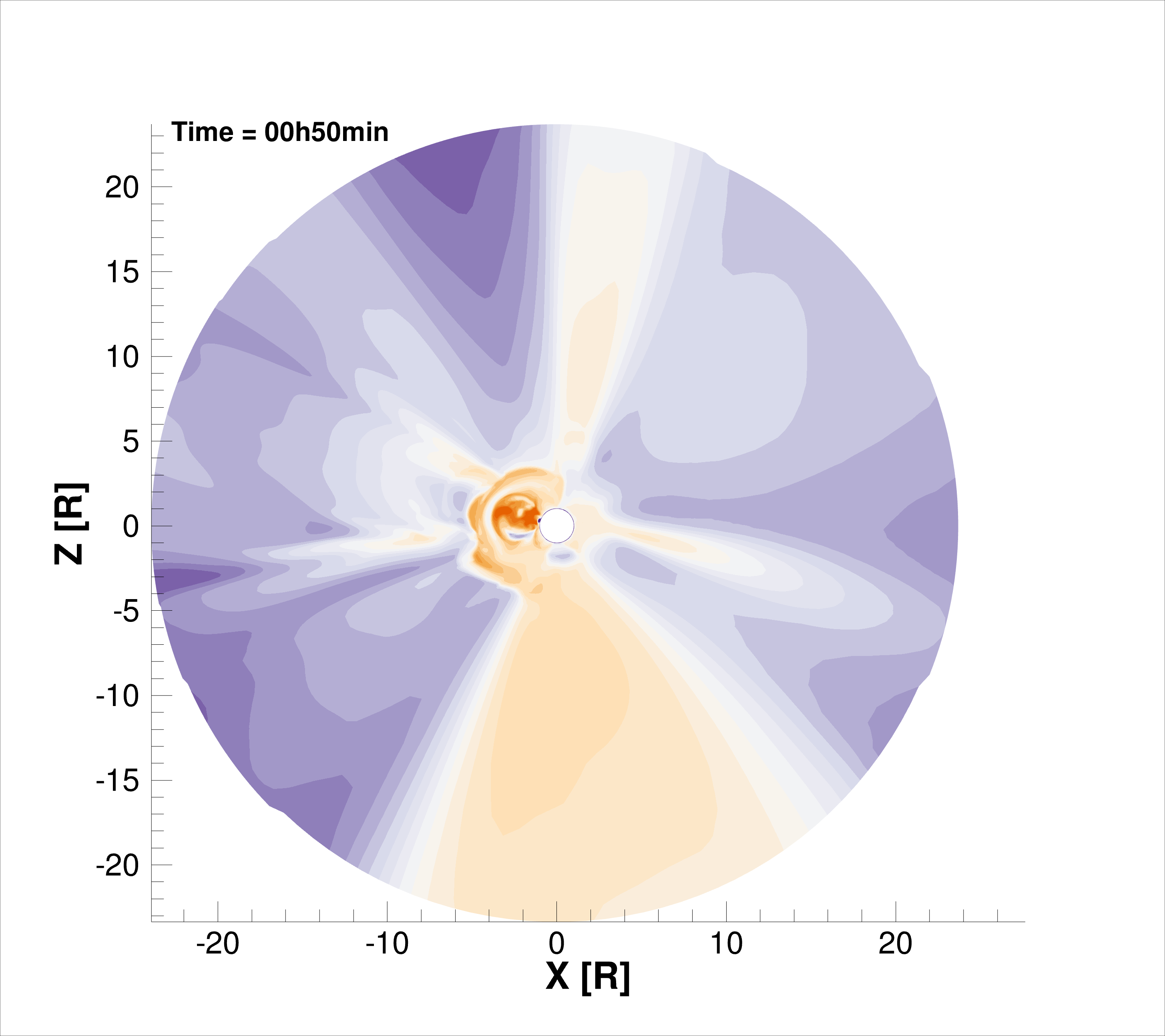}{0.245\textwidth}{}{\hspace{-1.5cm}}
          \fig{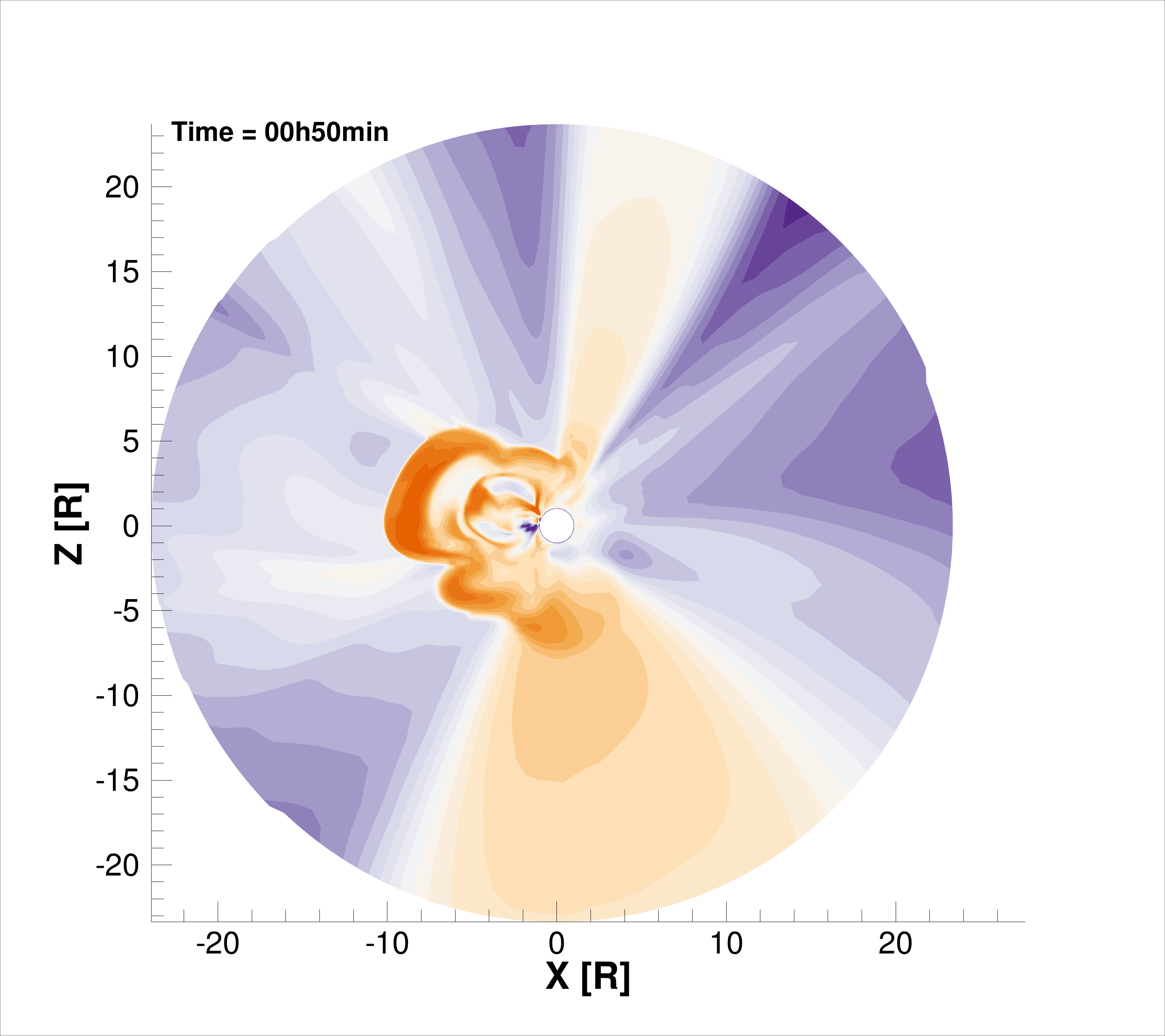}{0.245\textwidth}{}{\hspace{-1.5cm}}
          \fig{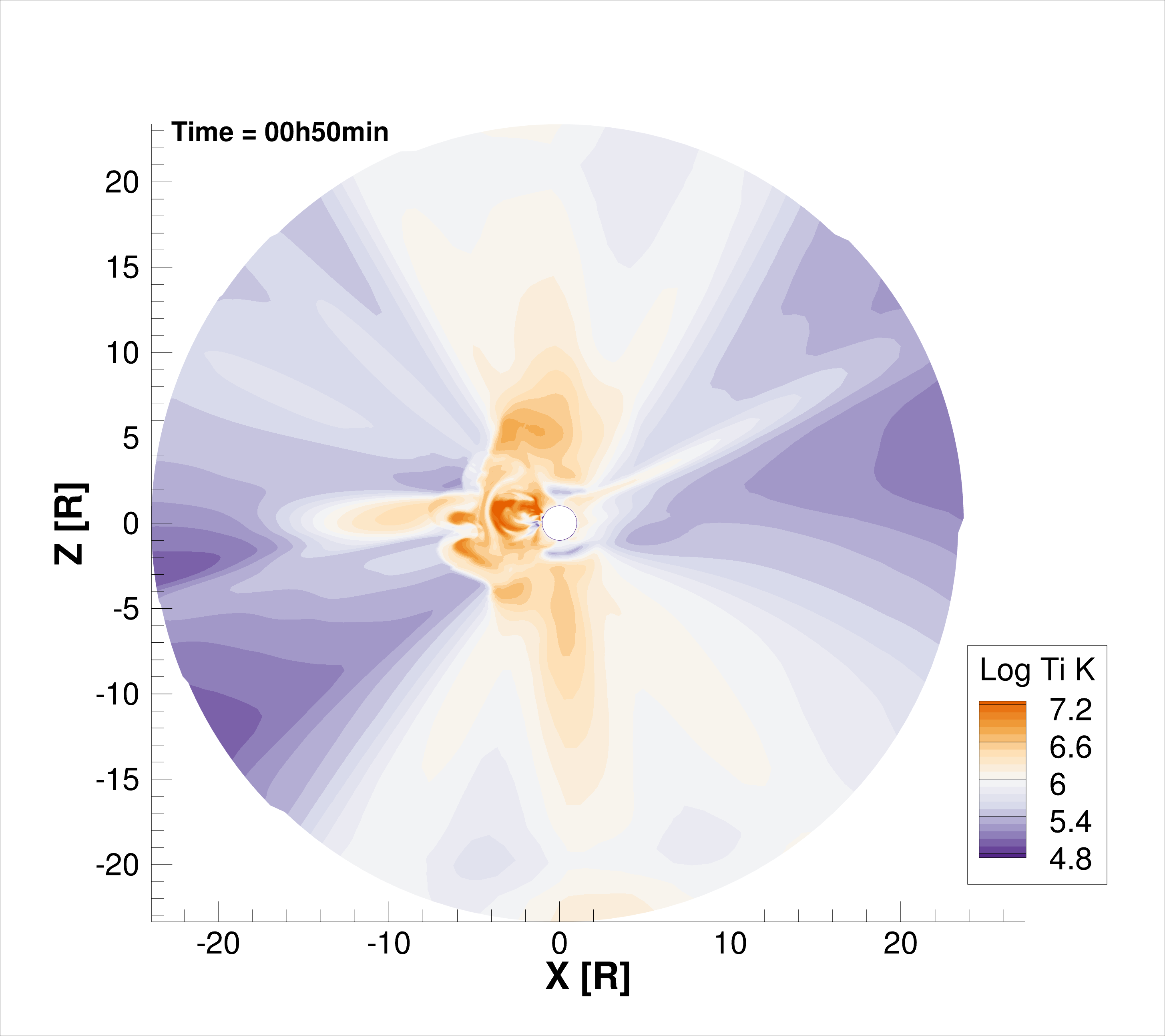}{0.245\textwidth}{}}
          \vspace{-1cm}
\gridline{\fig{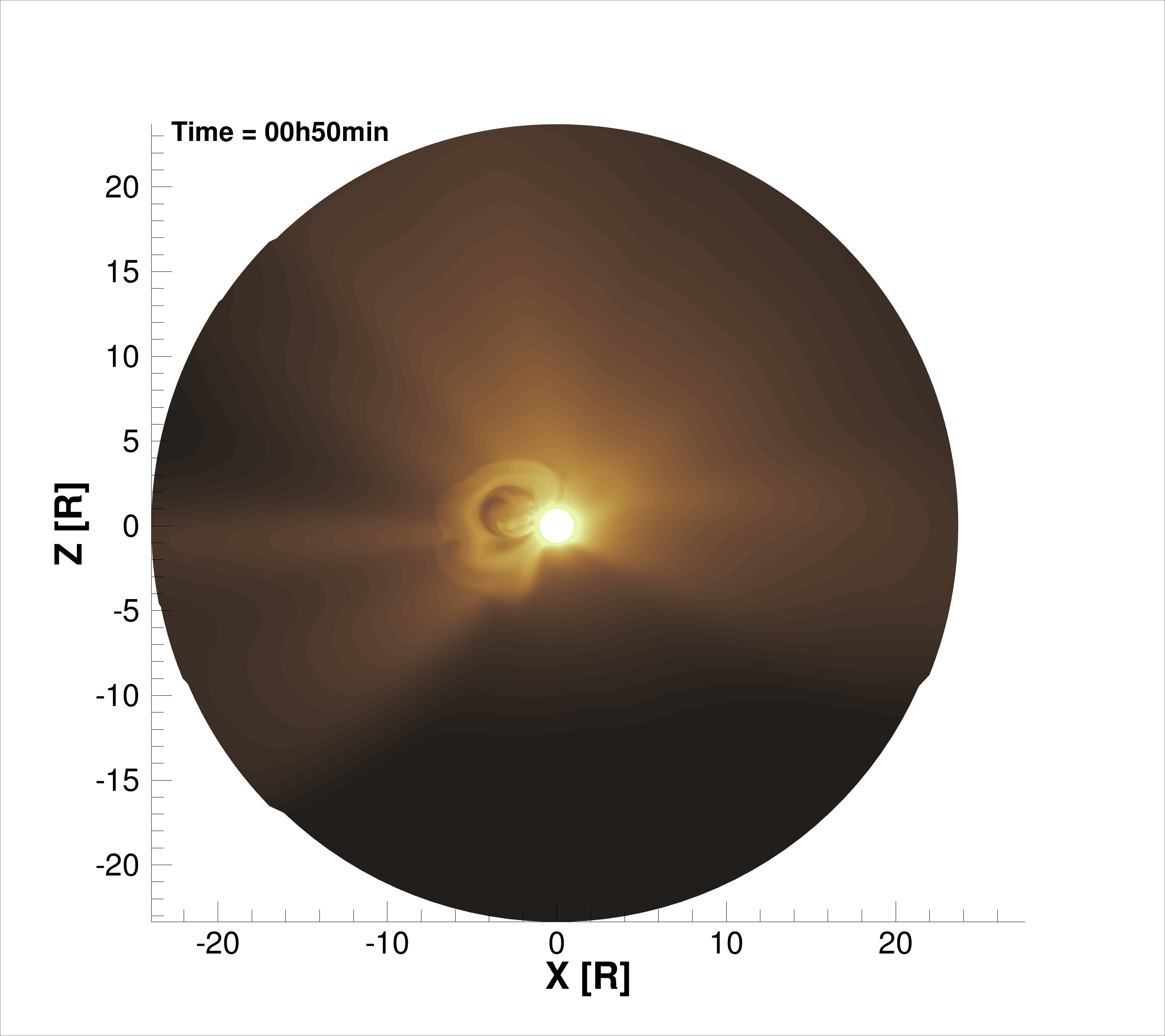} {0.245\textwidth} {}{\hspace{-1.5cm}}
           \fig{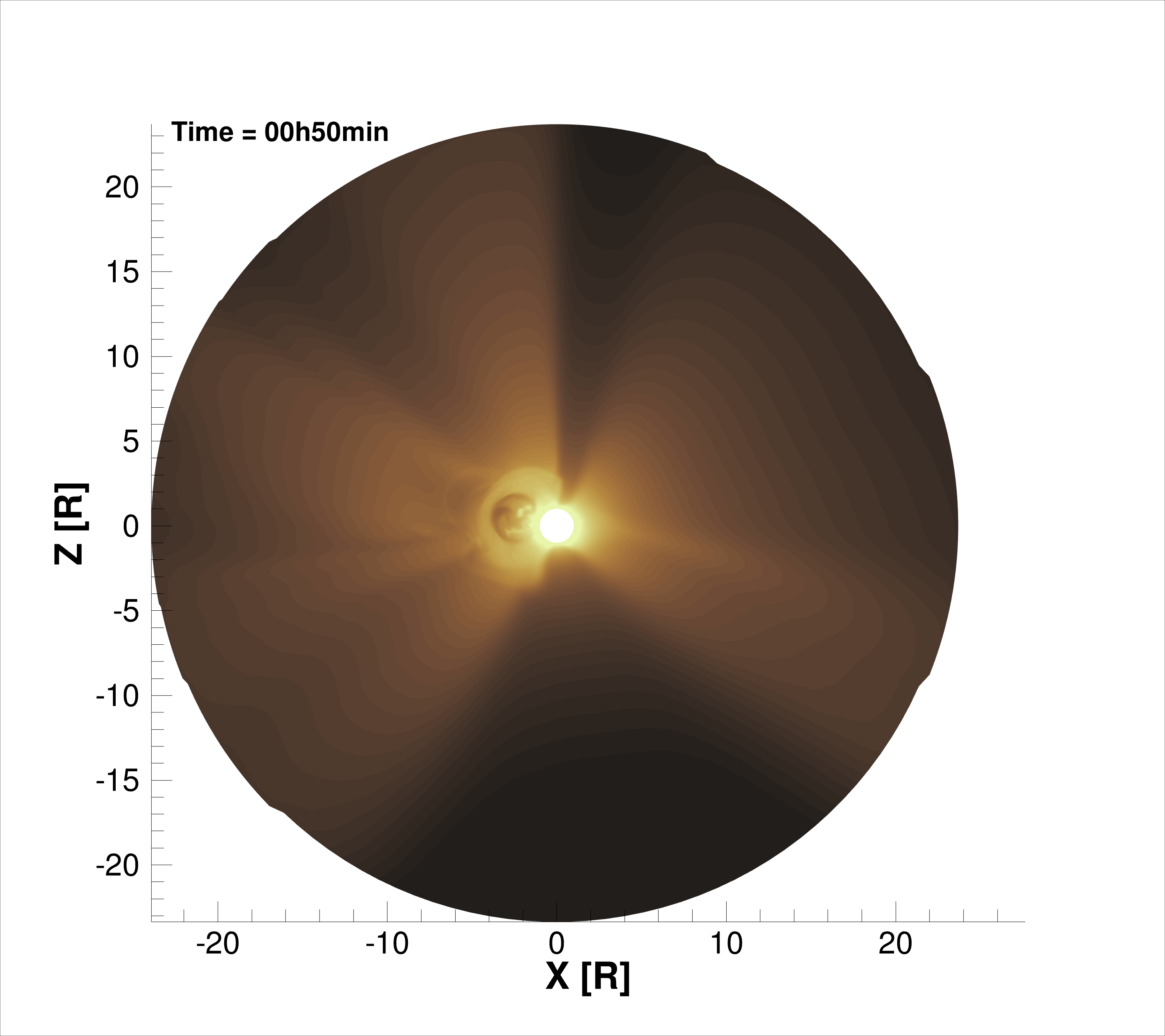}{0.245\textwidth}{}{\hspace{-1.5cm}}
          \fig{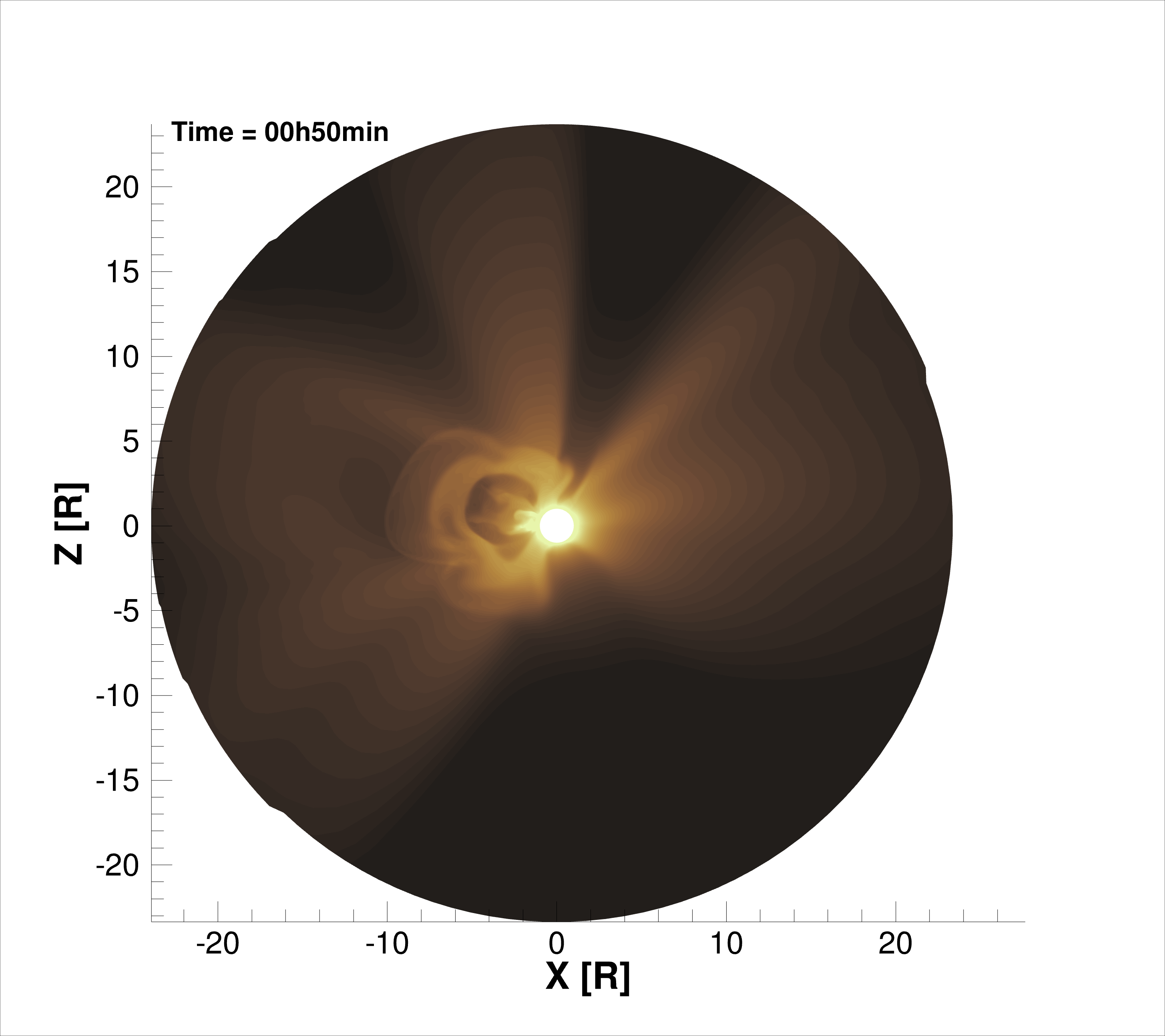}{0.245\textwidth}{}{\hspace{-1.5cm}}
          \fig{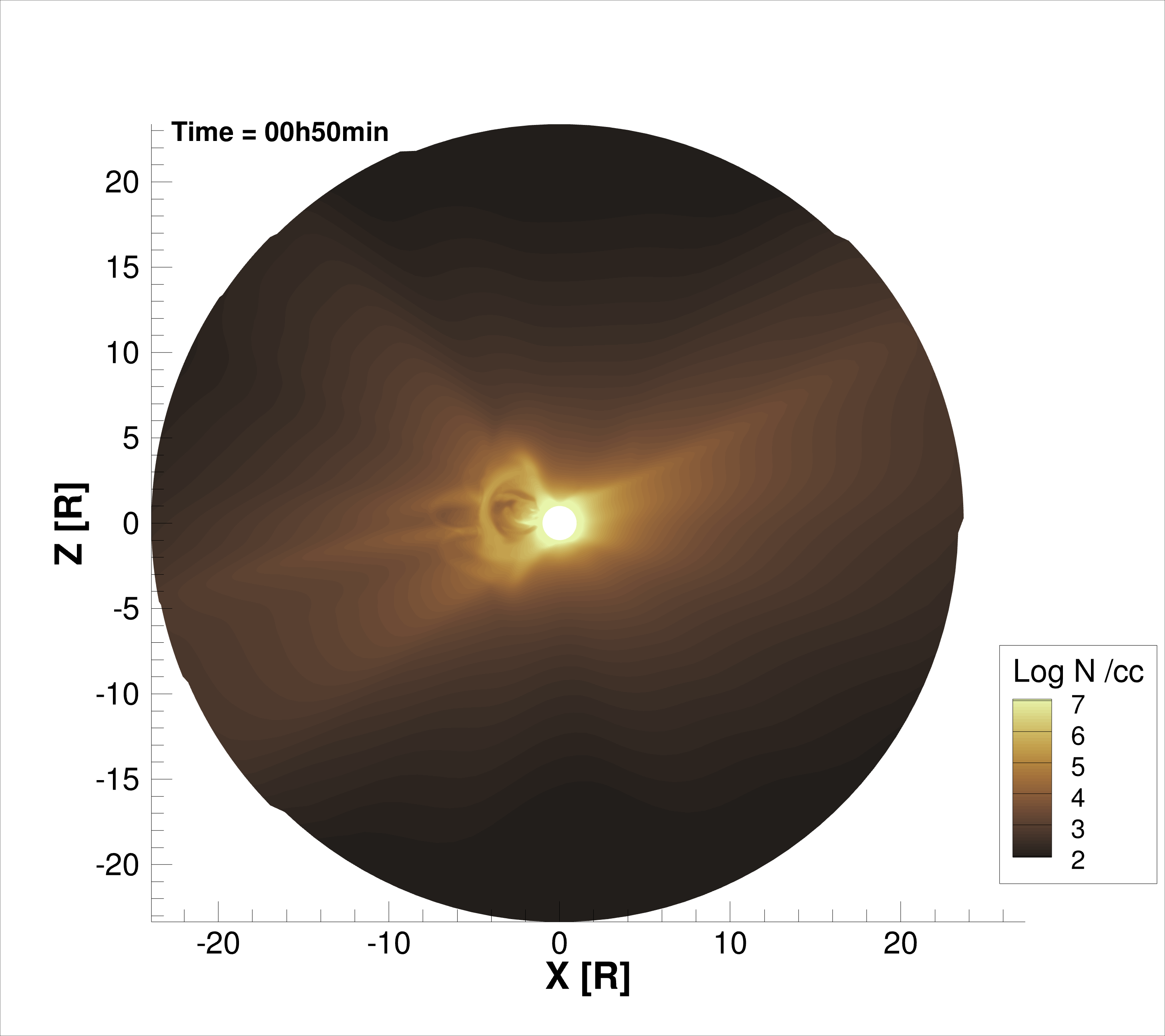}{0.245\textwidth}{}}
          \vspace{-1cm}
\gridline{\fig{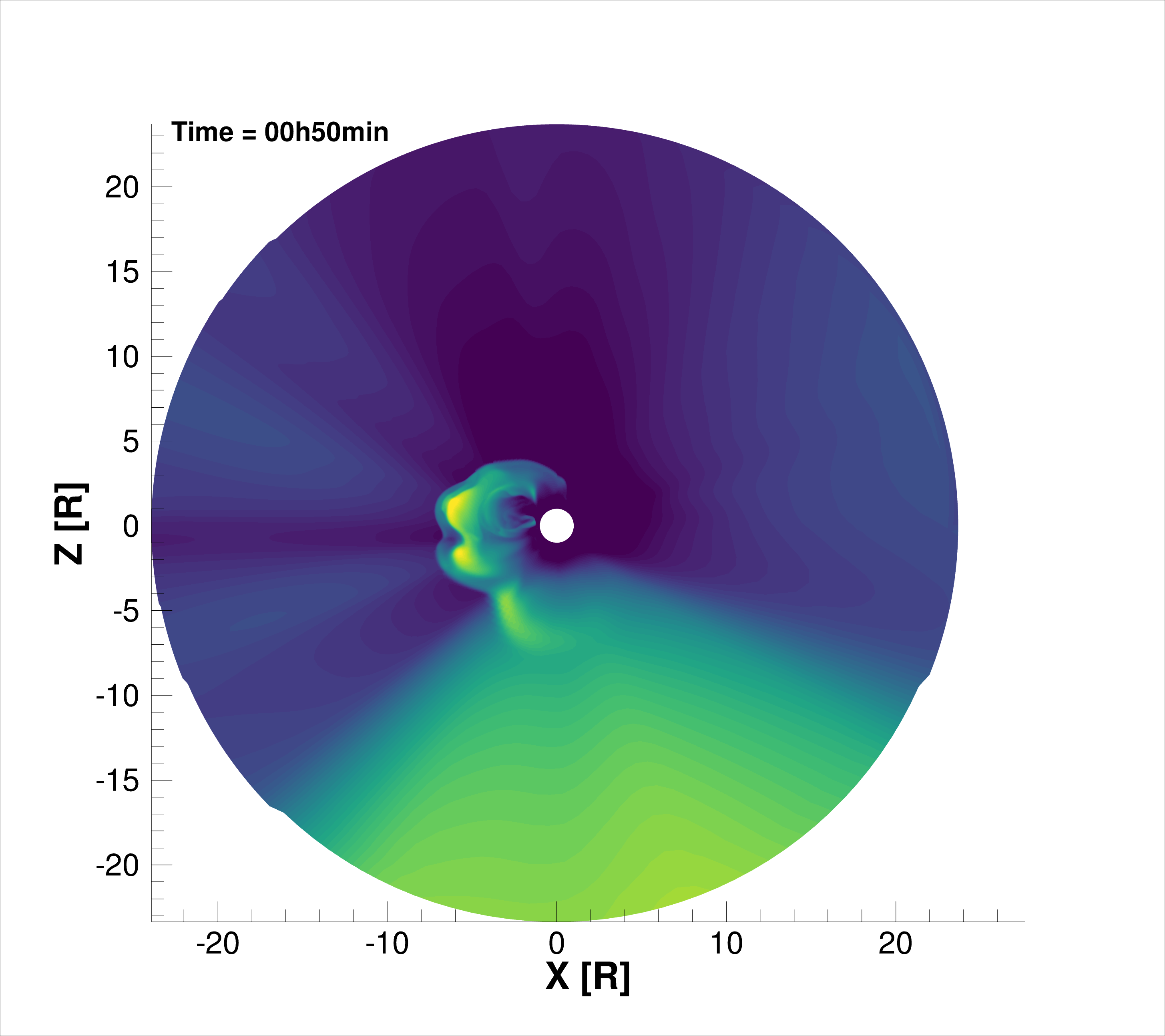}{0.245\textwidth}{}{\hspace{-1.5cm}}
        \fig{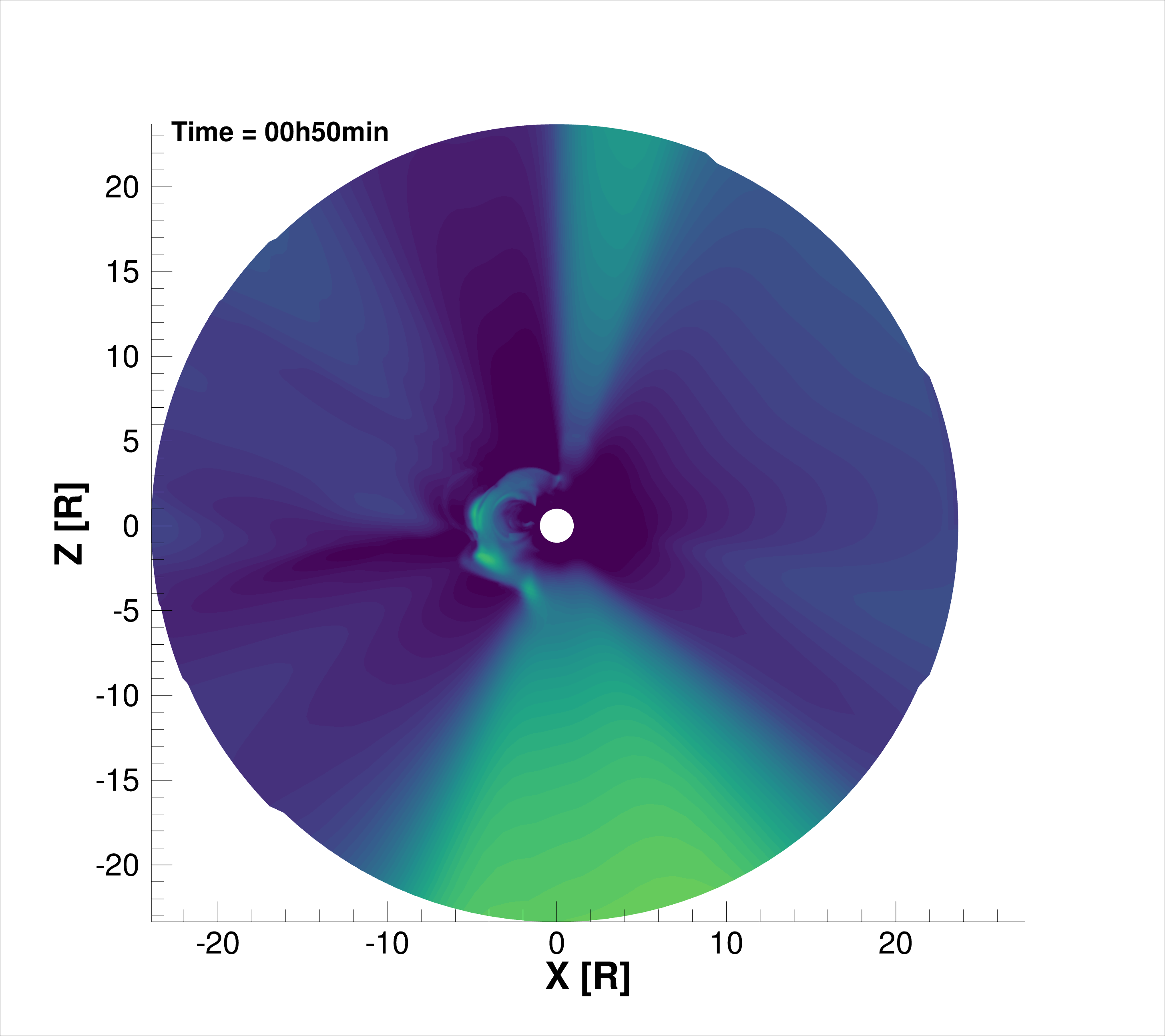}{0.245\textwidth}{}{\hspace{-1.5cm}}
        \fig{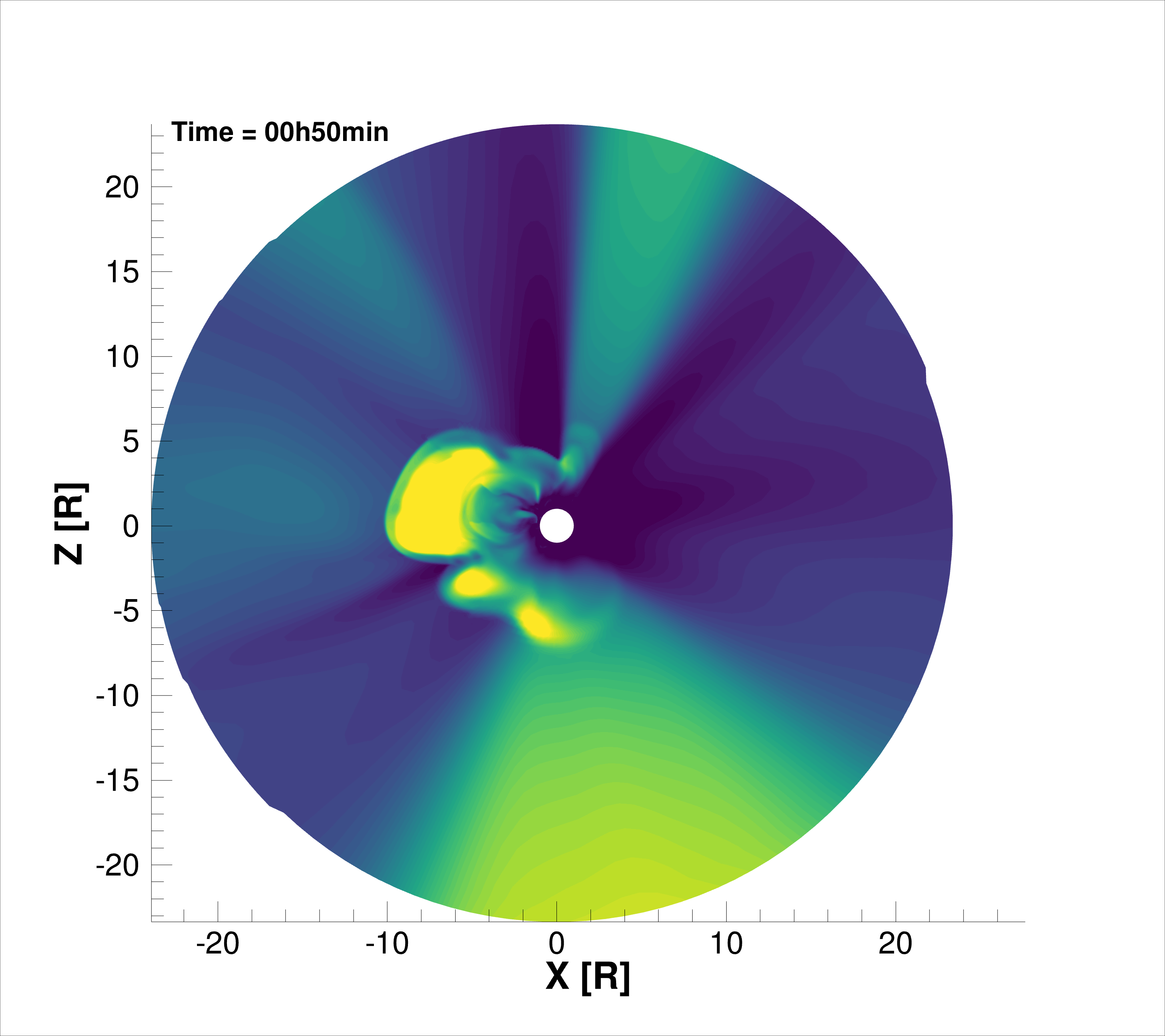}{0.245\textwidth}{}{\hspace{-1.5cm}}
        \fig{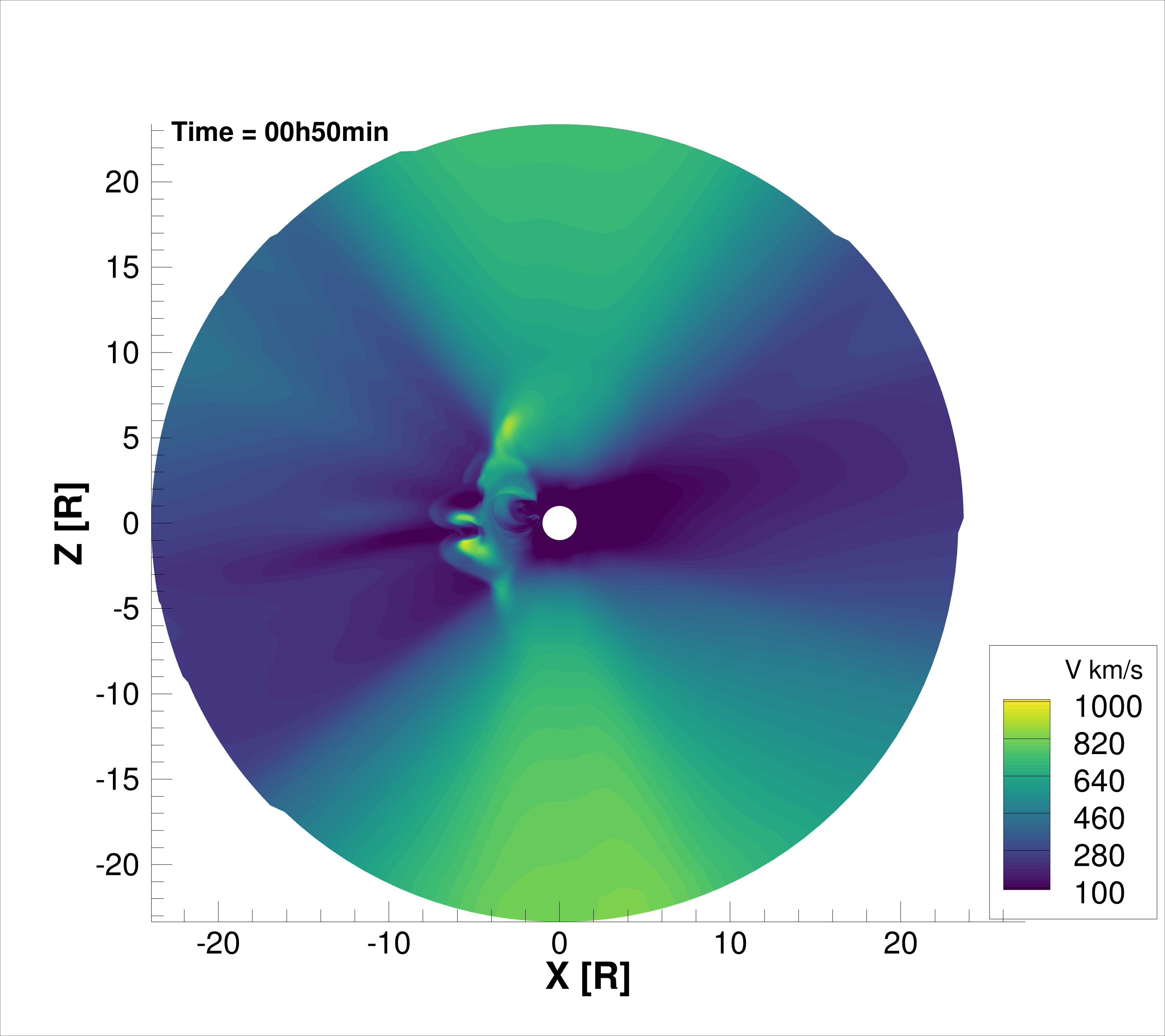}{0.245\textwidth}{}}
        \vspace{-1cm}
\gridline{\fig{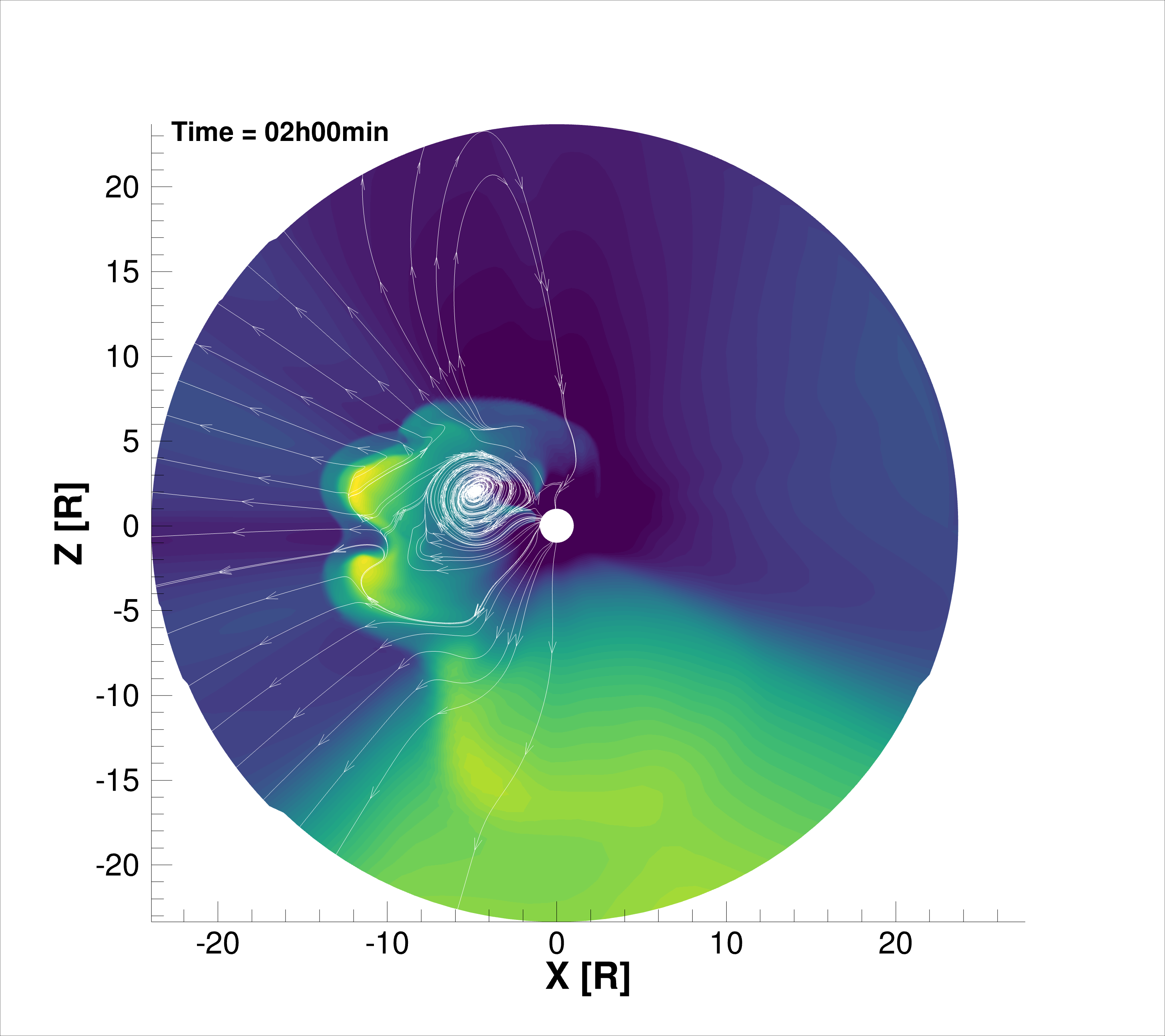}{0.245\textwidth}{(a) Case 1}{\hspace{-1.5cm}}
        \fig{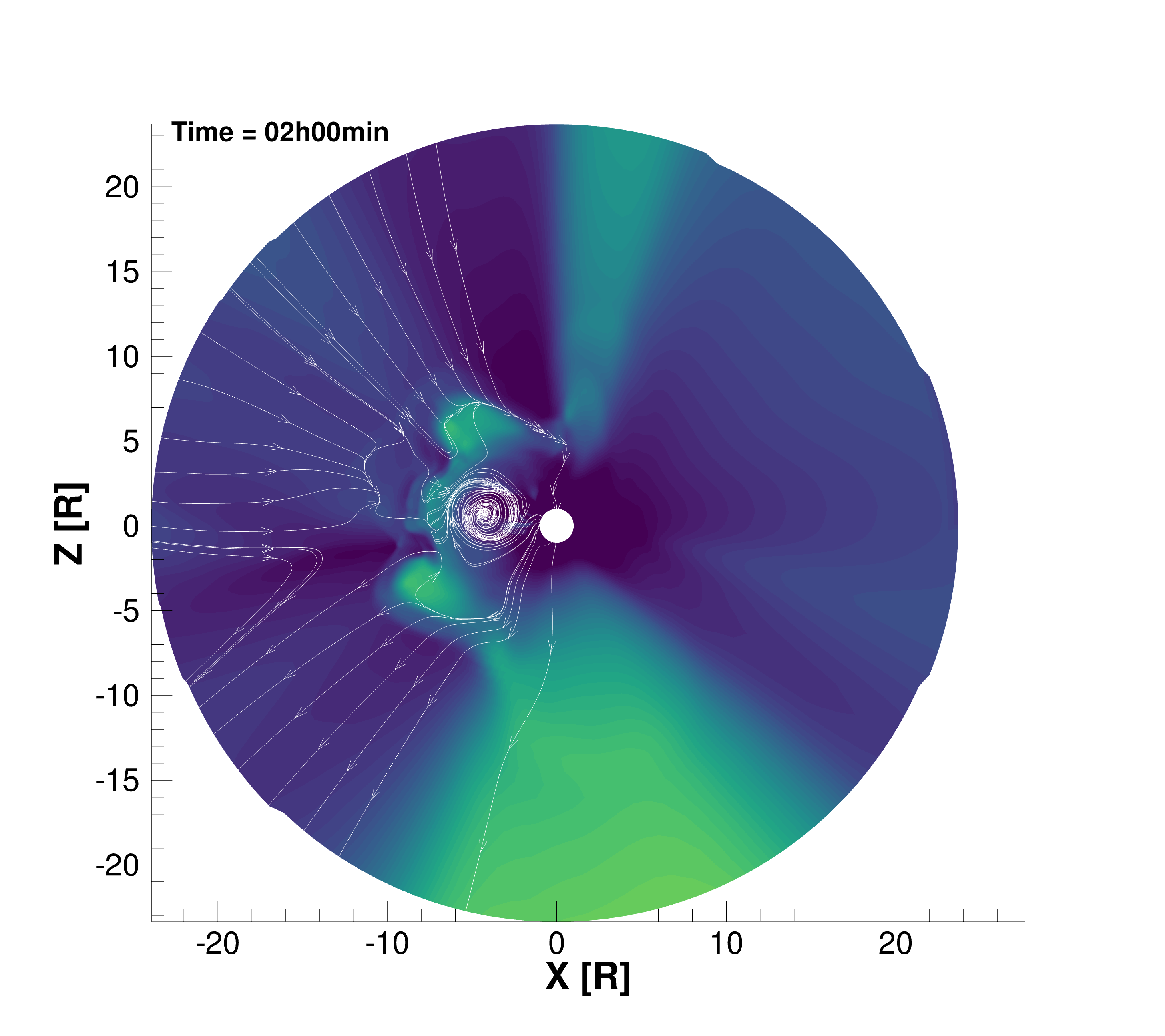}{0.245\textwidth}{(b) Case 2}{\hspace{-1.5cm}}
        \fig{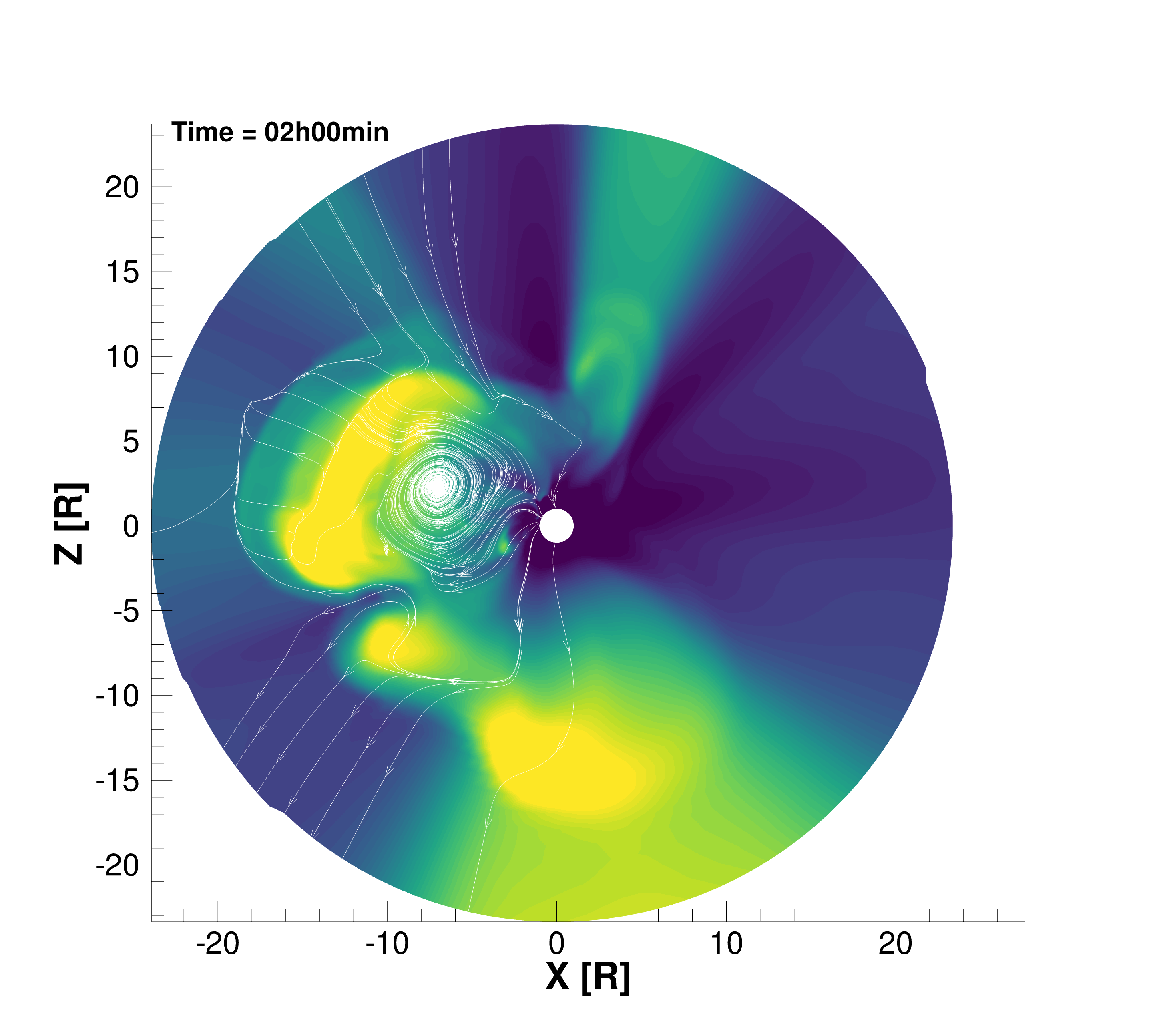}{0.245\textwidth}{(c) Case 3}{\hspace{-1.5cm}}
        \fig{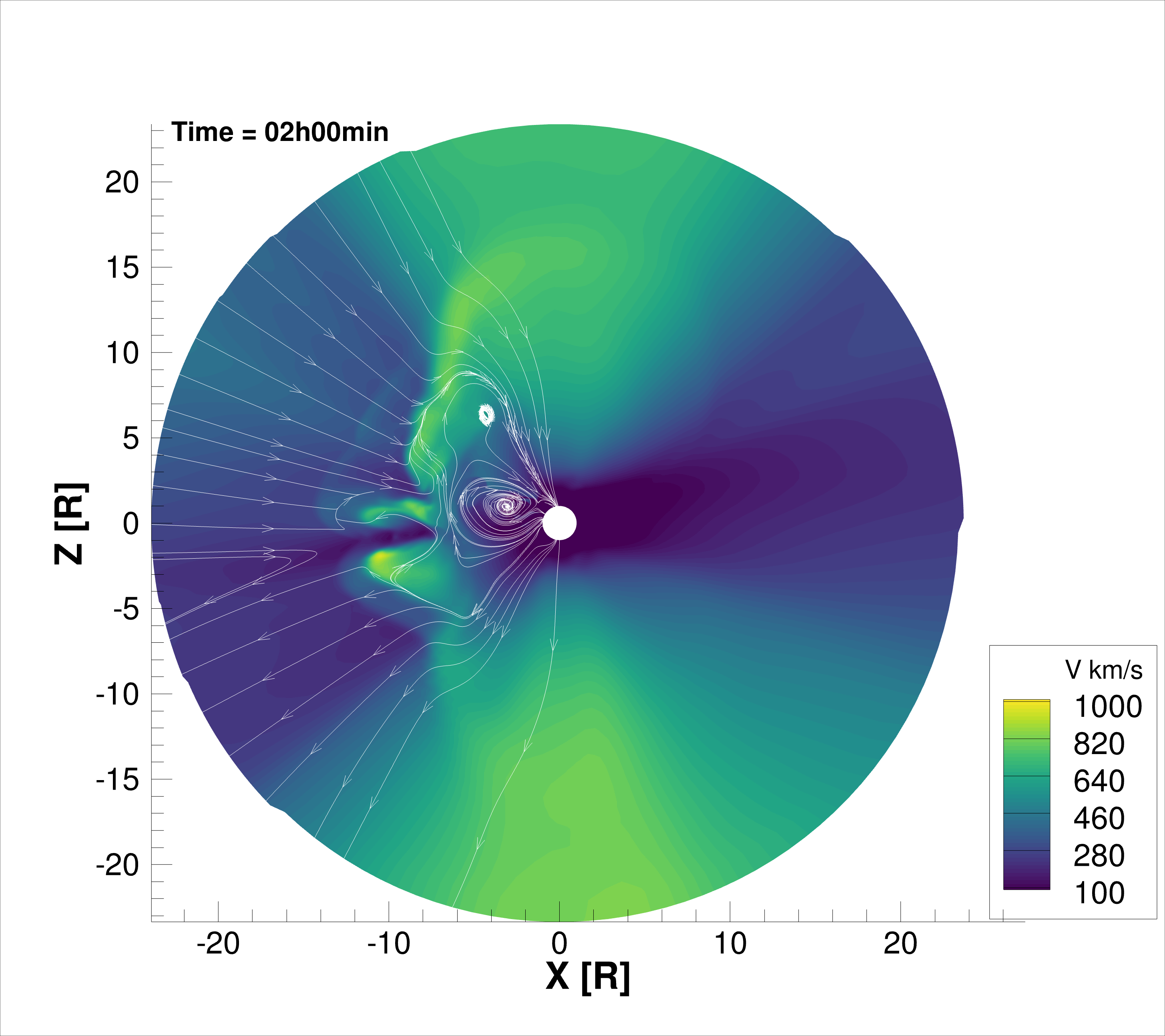}{0.245\textwidth}{(d) Case 4}}
 \vspace{-0.3cm}
\caption{Propagation of a CME in the meridional plane 
in the solar wind backgrounds driven by (1) ADAPT GONG, (2) ADAPT HMI, (3) GONG and (4) polar enhanced GONG maps. Top to bottom: The proton temperature (K, on log scale), density (cm$^{-3}$, on log scale) and speed (km/s) of the solar wind background and the CME traveling through it at time t=50 minutes after eruption. The last row shows the CME speed at time t=2\,hr\ 00\,min after eruption. The magnetic field lines are overplot to show the magnetic structure of the background and flux rope field.}\label{fig:y_cme}
\end{figure*}
\begin{figure*}[t!]
\centering
\includegraphics[trim=10 30 30 65, clip, width=0.243\textwidth]{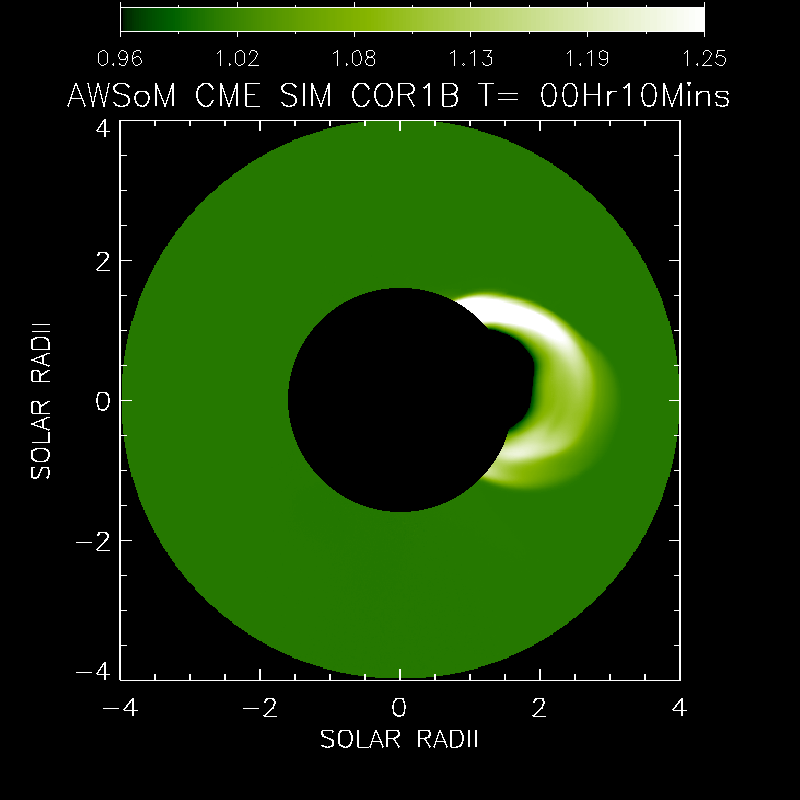}
\includegraphics[trim=10 30 30 65, clip, width=0.243\textwidth]{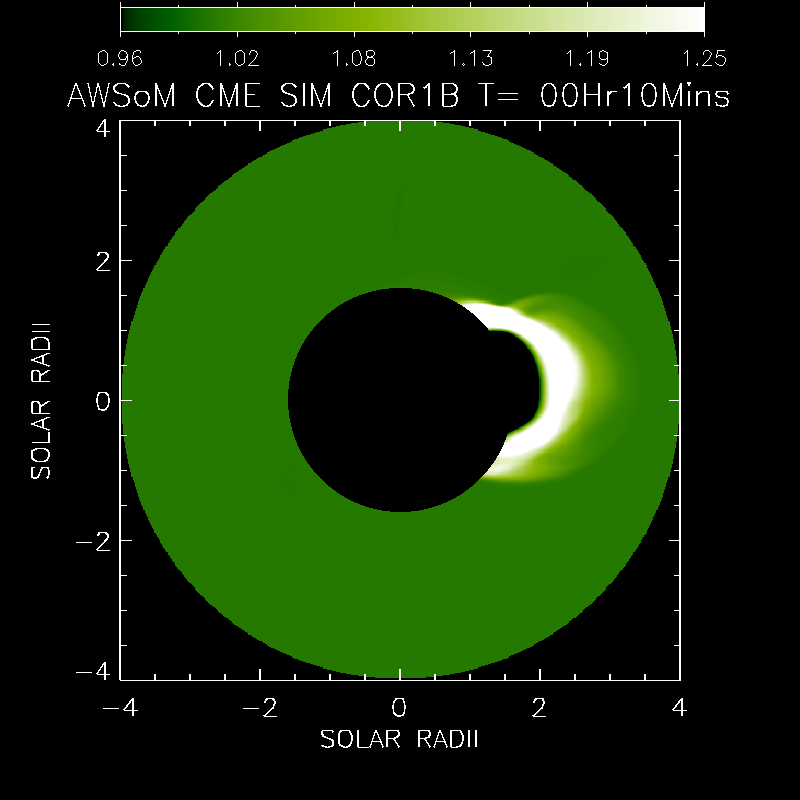}
\includegraphics[trim=10 30 30 65, clip, width=0.243\textwidth]{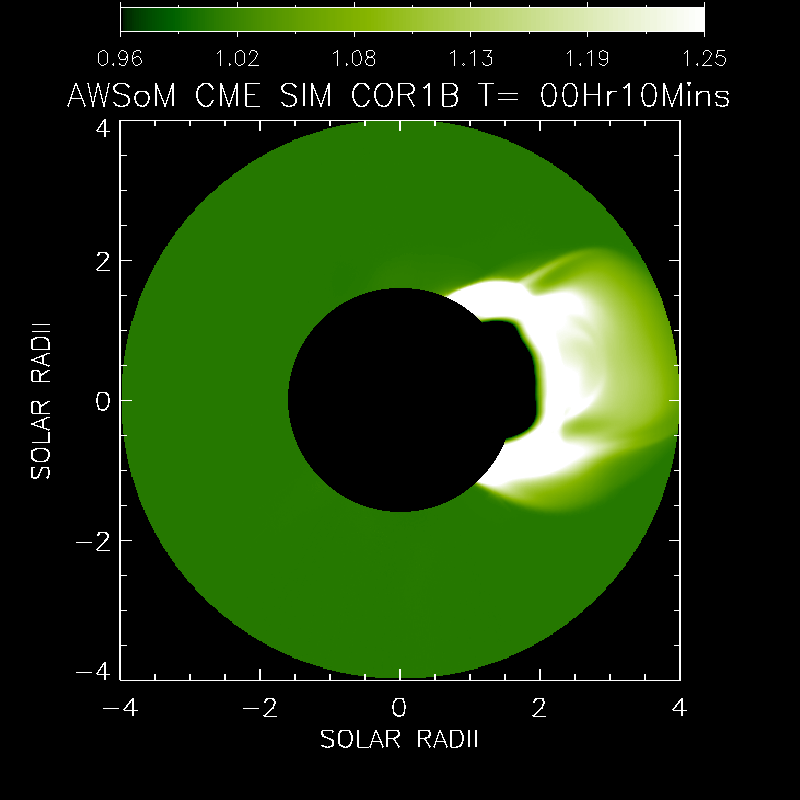}
\includegraphics[trim=10 30 30 65, clip, width=0.243\textwidth]{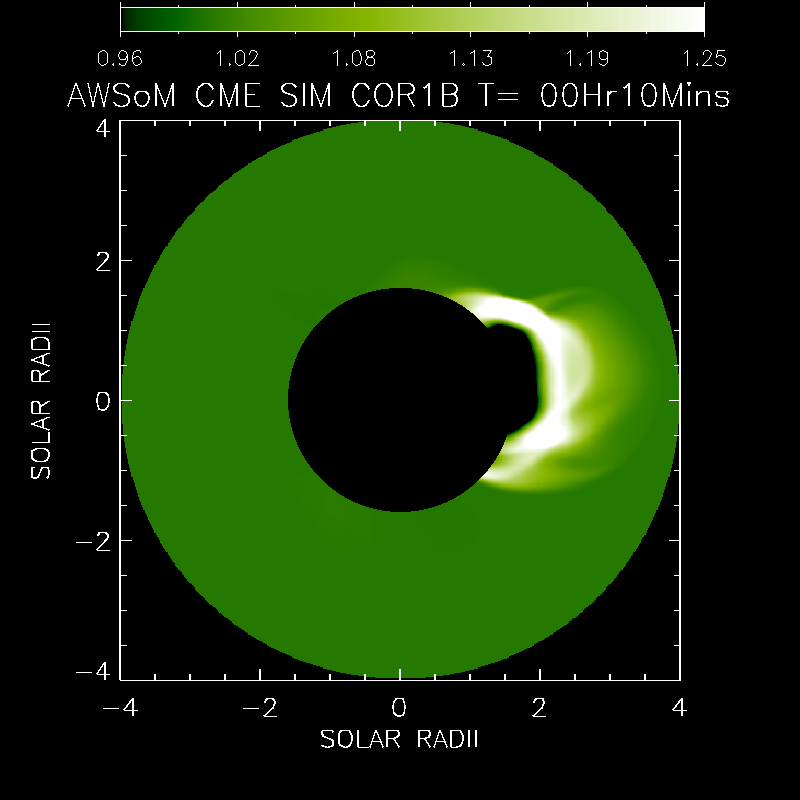}\\
\includegraphics[trim=10 30 30 65, clip, width=0.243\textwidth]{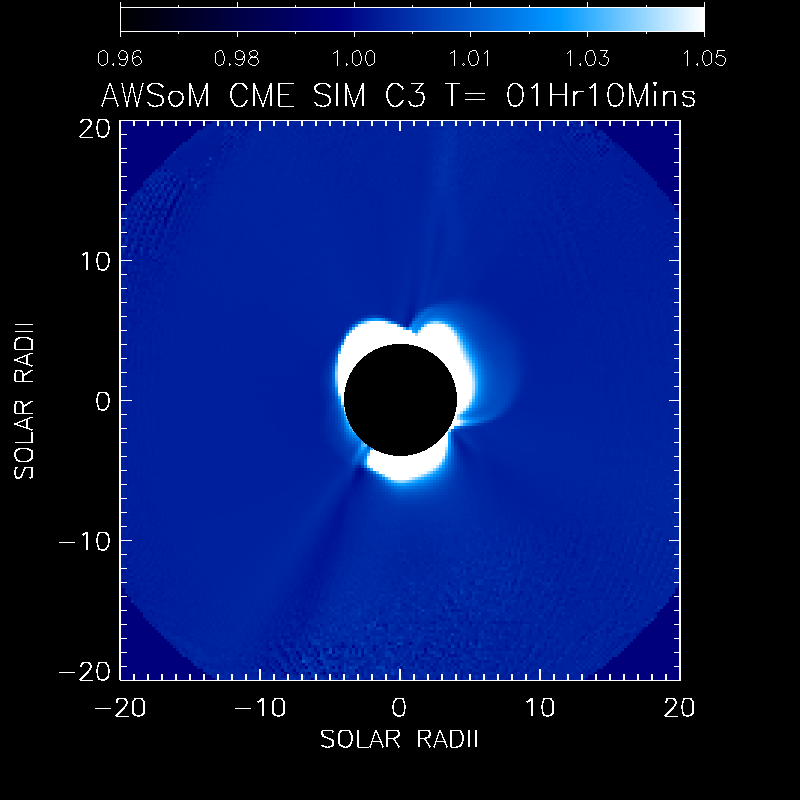}
\includegraphics[trim=10 30 30 65, clip, width=0.243\textwidth]{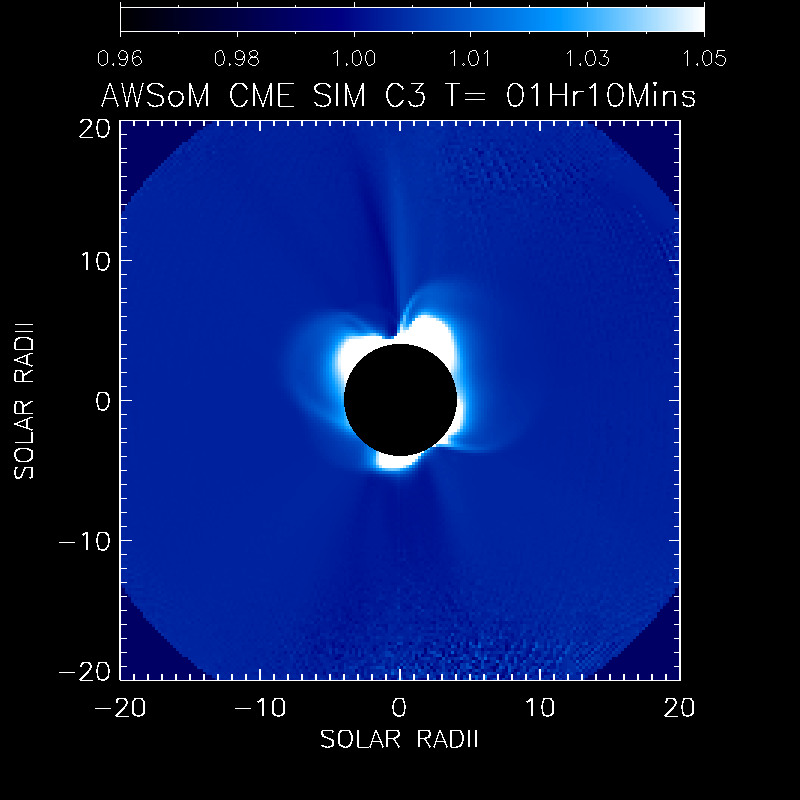}
\includegraphics[trim=10 30 30 65, clip, width=0.243\textwidth]{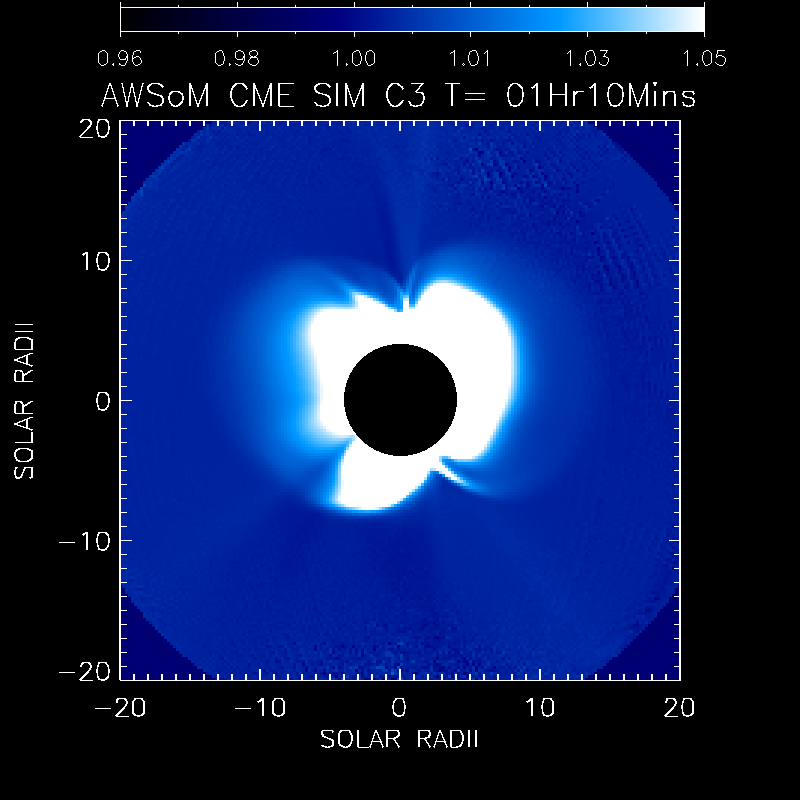}
\includegraphics[trim=10 30 30 65, clip, width=0.243\textwidth]{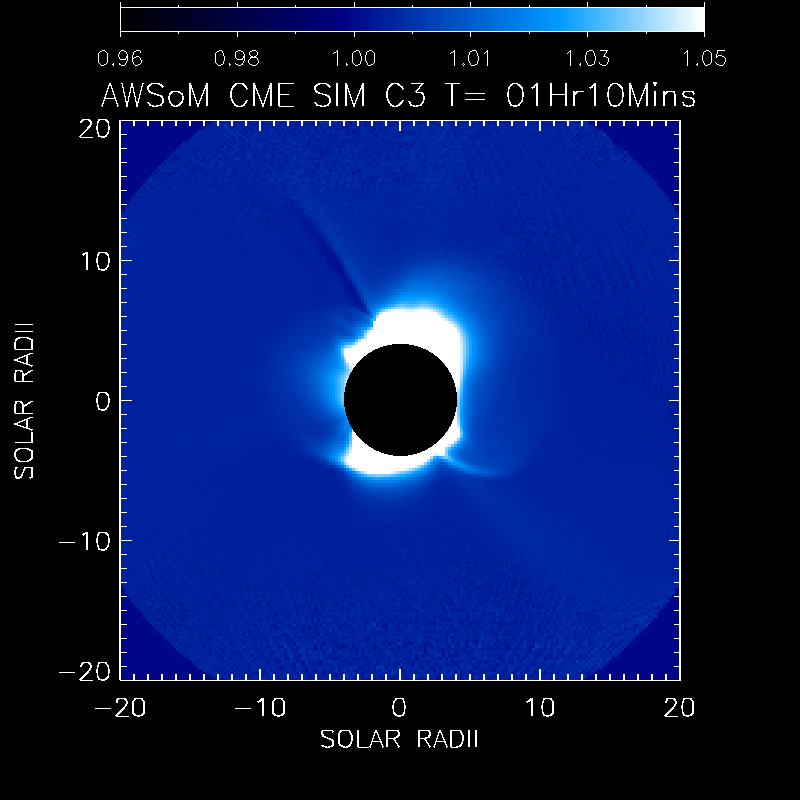} \\
\hspace{-0.5cm}(a) Case 1 \hspace{3.cm}(b) Case 2  \hspace{2.8cm}(c) Case 3 \hspace{2.8cm}(d) Case 4\\
\caption{Synthetic white-light coronagraph images. Top row shows the modeled line of sight image at 10 minutes corresponding to the STEREO-B COR 1 coronagraph field of view. Bottom row shows the modeled result at 1 hour 10 minutes corresponding to the SOHO/LASCO C3 coronagraph field of view. }\label{fig:WL}
\end{figure*}
Various magnetic field data products, including GONG, HMI, National Solar Observatory-HMI-Near-Real Time (NSO-HMI-NRT) and ADAPT maps have been used to drive AWSoM and validated against observations \citep{Huang:2023, Jin:2012, Sachdeva:2023}. In this work, we study the evolution of a GL flux rope CME in the solar wind solutions driven by four different magnetic field conditions for the same time period (CR2123). We utilize 3 sources of magnetic field data -  the ADAPT GONG, ADAPT HMI maps and the synoptic GONG map to prescribe the inner boundary conditions for AWSoM. Panels (a) and (b) of Figure~\ref{fig:Maps} show the ADAPT GONG map (Realization 01) and ADAPT HMI map (Realization 04) centered on 2012-05-12 20:00 respectively and panel (c) shows the synoptic GONG magnetogram for this CR. 
We apply a polar enhancement and an increase in the weak magnetic field regions in the GONG map (referred to as polar enhanced GONG map) and use it as the fourth input map (panel (d) in Figure~\ref{fig:Maps}). The polar magnetic field is increased using the following formula: 
$B_{r}' = B_{r}(1 + 2\sin^2\theta)$
where $\theta$ is the latitude. Furthermore, we increase the magnetic field  using: $B_{r} = \mathrm{sign}(B_{r,obs})\min(c|B_{r,obs}|, B_{r,min} +|B_{r,obs}|)$ where, $\mathrm{sign}$ is the sign function taking values $\pm1$,  $B_{r,obs}$ is the observed radial field from the magnetogram, $B_{r,min}= 5\,$G, and the multiplicative factor $c = 3.75$. This results in 4 sets of observation-based magnetic field maps used to drive the AWSoM solar wind solution - ADAPT GONG, ADAPT HMI, GONG map, and polar enhanced GONG map, hereafter referred to as cases 1, 2, 3, and 4 respectively in the text.
\section{Simulation set up}\label{sec:Sim}
The magnetic field configurations described in Section \ref{sec:Maps} are used to prescribe the radial magnetic field at the inner boundary of AWSoM. The tangential components have floating boundary conditions.
The proton number density at the inner boundary is overestimated to prevent chromospheric evaporation  \citep{Lionello:2009,vanderHolst:2014awsom}, and the initial electron and proton temperature (both perpendicular and parallel proton temperature) are set to 50,000 K. The Poynting flux $S_{A}$ that describes the energy flux of the outward propagating \alf waves \citep{Fisk:2001a,Fisk:1999b, Sokolov:2013} is proportional to the magnetic field strength at the inner boundary B$_{\odot}$. The proportionality factor that we refer to as the Poynting flux parameter, is set to $S_A/B_{\odot}=0.3\,$ MW\,m$^{-2}$T$^{-1}$ based on a previous study for this CR in \cite{Sachdeva:2021}. The \alf wave correlation length (L$_{\perp}$), transverse to the direction of the magnetic field is proportional to B$^{-1/2}$ \citep{Hollweg:1986} and proportionality factor is set to $L_{\perp}\sqrt{B}=1.5\times10^{5}$ m\,$\sqrt{T}$.

The SC component within SWMF extends from 1 to 24\,\Rs\ on a 3D spherical grid \citep{Toth:2012swmf,vanderHolst:2010} decomposed into blocks with $6\times8\times8$ grid cells .
To refine the computational grid, we use Adaptive Mesh Refinement (AMR) which provides an angular resolution of 1.4$^{\circ}$ below 1.7 \Rs and 2.8$^{\circ}$ in the remaining domain. Two additional levels of refinement (with an angular resolution of 0.35$^{\circ}$) are applied in a box extending from 1 to 24\,\Rs in the radial direction, and $\pm 20^{\circ}$ in longitude and latitude around the active region from where the flux rope CME is launched. This is done to make sure that the grid along the propagation direction of the CMEs is well-resolved. Additional AMR is performed below 1.7 \Rs along with the 5th order shock-capturing scheme \citep{Chen:2016} to produce high resolution line of sight synthetic extreme ultraviolet (EUV) images. There are $\approx$ 19 million cells in the SC domain. For the steady state background solar wind solution, the SC component is run for 80,000 iterations following which a flux rope is launched in the time-dependent mode for 3 hours.

In each simulation, the GL flux rope is inserted along the polarity inversion line (PIL) of AR 11476, circled in the ADAPT GONG map (panel (a) in Figure~\ref{fig:Maps}). We emphasize that this is not a real/observed CME and this strong active region was chosen because of its location near the central meridian. We prescribe the GL flux rope with five parameters describing the magnetic field strength, radius, helicity, distance of the torus center from the center of the Sun, and the stretch parameter that determines the flux rope shape. The initial position (longitude, latitude and orientation) of the flux rope is determined from the location of the active region on the magnetic field map and the helicity is determined based on the hemispheric rule: positive (dextral)/ negative (sinistral) for southern and northern hemispheres, respectively \citep{Borovikov:2017, Jin:2017a, Jin:2017b}. The GL flux rope parameters are: longitude= 183.5$^\circ$, latitude = -10.5$^\circ$ (of the AR), radius = 0.8\,\Rs, orientation = 0.15$^\circ$, stretch parameter = 0.6\,\Rs, apex height of the flux rope = 1\,\Rs, and helicity = -1. We refer the reader to \cite{Jin:2017a, Jin:2017b, Jin:2018} for a detailed description of these parameters. Upon insertion, the flux rope is in a state of force imbalance leading to an immediate eruption of the CME into the solar wind solution in the time-accurate mode.
For consistency, we specify the same model input parameter values for the ambient solar wind and GL flux rope in all four cases to asses the differences arising in the evolutionary properties driven by the four input maps. In this study, we focus on the plasma and magnetic properties in the coronal domain only ($R< 24$ \Rs).
\begin{figure*}[t!]
    \includegraphics[trim=10 0 0 30, clip, width=\textwidth]{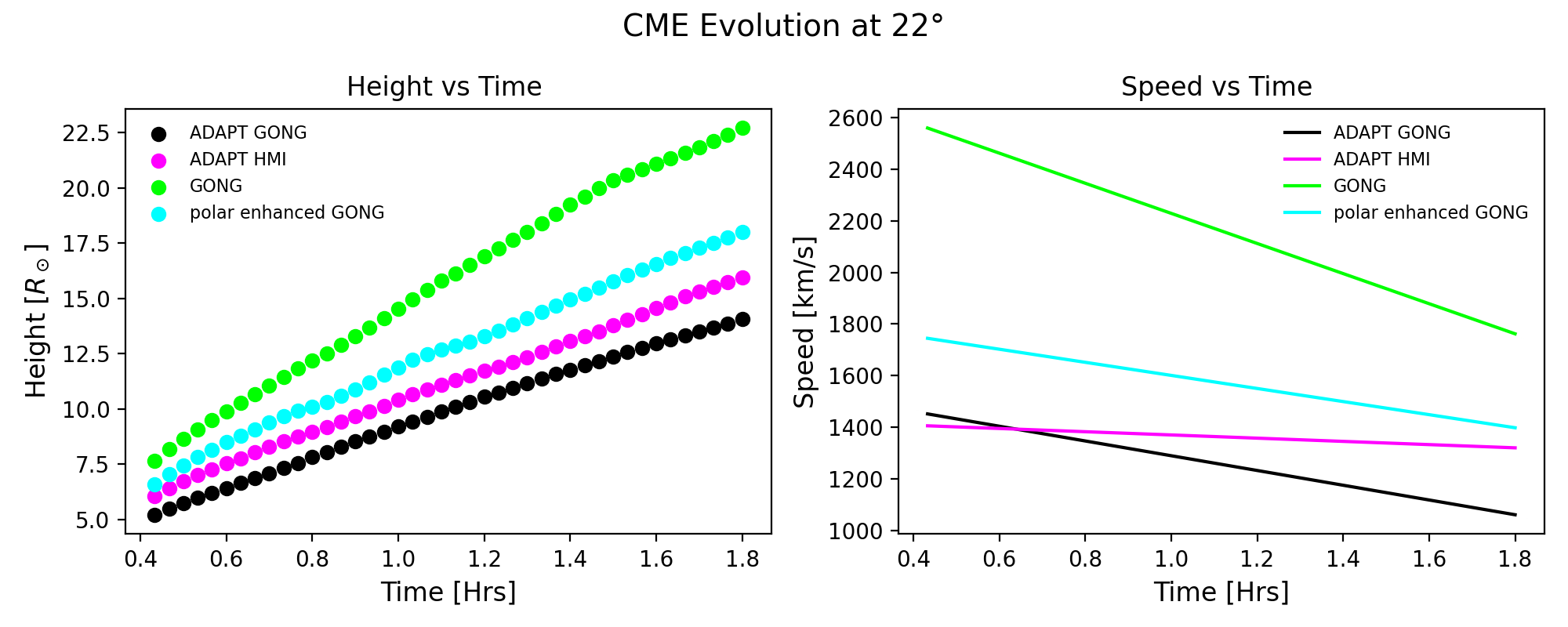}
    \caption{Summary of CME properties in the LASCO C3 field of view for the four cases. Left: Height \textit{versus} time plot for the CME leading edge extracted from the synthetic LASCO C3 coronagraph images.  Right: Speed (km/s) derived from the fitting a second-order polynomial to the height-time plot. Cases 1, 2, 3 and 4 correspond to plots in black, magenta, green and cyan colors, respectively.} \label{fig:Allmaps1}
\end{figure*}
\section{Results}\label{sec:Results}
First, we look at the input magnetic field maps utilized in this work to drive the AWSoM simulations. Figure~\ref{fig:Maps} shows the radial magnetic field ($B_{r}$) data for the ADAPT GONG, ADAPT HMI and GONG maps as available in the community. The polar regions in these maps show differences due to the limited observations, different data sources and methods used to create the maps, including enhancement of the polar and weak field regions of the GONG map in case 4 (polar enhanced GONG). The four input maps have a $360^\circ\times180\,^\circ$ resolution in longitude and latitude respectively, and are used to set the radial magnetic field as a boundary condition. We use the spherical harmonics solution to reconstruct the field \citep{Toth:2011}, and compare the resulting range of $B_{r}$ at 1 \Rs\ for the four cases in Table~\ref{tbl0}. It is worth noting that the polar enhanced map substantially overestimates the polar field strengths; however, we include it to illustrate the sensitivity of the modeled plasma and magnetic properties to such enhancements.
In the top row of Figure~\ref{fig:pfss_openclose_hcs}, we compare $B_{r}$ at 2.5\,\Rs\ calculated from the potential field source surface solution using the different input maps. For comparison, we calculate the total unsigned magnetic flux at 1\,\Rs, and at the source surface of 2.5\,\Rs for the four maps. This quantity is similar for cases 1 and 2, least for case 3 and largest for case 4. We also compute the dipole moment in the heliographic rotating (HGR) coordinate system. These quantities are listed in Table~\ref{tbl0} for the four solutions. The dipole moment for ADAPT GONG and ADAPT HMI (cases 1 and 2) indicate that these two maps share a similar global, lowest-order magnetic structure, despite differences in the data used (GONG \textit{versus} HMI). The GONG map (case 3) yields a much weaker dipole moment in comparison to case 1, where the same observational data (GONG) is used but processed through the ADAPT methodology. This highlights the sensitivity of the large-scale field to data assimilation. The polar enhanced GONG map (case 4) yields the largest dipole moment, resulting from the artificially strengthened polar fields and emphasizing the strong influence of polar flux on the magnetic heliospheric structure.

\begin{figure*}[t!]
\centering
\includegraphics[trim=10 105 10 65, clip, width=0.258\textwidth]{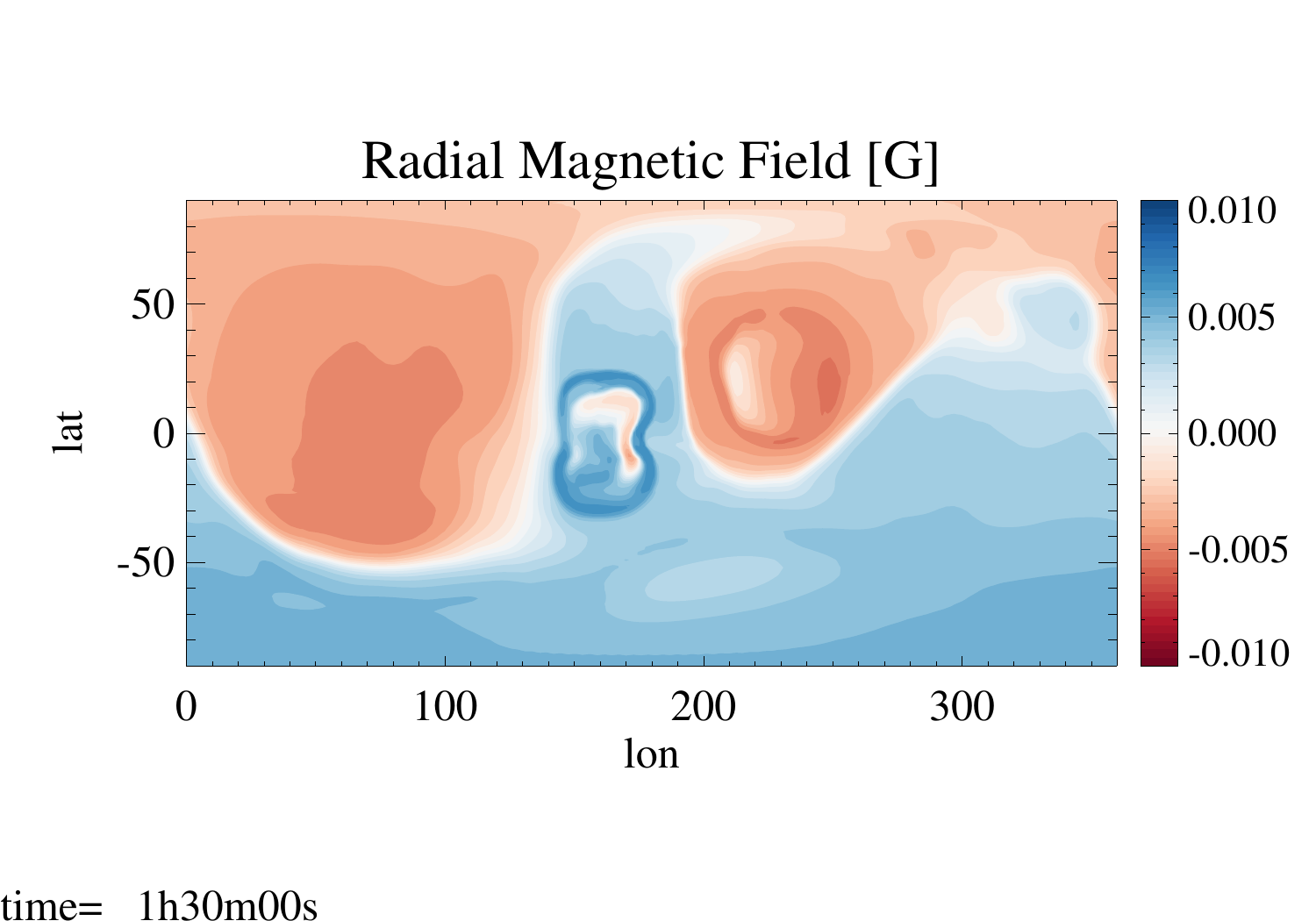}
\includegraphics[trim=60 105 10 65, clip, width=0.242\textwidth]{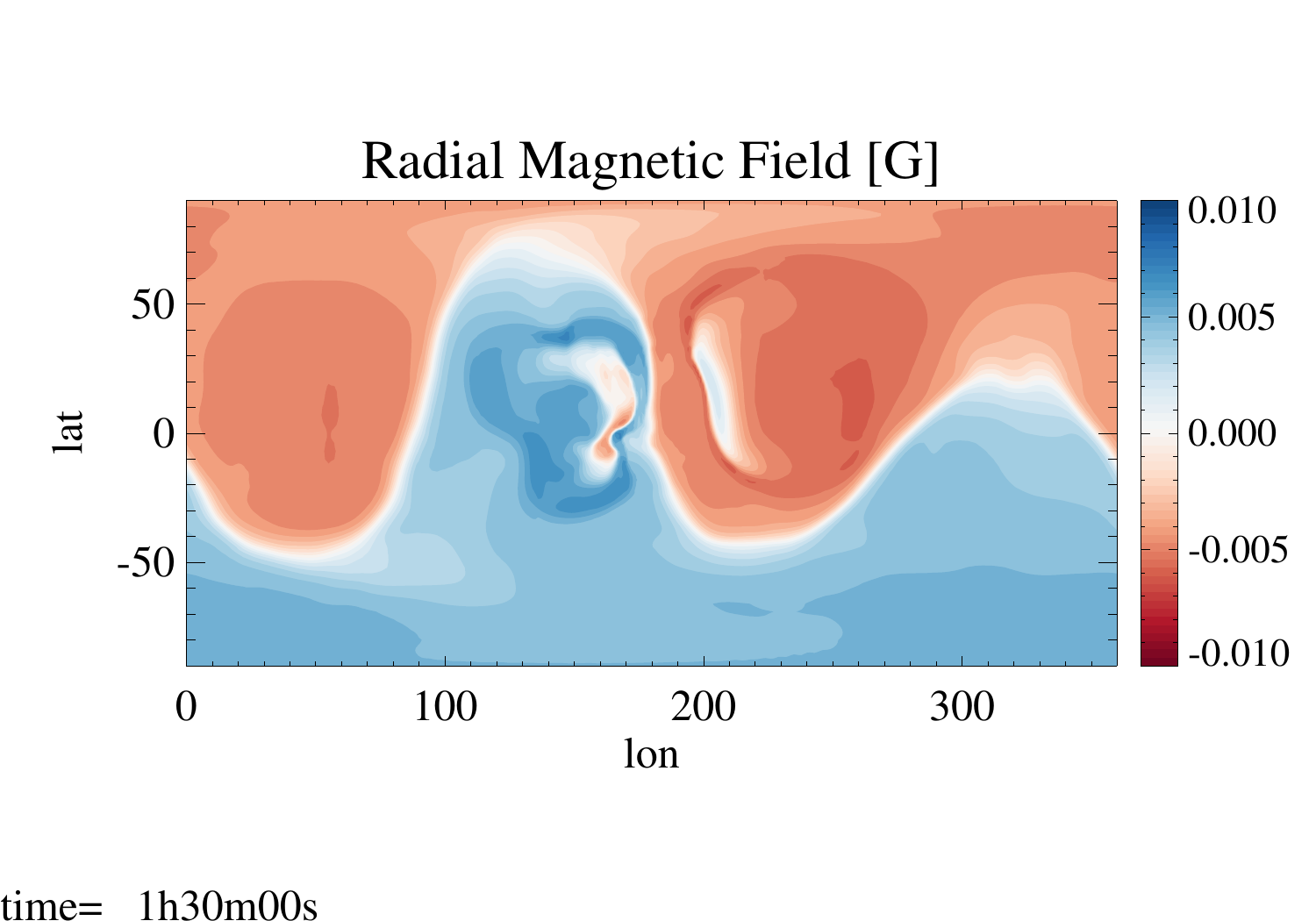}
\includegraphics[trim=60 105 10 65, clip, width=0.242\textwidth]{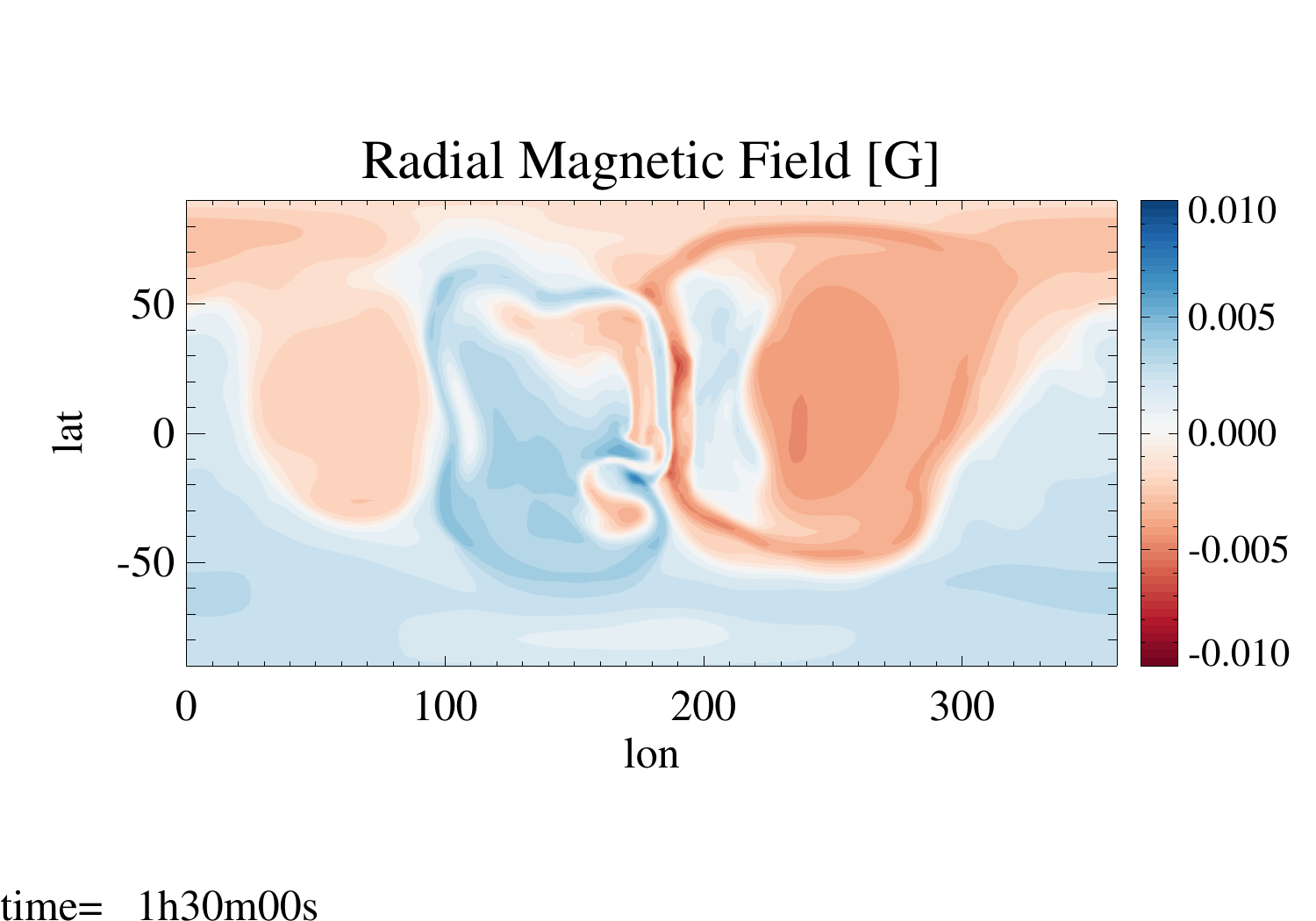}
\includegraphics[trim=60 105 10 65, clip, width=0.242\textwidth]{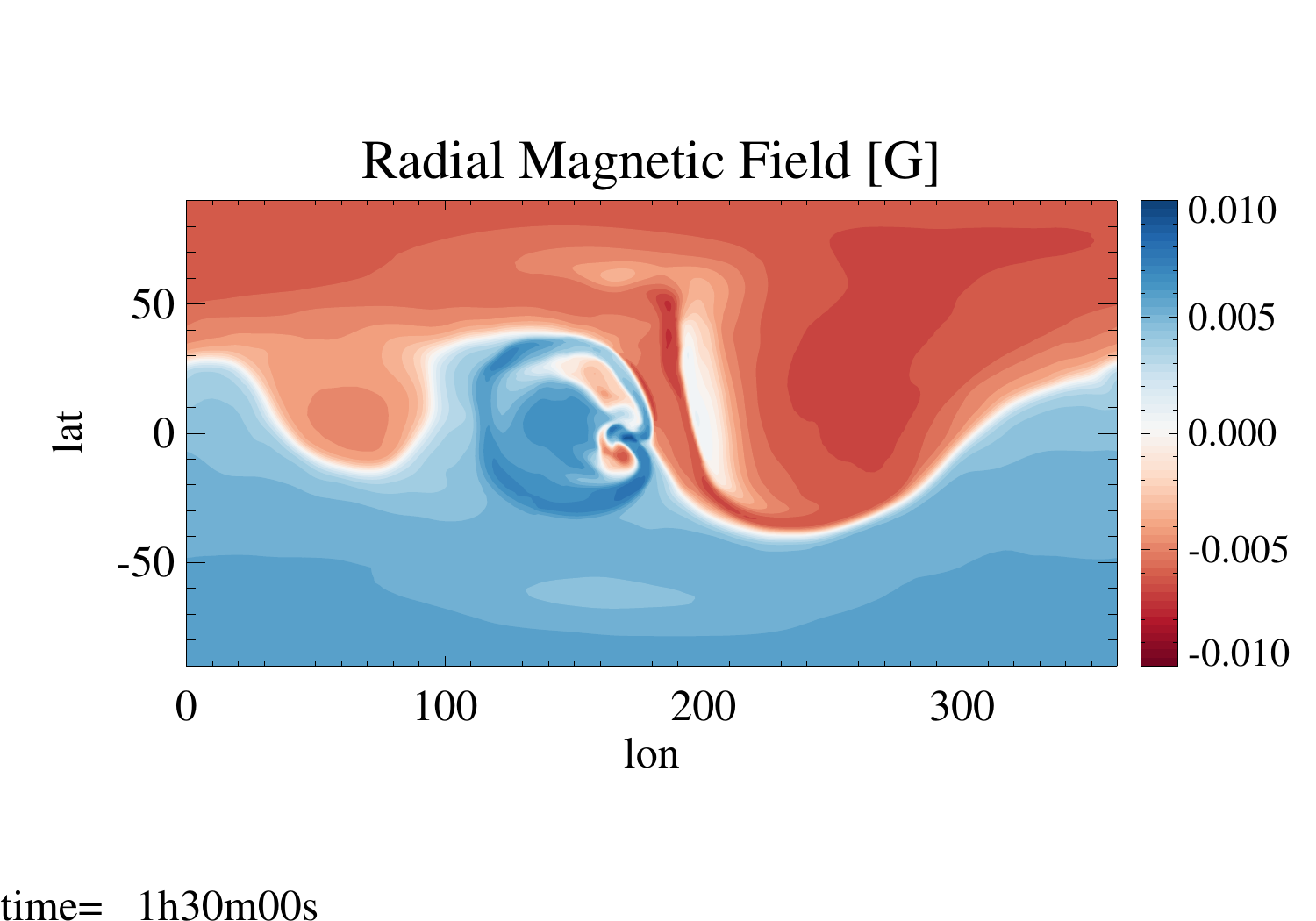}\\
\includegraphics[trim=10 50 10 65, clip, width=0.258\textwidth]{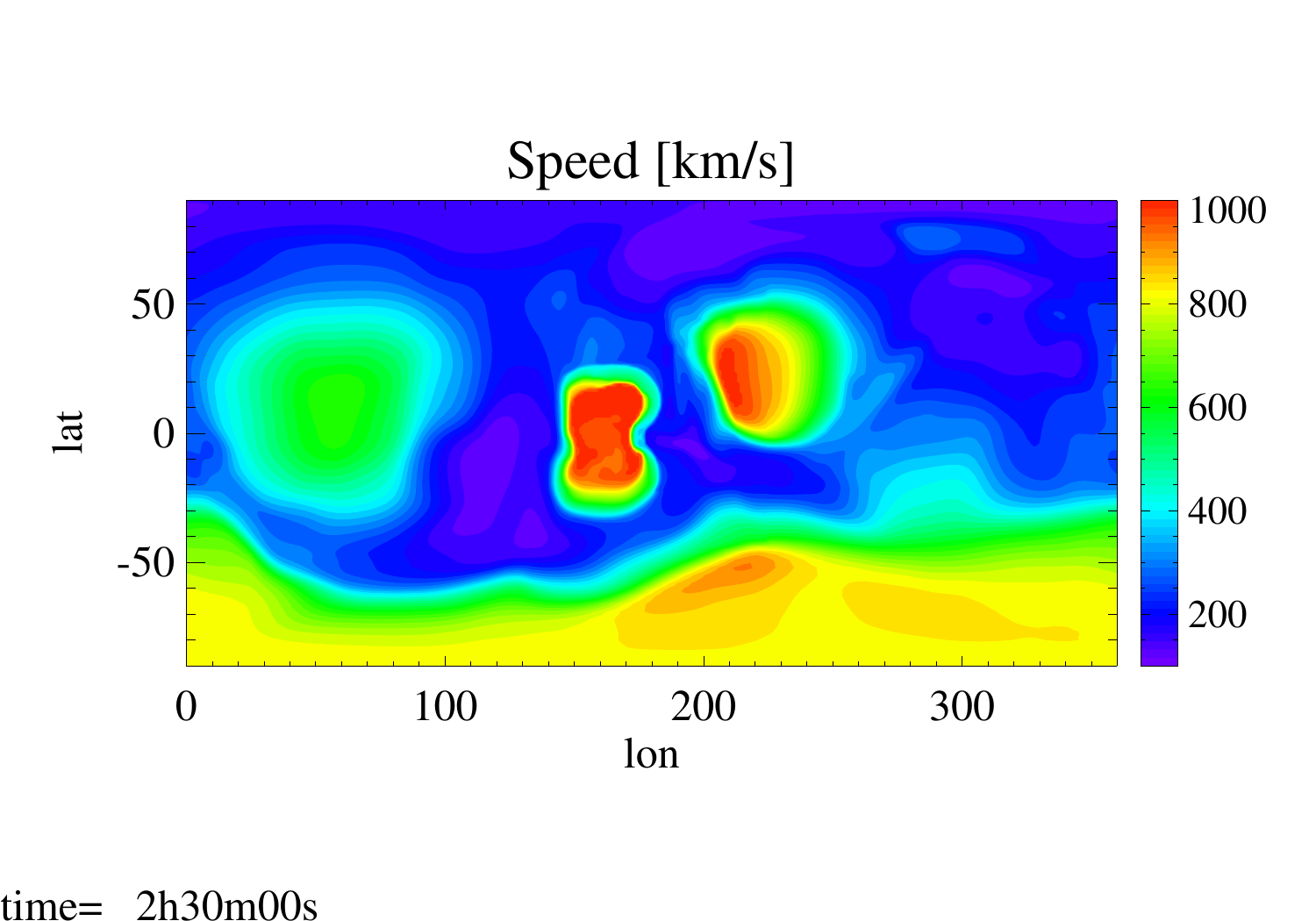}
\includegraphics[trim=60 50 10 65, clip, width=0.242\textwidth]{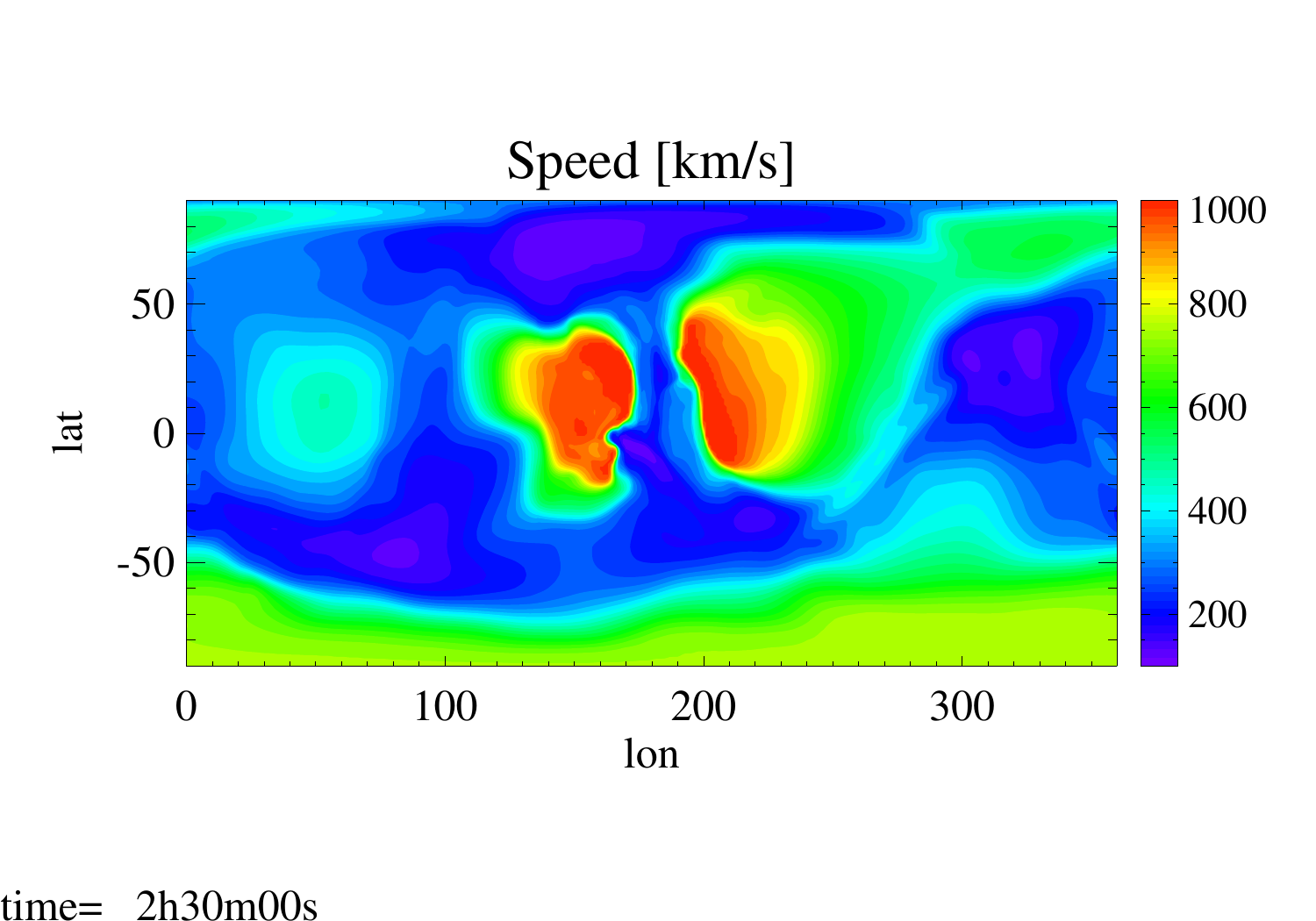}
\includegraphics[trim=60 50 10 65, clip, width=0.242\textwidth]{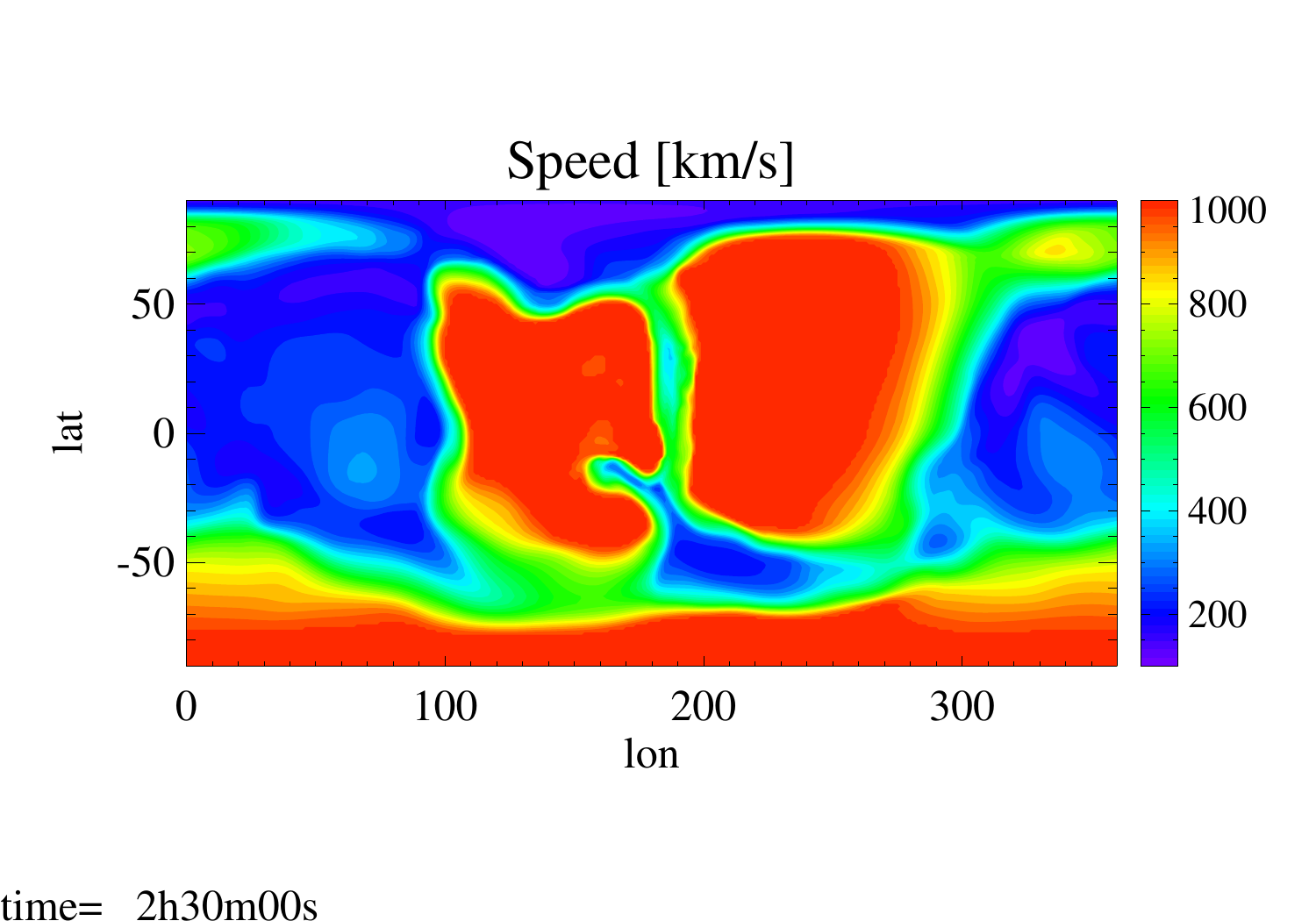}
\includegraphics[trim=60 50 10 65, clip, width=0.242\textwidth]{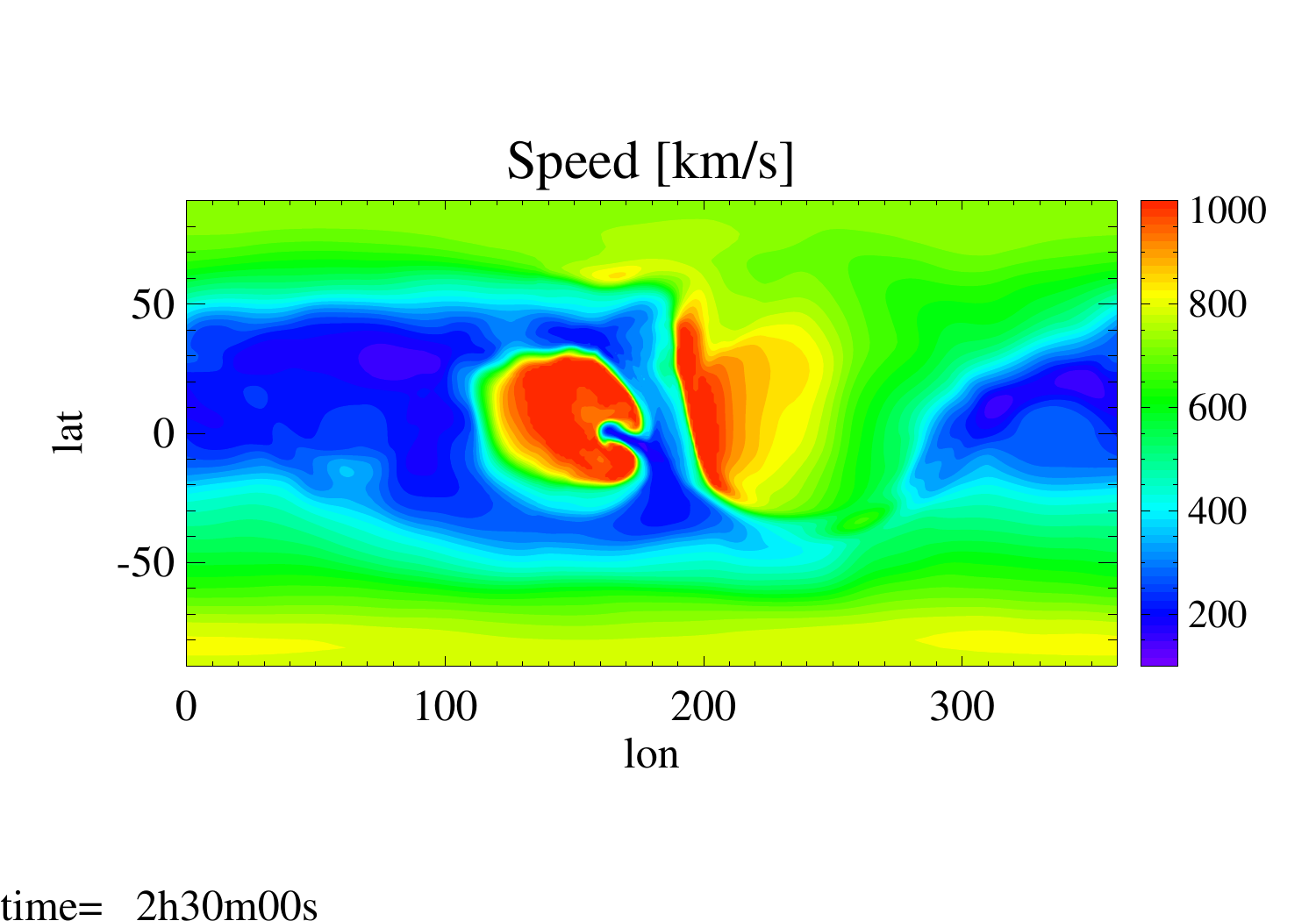} \\
\hspace{-0.5cm}(a) Case 1 \hspace{3.cm}(b) Case 2  \hspace{2.8cm}(c) Case 3 \hspace{2.8cm}(d) Case 4\\
\caption{Radial magnetic field and speed on longitude-latitude grid at different heliocentric distances and times during CME evolution. Top row: radial magnetic field (G) at 1 hour 30 minutes at radial distance of 10 \Rs. Bottom row: speed (km/s) at 2 hour 30 minutes at radial distance 15\,\Rs. 
}\label{fig:shl2d}
\end{figure*}
\subsection{Ambient Solar Wind Properties}
AWSoM provides the MHD properties of the solar wind in the 3D domain, that can be utilized to produce synthetic observables and study the global structure and evolution of the solar wind. For the four data-driven simulations, we compare the open and closed field regions at 1.01\,\Rs\ in the middle row of Figure~\ref{fig:pfss_openclose_hcs}, where open field regions are colored in deep blue. The northern polar region (as seen in the top row) differs quite significantly in the four solutions, as does the extent of the open field regions (coronal holes).  We also compare the heliospheric current sheet (HCS) in the bottom row of Figure \ref{fig:pfss_openclose_hcs}. The differences among the HCS structures are a direct consequence of the different magnetic field morphologies of the input maps used to drive AWSoM. The equatorial plane (z=0) is colored with the radial magnetic field. When comparing cases 3 and 4, we find a significant flattening of the current sheet in the polar enhanced GONG map (case 4) in comparison to the larger latitudinal extent of the HCS in the regular GONG map case. The flattening is a result of the enhancement of $B_{r}$ in the polar and weak field regions in case 4 (described in Section \ref{sec:Maps}). The heliospheric current sheet and the associated dense plasma sheet can strongly influence CME propagation.

We also synthesize the modeled EUV images from different spacecraft points of view for comparison. Top row in Figure~\ref{fig:los} shows the modeled EUV images in 193\,\AA\ from \textit{Solar Dynamics Observatory} (SDO, \citet{Pesnell:2012})/\textit{Atmospheric Imaging Assembly} (AIA, \citet{Lemen:2012}) point of view. 
The bottom row represents the synthetic line-of-sight EUV images from the \textit{Solar-Terrestrial Relations Observatory - Ahead} (STEREO-A, \citet{Howard:2008}) location at 195\,\AA\ for cases 1, 2, 3 and 4, respectively. We find that in the low corona, the location of the active regions and coronal holes agrees in all four cases, but their strength and extent differ. In the synthetic AIA 193\,\AA\ images, the coronal hole close to the bright central active region is found to vary in strength and size among the four cases. The coronal hole appears to be extended and darker for the GONG-driven case (column 3). Similarly, from the point of view of STEREO-A, we see that the overall brightness varies, along with differences in the size of the extended coronal hole (along the eastern limb) and the active region (in the south east). Coronal holes, and regions of strong magnetic fields, like active regions, can impact the direction of propagation and magnetic structure of CMEs erupting from a nearby active region.

Figure \ref{fig:SteadyState} shows the ambient solar wind plasma in the meridional plane when the GL flux rope is inserted at time t=0. This initial time configuration shows the properties of the solar wind that the CME will encounter as it propagates into the solar coronal domain. The proton temperature and density are shown in the top two rows on a log scale and the solar wind speed is shown in the bottom row. The scales are the same for all the cases to highlight the differences in the four solutions. 
Already the variation in the temperature, density, and speeds in the solutions driven by different maps is evident in this plane. In particular, in the -X direction (which is roughly the direction of CME propagation, as will be shown in the figures that follow) the solar wind speed ranges from a few hundred km/s in cases 1, and 2 to $\sim 600-800\,$km/s in cases 3 and 4. Similarly, the CME will encounter denser solar wind plasma in cases 2 and 4, and least dense in case 3. Next, we examine the CME properties under these four different data-driven solar wind backgrounds.
\begin{figure*}[th!]
\gridline{\fig{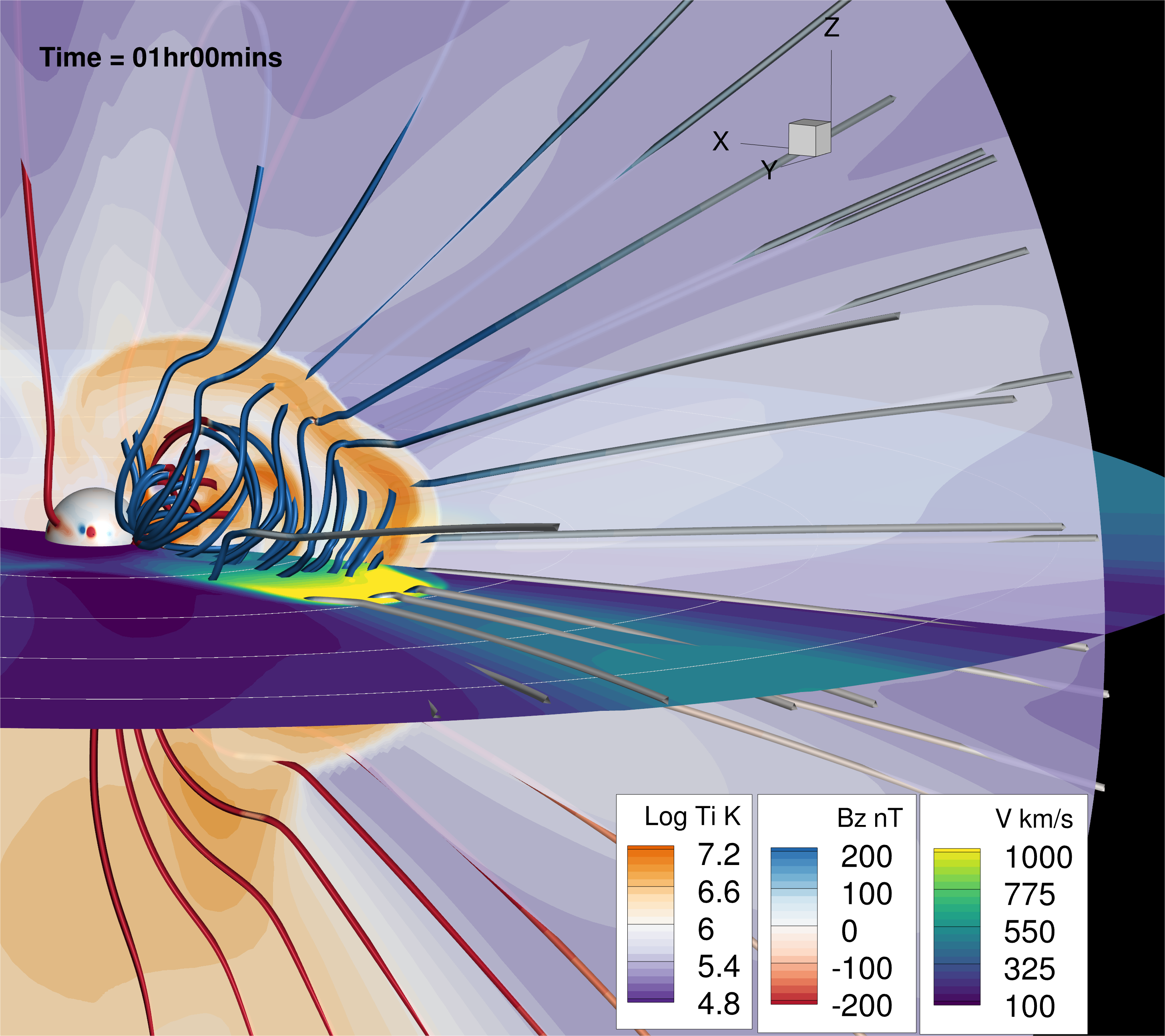}{0.25\textwidth}{}
         \fig{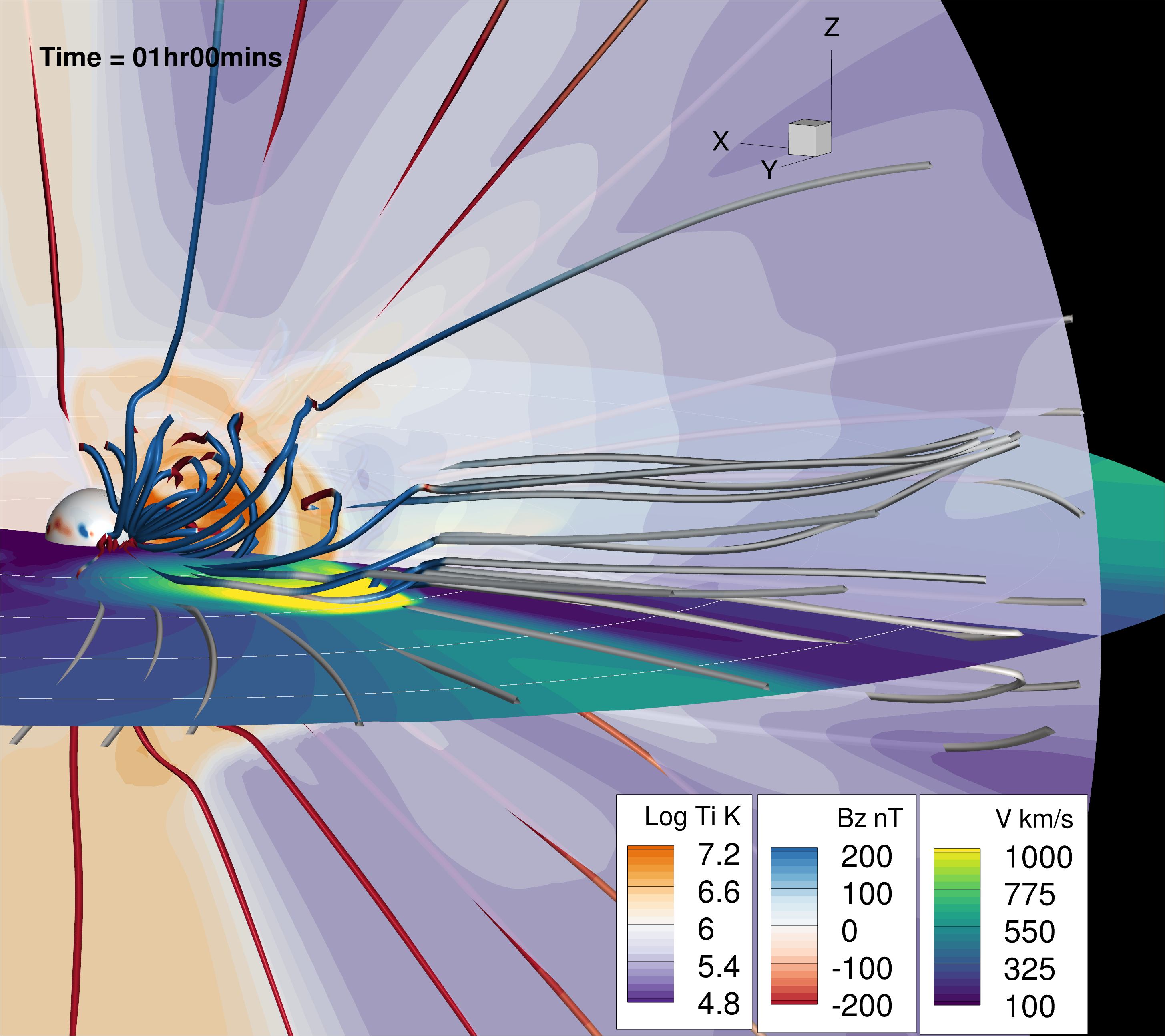}{0.25\textwidth}{}
         \fig{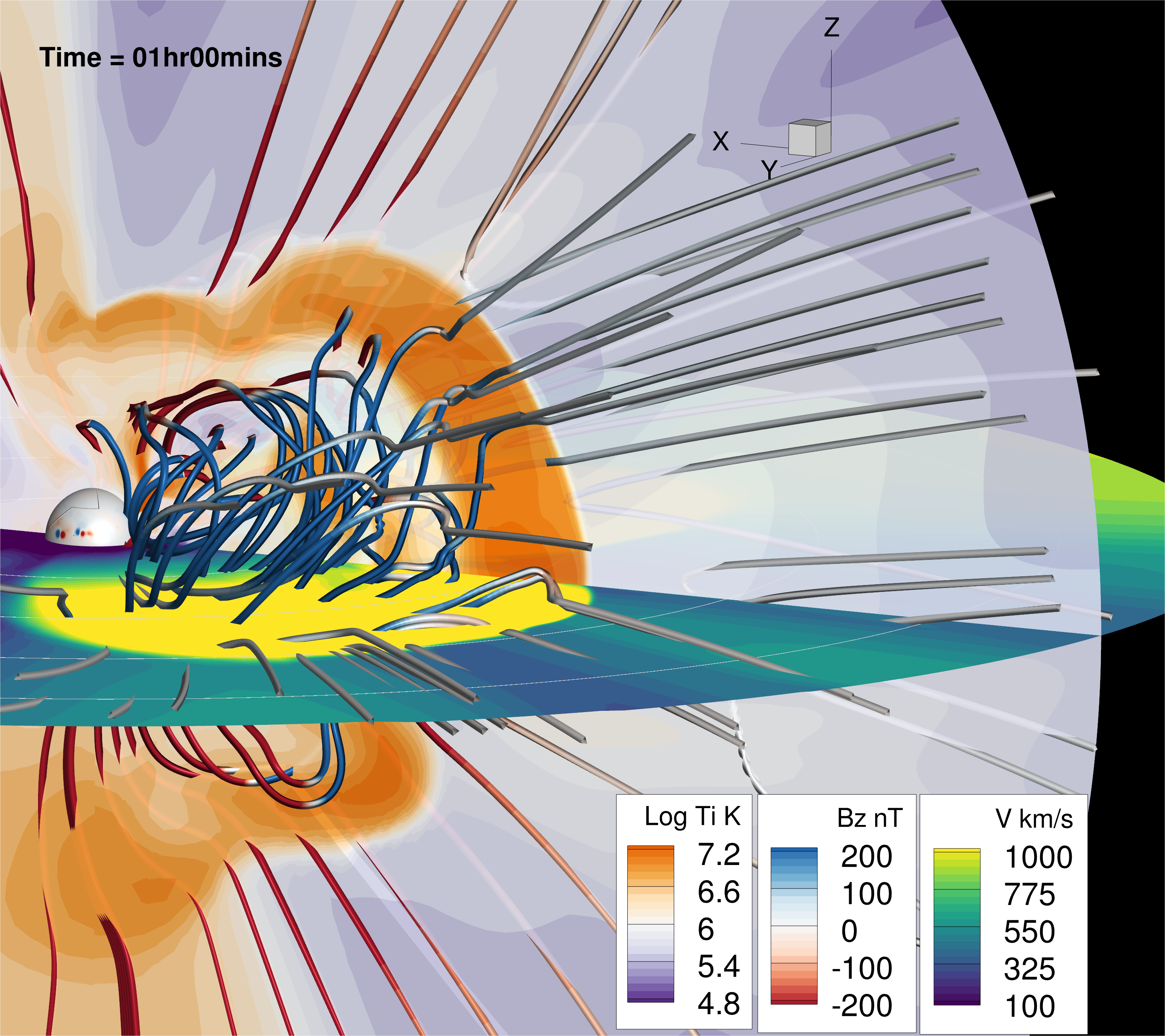}{0.25\textwidth}{}
         \fig{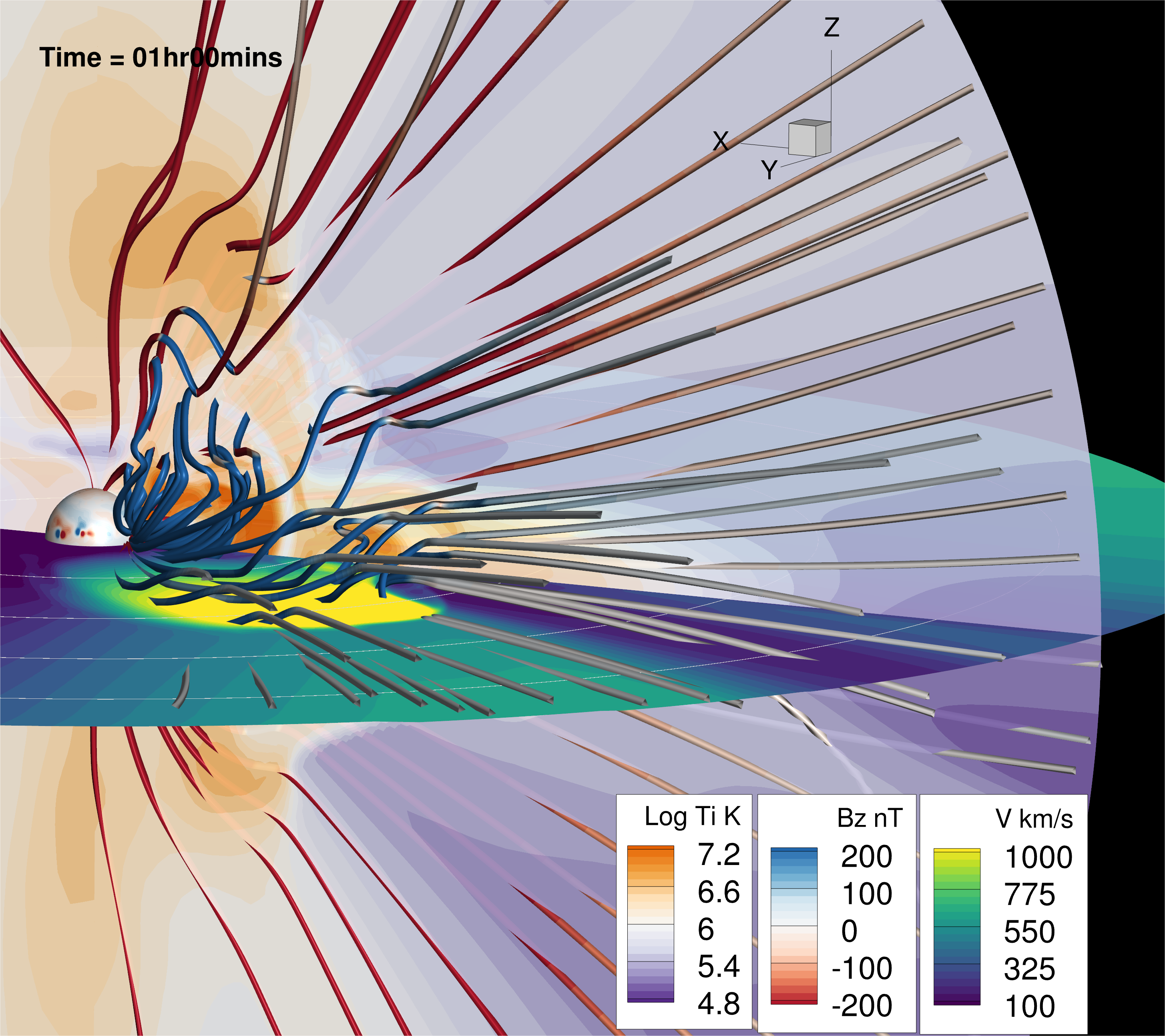}{0.25\textwidth}{}}\vspace{-0.8cm}
\gridline{\fig{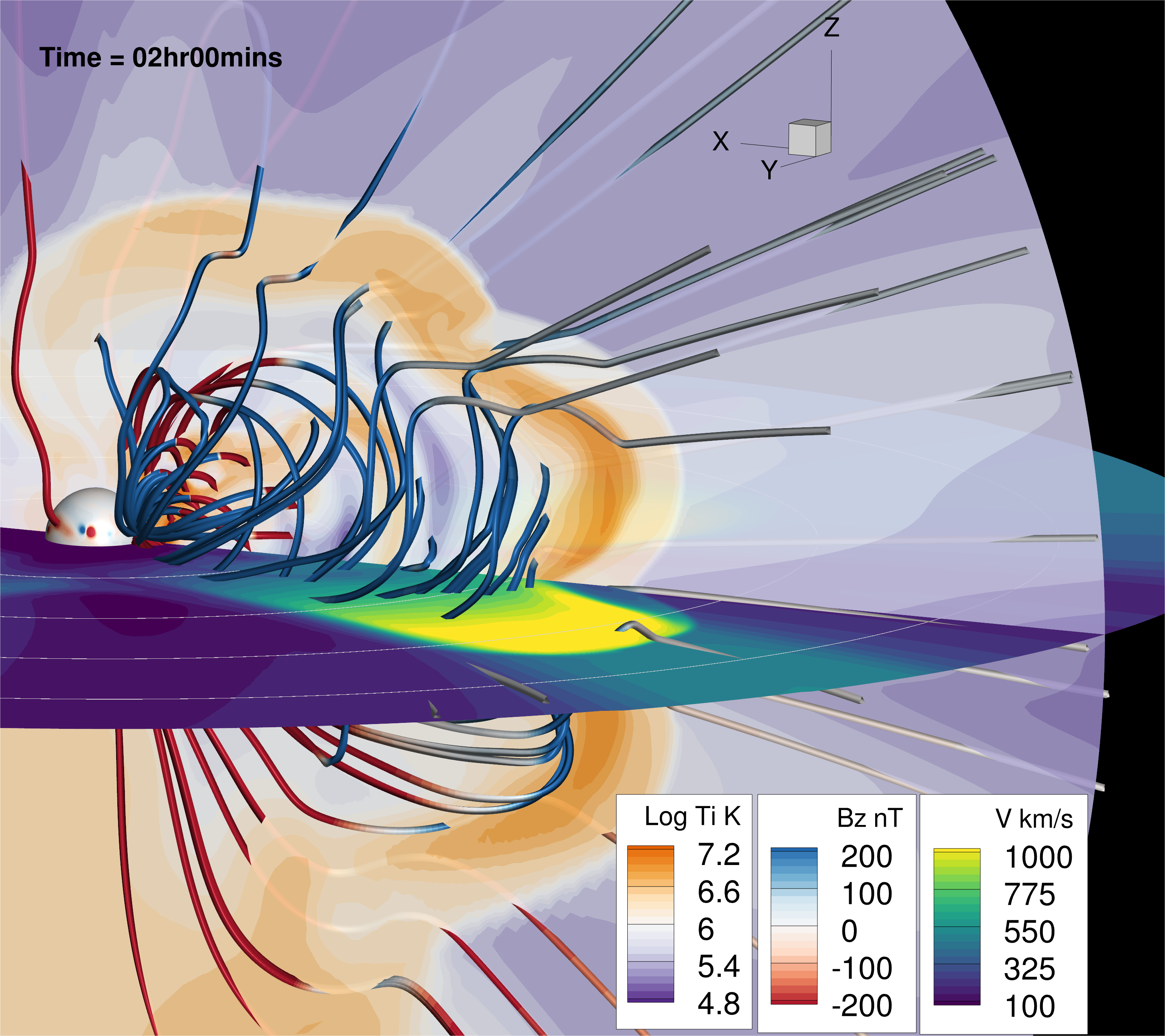}{0.25\textwidth}{}     
    \fig{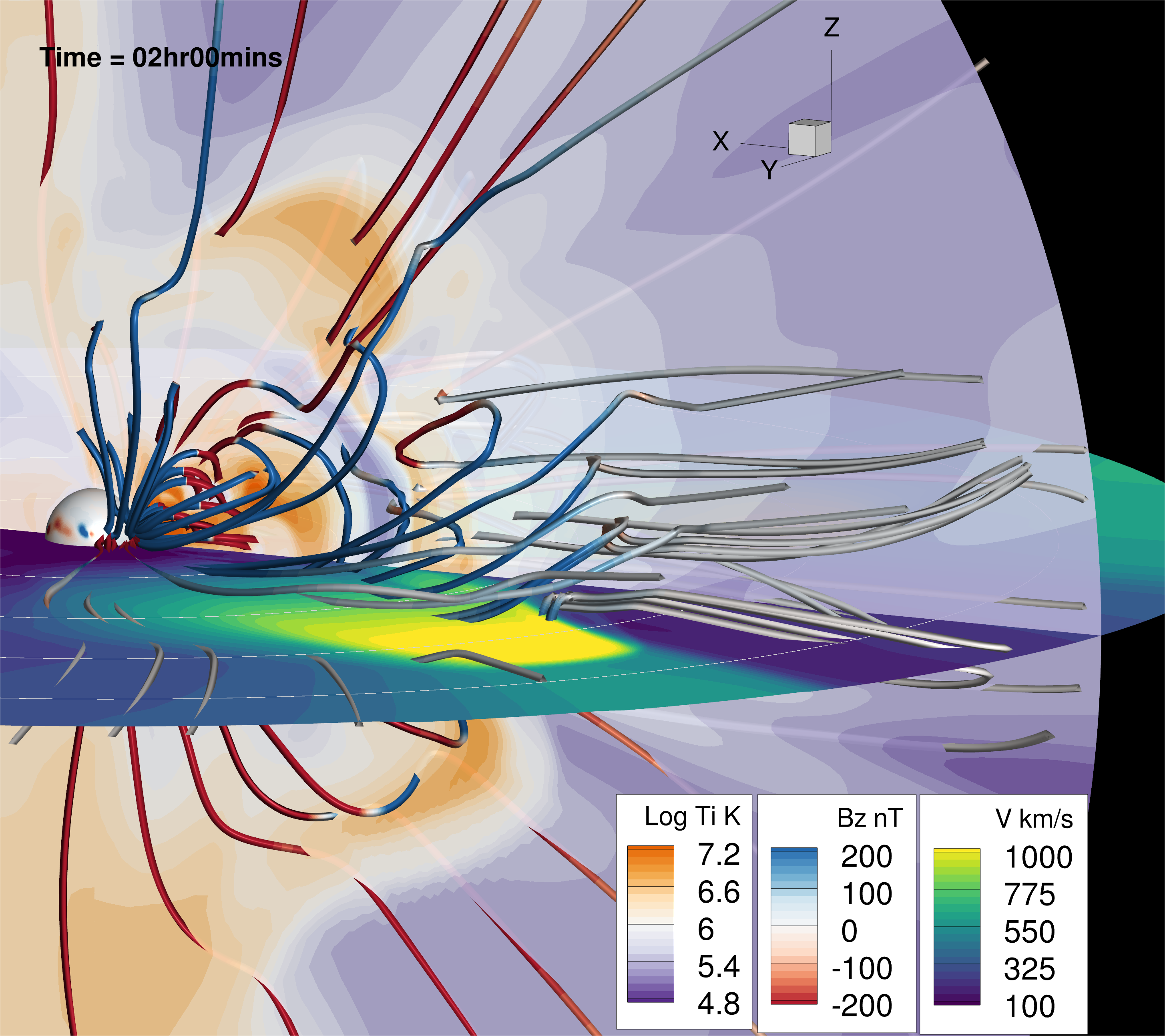}{0.25\textwidth}{}     
    \fig{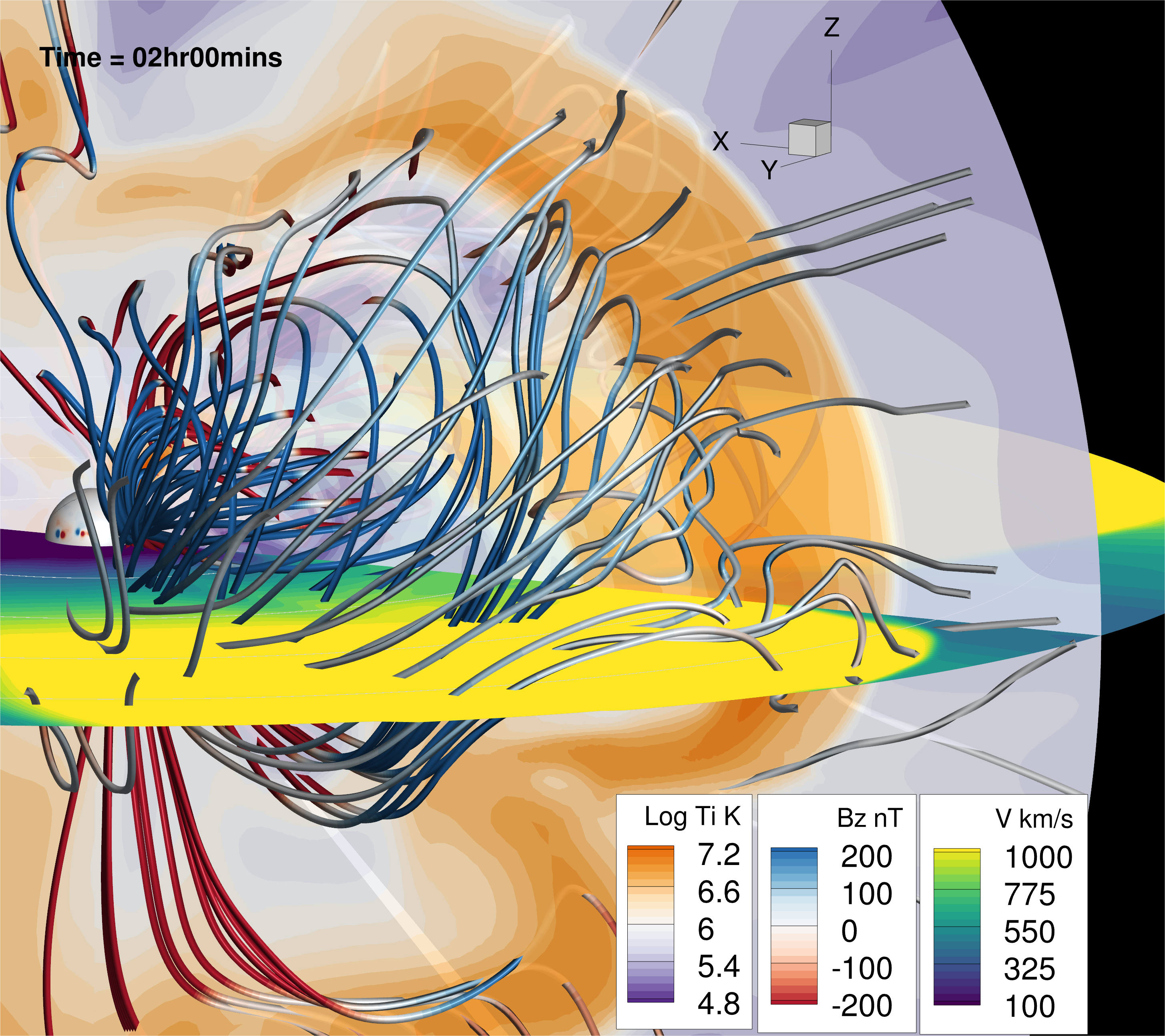}{0.25\textwidth}{}
    \fig{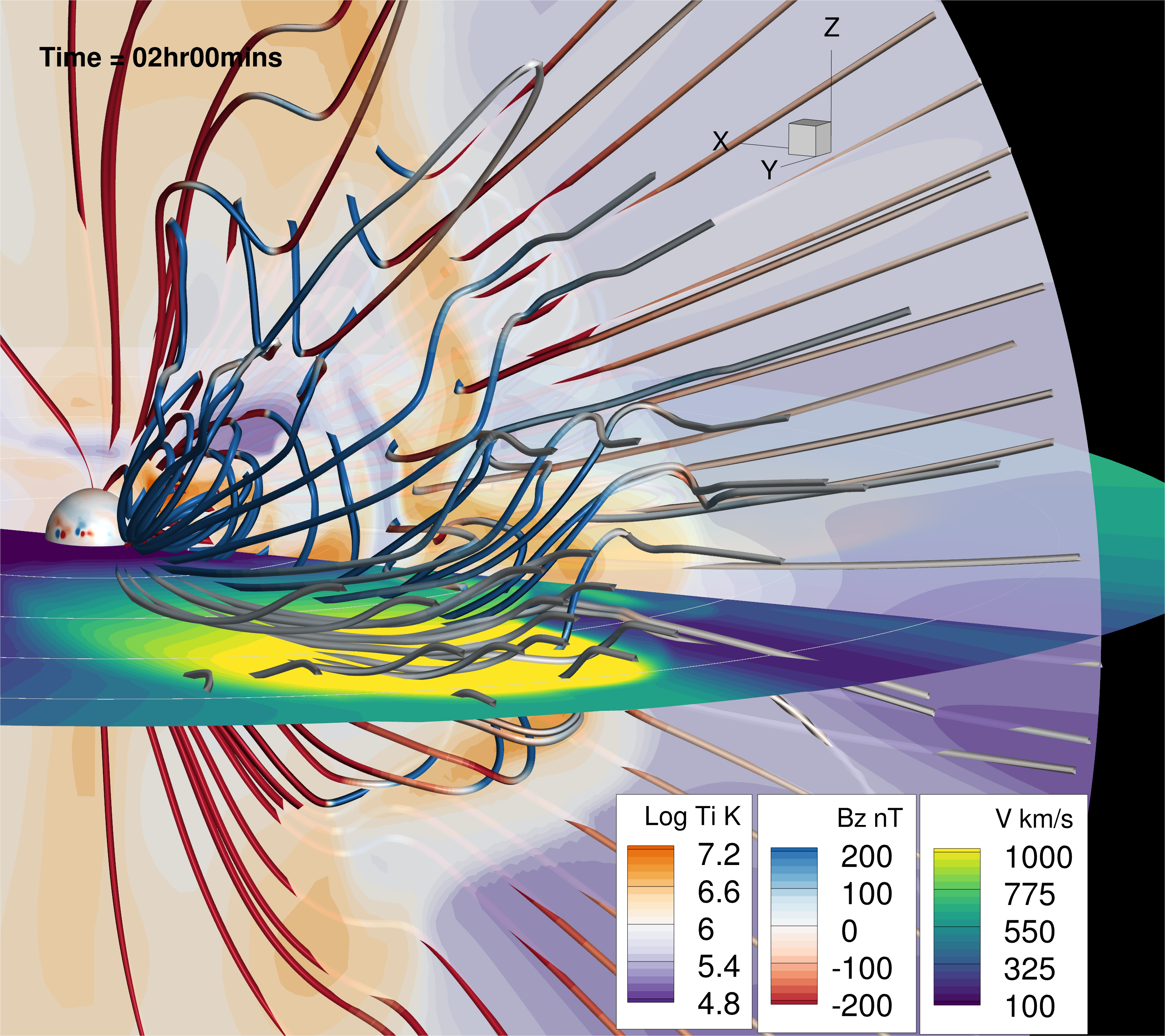}{0.25\textwidth}{}}\vspace{-0.8cm}
\gridline{\fig{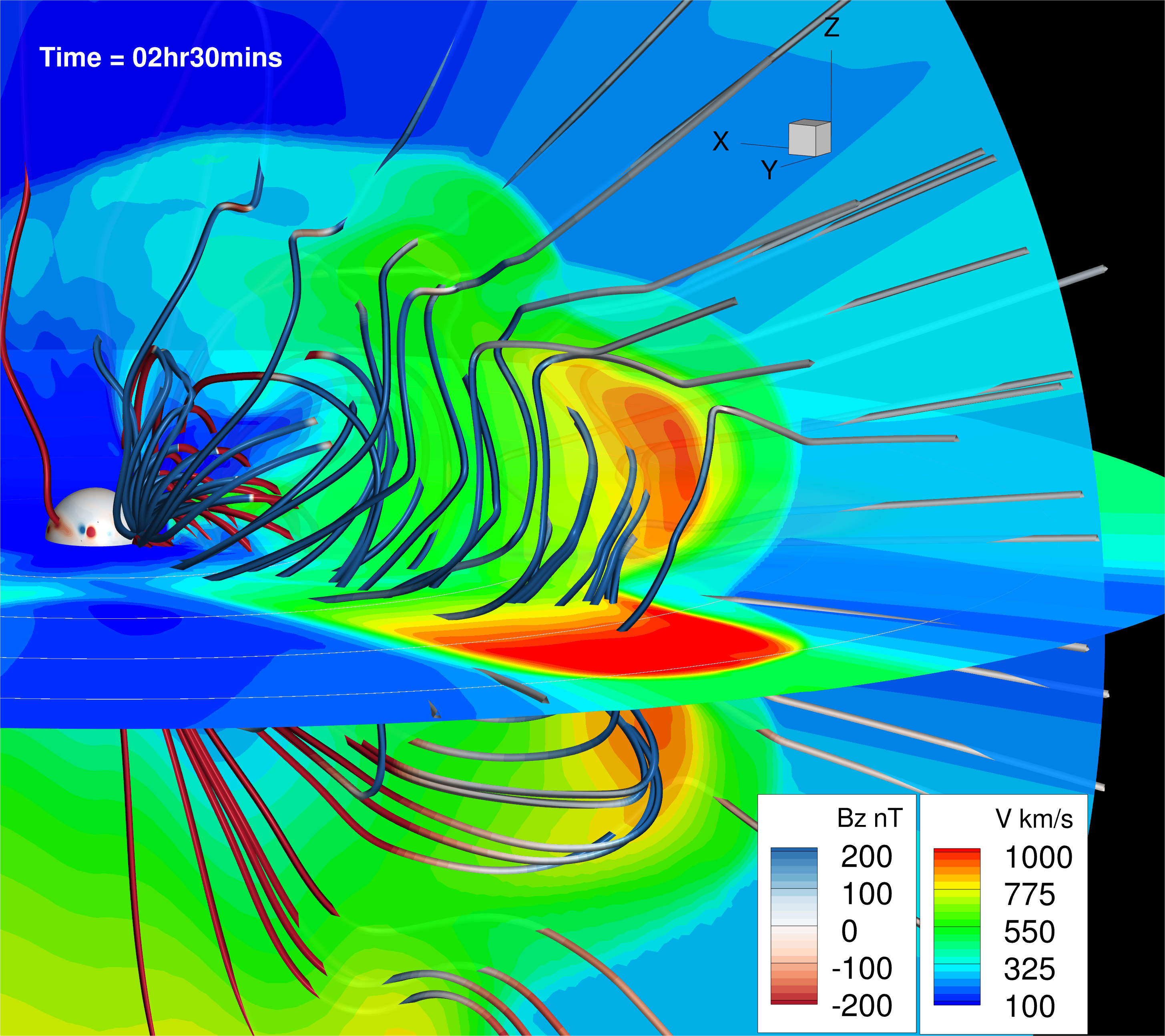}{0.25\textwidth}{\vspace{-0.2cm}(a) Case 1} 
         \fig{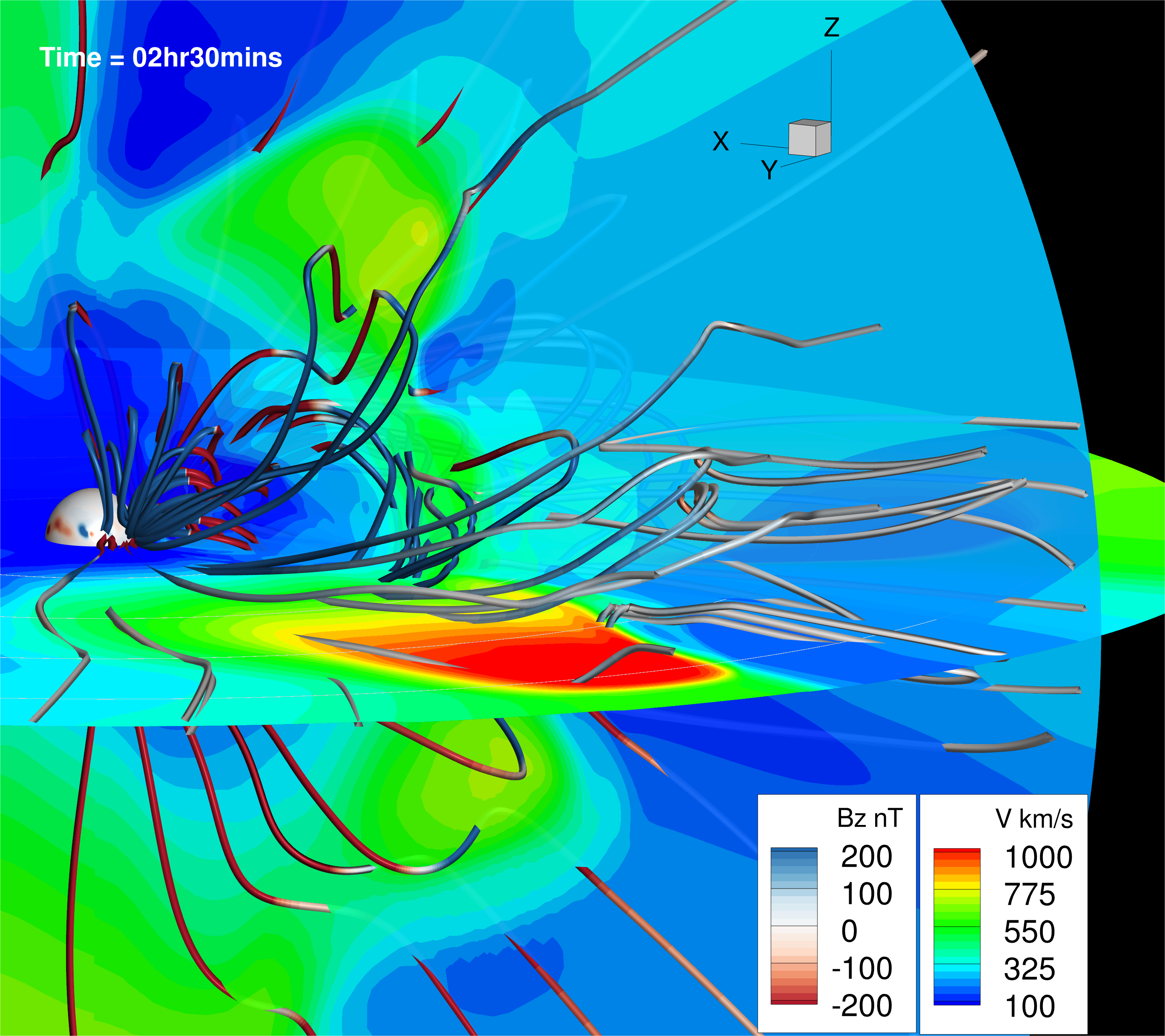}{0.25\textwidth}{\vspace{-0.2cm}(b)  Case 2}
         \fig{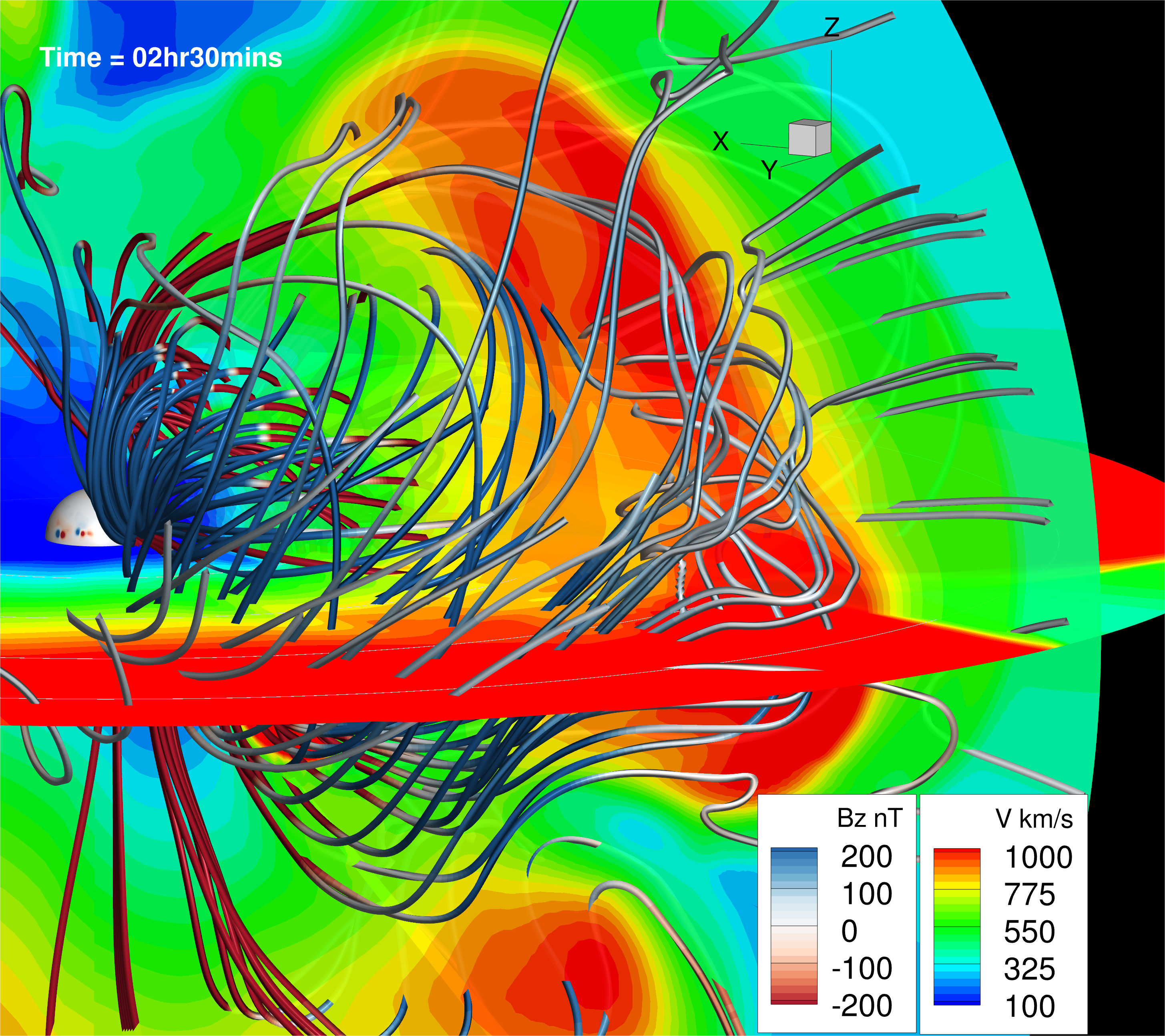}{0.25\textwidth}{\vspace{-0.2cm}(c) Case 3} 
         \fig{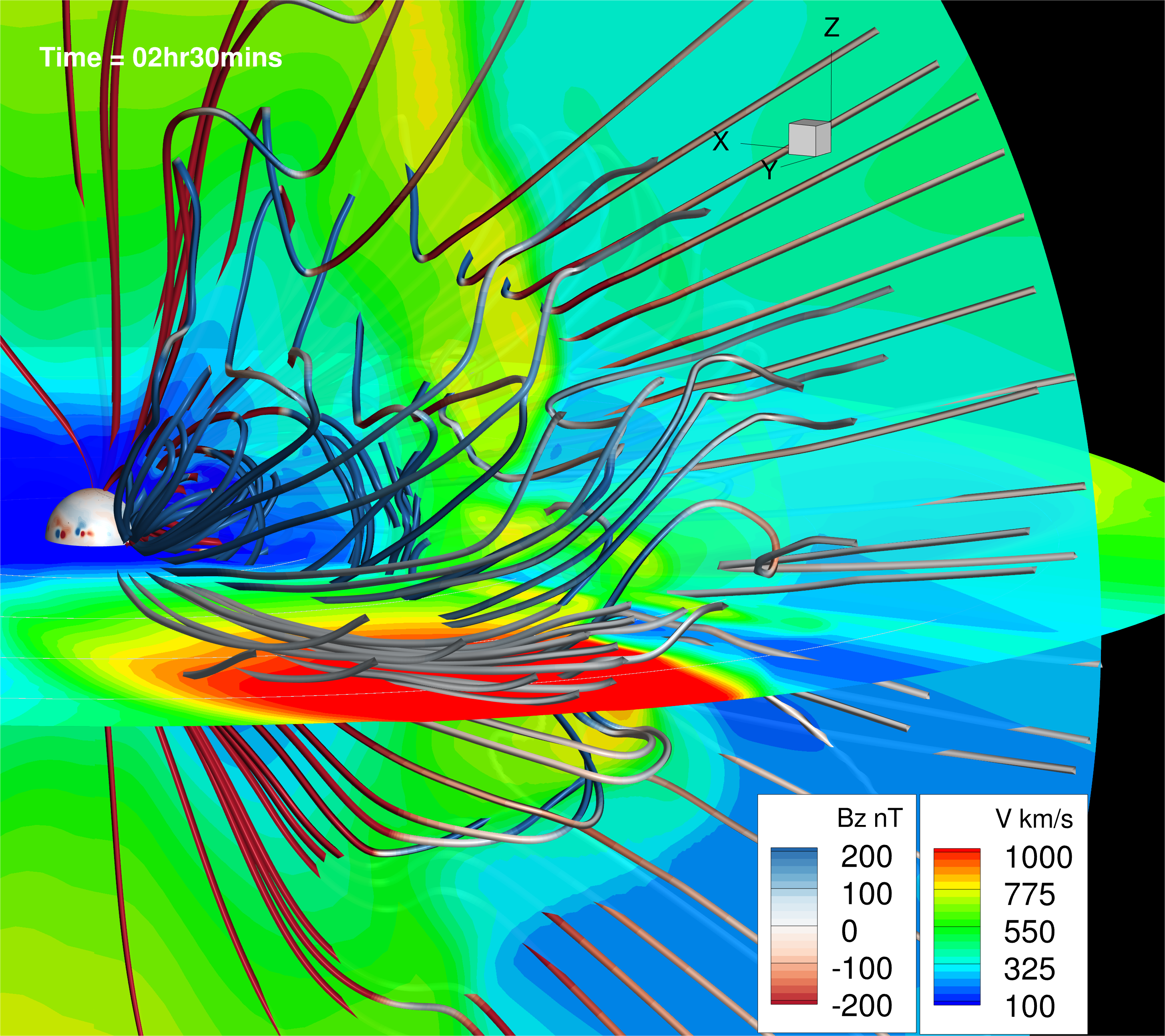}{0.25\textwidth}{\vspace{-0.2cm}(d)  Case 4}}
\caption{Evolution of the 3D magnetic flux rope CME at 1\,hour, 2\,hour and 2\,hour 30 minutes after eruption. The sphere at the left edge is at 1.05\,\Rs and the z=0 plane is colored with the speed. The magnetic field line structure shows the twisted flux rope extending out into the solar corona. Field lines are colored with the z-component of magnetic field (nT) to show the direction of the CME field. White concentric circles show the radial distances of 5, 10, 15 and 20 \Rs. Top two rows show the proton temperature (K, in log scale) on the semi-translucent meridional plane. Bottom row shows the speed profile in both y=0 and z=0 planes at time 2 hour 30 minutes.}\label{fig:3d_t=1h} 
\end{figure*}
\subsection{CME Propagation in the Ambient Solar Wind}
The out-of-equilibrium GL flux rope erupts immediately in the time-dependent simulation and propagates into the solar wind. Upon eruption, the flux rope expands and propagates in the solar corona interacting with the structured solar wind. Figure \ref{fig:y_cme} shows the solution in the meridional plane at 50 minutes after eruption. The proton temperature and density are on a log scale and clearly show the cold, dense plasma of the CME filament core (closer to the Sun), with a low density, cool filament cavity ahead of it and a hot ($>1$\,MK) and dense CME front/sheath. The third row shows the speed of the CME at this time, and the bottom row shows the CME speed at a later time of 2 hours. The CMEs in these simulations drive  MHD shocks ahead of them also evident in the density and temperature plots. The compressed and heated shocked front is followed by comparatively lower temperatures of the CME flux rope. In the speed plot (bottom row), the overlaying white lines show the magnetic field structure of the ambient solar wind and the twisted flux rope of the CME as it expands and propagates. The magnetic field lines are plot by integrating only the x- and z-components in this plane (y=0) and show bending while wrapping around the expanding CME flux rope indicating the shocked plasma structure ahead of the CME. Comparing the CME plasma properties 50 minutes after eruption show that the CME in case 3 expands the most, with a dense and hot CME front, in comparison to the other three cases.

The evolution of the CMEs in the four cases is subject to the varying plasma and magnetic field conditions of active region and the surrounding corona. A critical point to note is that the CME propagation strongly depends on the magnetic field of the active region, which contributes directly to the magnetic energy of the CME. Specifically, the initial background magnetic field of the active region ($B_{AR}$) and the flux rope magnetic field ($B_{GL}$) combine through the mutual term $B_{AR} \cdot B_{GL}$ to determine the magnetic energy of the CME. The strength of the AR in the four maps and the quadratic dependence of the magnetic energy on the field strength, lead to variations in the energy available to drive the eruption in the four cases, influencing CME propagation significantly. This is a direct consequence of the observed magnetic field map utilized for each case. Table~\ref{tbl2} lists the magnetic energy of the CME flux rope at the time of insertion (t=0) with the contribution from the ambient solar wind in the domain (before the insertion of the flux rope) excluded.

\begin{figure*}
    \includegraphics[trim=45 10 0 6, clip,width=0.45\textwidth]{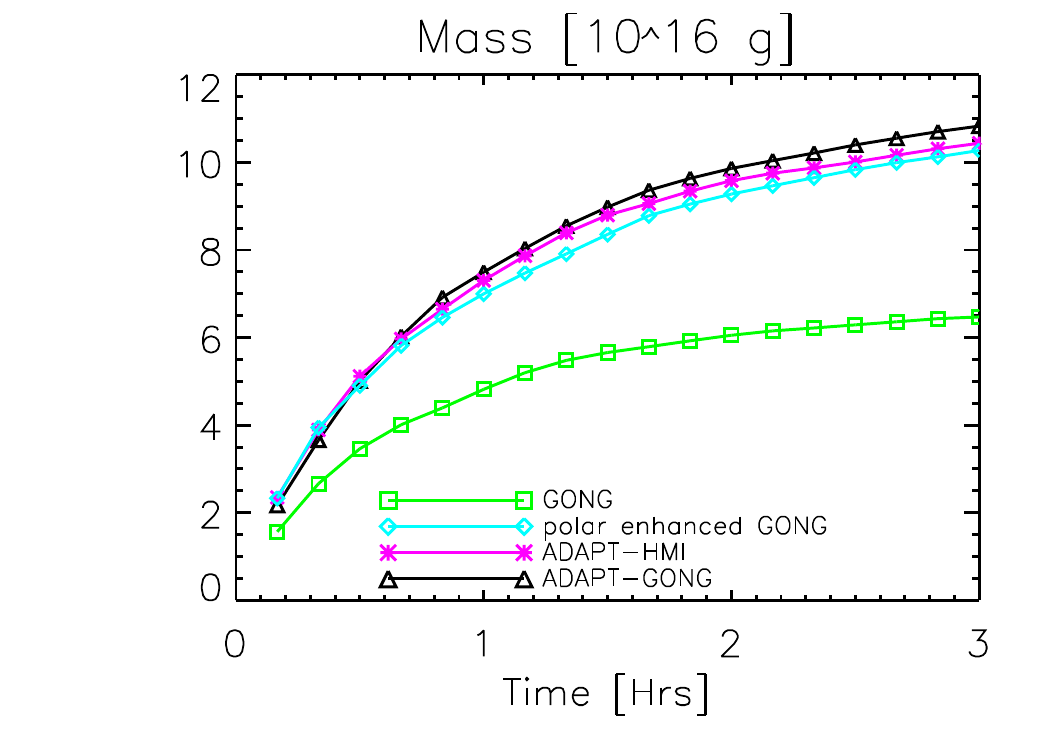}\hspace{-0.2cm}
    \includegraphics[trim=45 10 0 6, clip,width=0.45\textwidth]{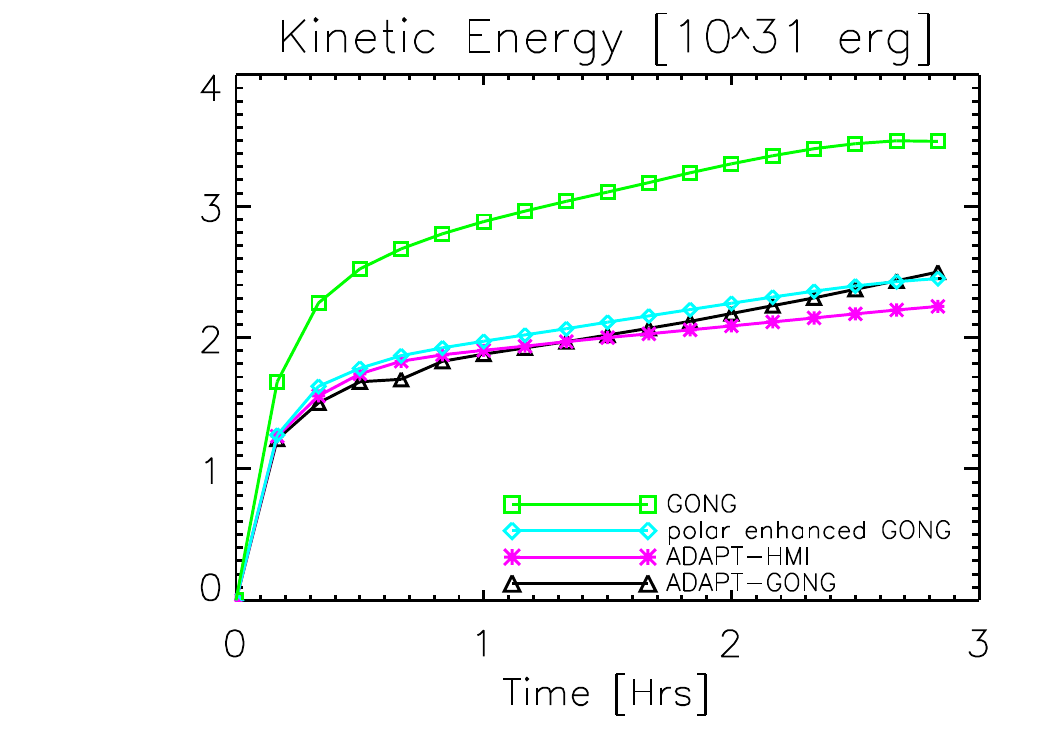}
    \caption{Left: CME Mass ($10^{16}\,$g), and Right: Kinetic energy ($10^{31}\,$erg). Color scheme is the same as Figure~\ref{fig:Allmaps1}.}\label{fig:Allmaps2}
\end{figure*}

In Figure~\ref{fig:y_cme}, the CME in case 3 travels faster ($> 1000\,kms^{-1}$) and reaches $\sim 20$ \Rs\, within about 2 hours, in comparison to the CME traveling in the ADAPT HMI map driven solution (case 2) that only extends up to about $\sim 10$ \Rs\, in this plane at this time. In case 2 (ADAPT HMI), the ambient solar wind plasma ahead of the CME is relatively denser and slower compared to case 3 (GONG). This causes the case 2 CME to encounter more resistance leading to less expansion and slower CME speed. In contrast, for case 3 the faster and less dense solar wind allows the CME to propagate fast. This variation in the background wind results in different speeds and expansion profiles of the CMEs for the four cases driven by different magnetograms. The CME expansion also dictates the temperature profile of the CME plasma. For case 3, as the flux rope expands, the CME plasma cools down much faster and to relatively lower temperatures in the cavity in comparison to the other three cases. The CME speeds (in the y=0 plane) are slowest for the ADAPT HMI case (preceded by the dense, slow solar wind), and faster for the ADAPT GONG and GONG map cases. 

\subsubsection{Synthetic Coronagraph images}
AWSoM provides synthetic white-light coronagraph images representative of the SOHO/LASCO C2, C3, STEREO-A/B COR1 and COR 2 coronagraphs that we use to capture the CME evolution. For brevity, we show the model results in Figure~\ref{fig:WL} corresponding to the STEREO-B COR 1 and LASCO C3 fields of view at 10 minutes and 1 hour 10 minutes, respectively. These are representative of the white-light coronagraph images that can be utilized for a direct comparison with observations from the aforementioned satellites. The LASCO C3 images show a halo CME structure of a CME as expected from an eruption from near the central meridian. At this time, STEREO B was at 117.8$^{\circ}$ from Earth. The CMEs in these instrument domains for the four cases also show the difference in evolution from different points of view that are usually available for studying the CME propagation via observations. In particular, from the STEREO-B COR 1 point of view, the CME in case 3 is fastest and expands the most. \citep{Chen:2025} described a method of tracking the leading edge of the CME structure in synthetic white-light coronagraph images. We utilize this methodology to obtain the height \textit{versus} time evolution of the leading edge of the CME in the synthetic LASCO C3 images (Left panel of Figure~\ref{fig:Allmaps1}).
\subsubsection{CME Kinematics and Energetics}
Momentum coupling between the solar wind and the CME determines the relative drag that either slows down a fast CME or speeds up a slow CME traveling in the ambient plasma. In addition, the amount of magnetic energy available to convert to kinetic and thermal energy determines the propagation speed and direction of the CME. The right panel of Figure \ref{fig:Allmaps1} shows the speed of the CME leading edge in the LASCO C3 field of view derived by fitting  a second-order polynomial to the extracted height \textit{versus} time points. We also calculate the deceleration for each case (listed in Table~\ref{tbl2}). In the 3D domain, we calculate the angle between the radial vector (from the active region) and the momentum vector for the four CMEs to quantify the deflection of the CME from the radial direction (of the selected erupting active region). The momentum vector is chosen because CMEs expand non-uniformly as they propagate, with some portions of the CME traveling faster than others. We compare the average deflection over 2 hours of evolution and list the values in Table \ref{tbl2}. The average deflection is largest for CMEs in case 1 and 4, while cases 2 and 3 show the least deflection. In case 4 (polar enhanced GONG map) the propagation of the CME towards the heliospheric current sheet is likely the cause of this change in direction while the fastest CME in case 3 shows smallest deviation from the radial trajectory. 

A major advantage of global MHD simulations is the ability to visualize and study CME properties at any location in the 3D domain. While observation are limited by the field of views of the available instruments, in the simulation we can examine the synthetic output at various heliocentric distances and at any longitudinal/latitudinal positions. 
This is analogous to placing virtual satellites at different locations in the computational domain. To demonstrate this, we plot the radial magnetic field, and speed on a longitude-latitude grid at two different radial distances and times highlighting the differences in the CME structures in the four solutions in Figure \ref{fig:shl2d}. Top row in Figure \ref{fig:shl2d} shows the radial magnetic field at a distance of 10\, \Rs at 1 hour 30 minutes of solution time, and the bottom row shows the speed at 15\,\Rs at 2 hour 30  minutes after eruption.
The figure shows the differences in both the extent and strength of the CMEs as they evolve in different solar wind conditions. We can see the variation in the magnetic field strength at 10\,\Rs, while the speed plot shows the difference in speeds as well as lateral and longitudinal expansion of the CMEs in the four solutions at this time and distance. In other words, these longitude-latitude plots are like slices of the 3D structure at different radial distances. 
To appreciate the 3D evolution of the CMEs, Figure \ref{fig:3d_t=1h} shows the 3D structure of the magnetic flux rope at 1 hour (top), 2 hours (middle) and 2 hours 30 minutes (bottom) after eruption in the four cases. The z=0 plane is colored with speed, and the white concentric circles show radial heliospheric distances in this plane (5, 10, 15, and 20 \Rs) indicating the radial extent of the CMEs in the four cases at different times. The 3D magnetic field lines are colored with the z-component of the magnetic field (in nT) showing the complete magnetic morphology of the twisted flux ropes and the open field lines represent the background magnetic field. The top two rows show the proton temperature on a log scale in the meridional plane demonstrating the expanding CME structure with a hot plasma front that cools as the CME expands in the solar wind. The bottom row shows speed in both planes highlighting the speeds attained by the CMEs in the four cases. 
\begin{table*}[t!]
\begin{center}
\caption{CME properties for the four cases.}\label{tbl2}
\begin{tabular}{|c|c|c|c|c|c|c|}
\hline
\multicolumn{2}{|c|}{} & \multicolumn{2}{c|}{At t = 0} & \multicolumn{1}{c|}{} & \multicolumn{2}{c|}{At t = 2 hr}\\
\hline
Case & Input & Initial Magnetic & Initial Mass & {Deceleration} & Loss in        & Mean \\
     &  Map  &  Energy          &              &                   & Magnetic Energy&  Deflection \\
\hline
     &       & [$10^{32}$\,erg] & [$10^{15} g$] & [m/s$^2$]        &   [$\%$]     & [degrees] \\
\hline
1 & ADAPT GONG & 3.1 & 1.7 & -79.4 & 26 & 9 \\
\hline
2 & ADAPT HMI  & 3.4 & 1.8 & -17.3 & 24 & 6 \\
\hline
3 & GONG       & 2.4 & 1.5 & -162.0 & 37 & 5 \\
\hline
4 & polar enhanced GONG & 2.6 & 1.7 & -70.4 & 28 & 11 \\
\hline
\end{tabular}
\end{center}
\end{table*}

It should be noted that we do not model CME initiation or energy buildup, instead utilize analytical flux ropes to launch a CME. The added magnetic energy of the flux rope converts to kinetic, thermal and gravitational energy that expands, heats and accelerates the CME. The magnetic energy of the system decreases, while the kinetic and thermal energy increase initially, followed by a decrease due to the CME expansion and deceleration \citep{Manchester:2004} and the magnetic pressure is responsible for driving the flux rope out. 
We list the initial magnetic energy and mass added with the insertion of the GL flux-rope in the four cases in Table \ref{tbl2}. The mass added is similar, while the differences in the magnetic energy (of the flux rope) is a representation of the difference in the magnetic strength of the active region, and the surrounding corona as described previously. In the 3D domain, we isolate the CME structure by identifying the regions of density enhancement over the background steady-state plasma environment. We plot the mass of the CMEs in the left panel of Figure \ref{fig:Allmaps2} which shows the mass pile up as the CME sweeps the plasma from the surrounding solar wind \citep{Lugaz:2005a}. As the CME in case 3 propagates into a relatively sparser plasma background, the mass pile up is less compared to the other three cases in which the mass increase is similar. The right panel in Figure~\ref{fig:Allmaps2} shows the kinetic energy profiles for the four cases. Table \ref{tbl2} also lists the decrease in magnetic energy over the first two hours of CME evolution from initial time. The case 3 CME shows maximum loss in magnetic energy as it propagates with higher speeds in the coronal domain. 

\section{Conclusions and Discussion\label{sec:conc}}
This work demonstrates how the properties of the solar wind and CMEs propagating in the solar wind vary, studied using a global solar coronal model driven by different magnetic field conditions. We use four magnetic field maps for the same time period that differ in either the observation source or the methodology used to produce synoptic/synchronic maps. We utilize ADAPT GONG, ADAPT HMI, GONG synoptic and polar enhanced GONG maps for CR2123 as described in section \ref{sec:Maps}. These four maps are used to drive the 3D global solar wind model AWSoM to obtain four ambient solar wind solutions while all free parameters in the model are kept the same. We highlight the differences in the maps and the ambient solar wind properties in the four cases by comparing the open-closed field regions, total unsigned flux, synthetic EUV images, and the magnetic and plasma structure of the solar wind solutions. We find that the solar wind solutions differ significantly in the global structure of the HCS, and the strength and brightness of coronal holes and active regions. The ambient plasma density, speed and temperature differ between the four cases providing varying background conditions into which CMEs will propagate. Next, we inserted the same Gibson-Low flux rope into the four solar wind solutions to launch a CME, and compared the evolution properties in the four cases. It should be noted that while the active region where the CME was launched from was selected because it is central and strong, the CME does not correspond to an observed event. Instead, our goal is to show that if there was a CME erupting from this active region, then the CME evolution would differ quite noticeably based on the magnetic field map used. The magnetic field of the active region obtained via the input maps also contributes to the free magnetic energy of the CME that impacts its kinematics. Therefore, the input maps directly influence the properties of CME propagation.


We analyzed the properties of the CMEs in the four cases and discuss the differences in the propagation direction, speed, density, thermodynamic and magnetic structures as the CMEs evolve and interact with the solar wind background. The CME propagating in the GONG driven solar wind solution (case 3) shows least amount of mass pile up as it encounters a less dense solar wind plasma ahead of it, and achieves faster propagation speeds. The case 4 CME on the other hand shows largest deflection from the radial direction of propagation.
From the 3D simulated solutions, we synthesize white-light coronagraph images for the four cases, and compare their height-time evolution in the LASCO C3 field of view using the edge detection technique developed by \cite{Chen:2025}. The CME mass, kinematics and energies are also compared to both similarities and differences in the propagation characteristics in the four cases. The 3D solution provides an opportunity to study the CME evolution at different locations and directions not covered by observational field of views. This also suggests the critical need of multi-viewpoint observations, and that CMEs do not evolve uniformly while interacting with the structured solar wind that can cause deflection, deformation and non-uniform expansion, as evident from the cases studied in this work. Although these are modeled CMEs erupting from the same active region, it can be seen that the evolution can vary from an above-average speed event in case 4 (driven by the polar enhanced GONG map) to a fast CME in case 3 (driven by a GONG map) (see, figure \ref{fig:Allmaps1}). This strongly suggests that the influence of the magnetogram data driving MHD simulations is significant, and it is crucial to consider and understand their impact. This is particularly important for operational and predictive efforts for space missions.

We also point out that in this study we focused on varying only the input magnetic field maps to analyze their influence on the MHD model simulation of the solar wind and CME evolution in the coronal domain. However, there are other parameters associated with the MHD model that can influence these results. The Alfv\'en wave energy density and dissipation length scale can impact the thermodynamics and magnetic structure of the corona. The parameters associated with the CME flux rope model, specially magnetic field strength and helicity can also substantially impact the properties of CME evolution. A subset of these have been studied in \cite{Jivani:2023, Chen:2025} to identify the most important model parameters that impact the ambient solar wind and CME solutions. However, here we do not vary multiple parameters simultaneously and only change the input driving condition by using different magnetic field maps. In addition, other flux rope models used for modeling CME eruptions can also influence the CME dynamics. In this article, we have only analyzed the first few hours of CME evolution to show how the early propagation properties differ in these four cases. Future work will continue the CME out into the inner heliosphere to study their evolution, arrival and impact at 1 au, representative of CMEs arriving at Earth.


\begin{acknowledgments}
This work is primarily supported by NASA LWS grant 80NSSC24K1104. N. Sachdeva is also supported by NASA LWS SC 80NSSC22K0892. H. Chen is supported by NSF grant PHY-2027555 (SWQU). Z. Huang is supported by the NSF Solar Terrestrial grant No. 2323303, and the NASA grants No. 80NSSC23K0450.
\end{acknowledgments}





%


\bibliography{csem,ref}{}
\bibliographystyle{aasjournalv7}
\end{document}